\documentclass[11pt,twoside,letterpaper]{article} 
\usepackage{times,fancyhdr}
\usepackage[dvips]{graphicx}


\makeatletter
\def\cleardoublepage{\clearpage\if@twoside \ifodd\c@page\else%
    \hbox{}%
    \thispagestyle{empty}%
    \newpage%
    \if@twocolumn\hbox{}\newpage\fi\fi\fi}
\makeatother

\def\be{\begin{equation}}
\def\ee{\end{equation}}
\def\bea{\begin{eqnarray}}
\def\eea{\end{eqnarray}}
\def\beaN{\begin{eqnarray*}}
\def\eeaN{\end{eqnarray*}}
\def\ed{\end{document}}
\def\bit{\begin{itemize}}
\def\eit{\end{itemize}}

\def\sig{\sigma}
\def\Sig{\Sigma}
\def\lam{\lambda}

\def\Del{\Delta}
\def\del{\delta}
\def\Bg{\Bar g}
\def\hg{\hat g}
\def\k{\kappa}
\def\alf{\alpha}
\def\ga{\gamma}
\def\Ga{\Gamma}
\def\BD{\Bar D}

\def\di{\partial}

\def\Lix{\pounds_\xi}

\def\half{{\textstyle{1 \over 2}}}
\def\~{\tilde}
\def\lag{{\hat{\cal L}}}
\def\m{\label}
\def\l{\left}
\def\r{\right}

\def\goto{\rightarrow}
\def\Bar{\overline}
\def\const{\rm const}

\begin{document}
\title{
{\begin{flushleft}
\vskip 0.45in
{\normalsize\bfseries\textit{Chapter~1}}
\end{flushleft}
\vskip 0.45in \bfseries\scshape Nonlinear Perturbations and
Conservation Laws on Curved Backgrounds in GR and other Metric
Theories}}
\author{\bfseries\itshape A.N. Petrov\thanks{E-mail address: anpetrov@rol.ru;
Telephone number: +7 (495) 7315222.}\\
Relativistic Astrophysics group, Sternberg Astronomical institute,\\
Universitetskii pr., 13, Moscow, 119992, RUSSIA}
\date{}
\maketitle
\thispagestyle{empty}
\setcounter{page}{1}
\thispagestyle{fancy}
\fancyhead{}
\fancyhead[L]{In: Classical and Quantum Gravity Research \\
Editor: Mikkel N. Christiansen\\ and Tobias K. Rasmussen, pp. {\thepage-\pageref{lastpage-01}}} 
\fancyhead[R]{ISBN 978-1-60456-366-5  \\
\copyright~2008 Nova Science Publishers, Inc.} \fancyfoot{}
\renewcommand{\headrulewidth}{0pt}

\begin{abstract}
In this paper we review the field-theoretical approach. In this
framework perturbations in general relativity as well as in an
arbitrary $D$-dimensional metric theory are described and studied. A
background, on which the perturbations propagate, is a solution
(arbitrary) of the theory. Lagrangian for perturbations is defined,
and field equations for perturbations are derived from the
variational principle. These equations are exact, equivalent to the
equations in the standard formulation and have a form that permits
an easy and natural expansion to an arbitrary order. Being
covariant, the field-theoretical description is also invariant under
gauge (inner) transformations, which can be presented both in exact
and approximate forms. Following the usual field-theoretical
prescriptions, conserved quantities for perturbations are
constructed. Conserved currents are expressed through divergences of
superpotentials --- antisymmetric tensor densities. This form allows
to relate a necessity to consider local properties of perturbations
with a theoretical representation of the quasi-local nature of
conserved quantities in metric theories. Properties of the conserved
quantities under gauge transformations are established and analyzed,
this allows to describe the well known non-localization problem in
explicit mathematical expressions and operate with them.
Applications of the formalism in general relativity for studying 1)
the falloff at spatial infinity in asymptotically flat spacetimes,
2) linear perturbations on Friedmann-Robertson-Walker backgrounds,
3) a closed Friedmann world and 4) black holes presented as
gravitationally-field configurations in a Minkowski space, are
reviewed. Possible applications of the formalism in cosmology and
astrophysics are also discussed. Generalized formulae for an
arbitrary metric $D$-dimensional theory are tested to calculate the
mass of a Schwarzschild-anti-de Sitter black hole in the
Einstein-Gauss-Bonnet gravity.
\end{abstract}


\noindent \textbf{PACS} 04.20.Cv, 04.25.Nx, 04.50.+h, 11.30.-j.


\section{Introduction and preliminaries}
\m{Intro}

\subsection{Perturbations in gravitational theories,
cosmology and relativistic astrophysics} \m{Cosmology-and-RA}

Much of research in general relativity (GR) is frequently carried
out under the assumption that perturbations of different kinds
propagate in a given (fixed) background spacetime (exact solution to
the Einstein equations) \cite{LL} - \cite{DeWitt-book}. A majority
of cosmological and astrophysical problems are also studied in the
framework of a perturbation approach. It is quite impossible to give
a more or less full bibliography on this topic. Nevetherless, to
stress the importance of such studies, we shall outline shortly some
of the related directions.

\pagestyle{fancy} \fancyhead{} \fancyhead[EC]{A.N. Petrov}
\fancyhead[EL,OR]{\thepage} \fancyhead[OC]{Nonlinear Perturbations
and Conservation Laws\dots} \fancyfoot{}
\renewcommand\headrulewidth{0.5pt}

Cosmological perturbations on Freidmann-Robertson-Walker (FRW)
backgrounds have been considered beginning from the famous work by
Lifshitz \cite{Lifshitz46}. The Lifshitz principles were developed
by many authors in a variety of approaches (see, e.g., the popular
review \cite{Mukhanov}, and a recent review \cite{Lukash-rev}). As
well known examples, one can note the following. The effect of
amplification of gravitational waves in an isotropic world was
discovered by Grishchuk \cite{Grishchuk74}; the gauge invariant
theory of perturbations was formulated by Lukash \cite{Lukash1} and
Bardeen \cite{Bardeen80}. Lately, non-trivial perturbations in FRW
worlds are being considered more frequently. For example, note the
recent papers \cite{KKS02,KKMS03} where,  using the quasi-isotropic
expansions, the authors describe non-decreasing modes of adiabatic
and isocurvature scalar perturbations, and gravitational waves close
to cosmological singularity. Deviations of a space-time metric from
the homogeneous isotropic background become large; while locally
measurable quantities, like Riemann tensor components, are still
close to their FRW values. An approach, where integrals of FRW
models are presented in a time independent form both for ``vacuum''
and for ``usual'' matter, has been developed \cite{Chernin1}. In the
two last decades, cosmological perturbations are considered not only
in the linear approximation, but including the second order also
(see the recent review \cite{Bartolo}).

The evolution of quantized fields, including gravitational and
electromagnetic fields, is intensively studied on curved backgrounds
of various  general classes, such as  globally hyperbolic, static
with symmetry groups,  {\it etc.} In this framework, exact
solutions, like FRW, anti-de Sitter (AdS) and Bianchi of different
types, are also exploited as backgrounds. As an example, note the
theory which is being developed by Grishchuk (see the reviews
\cite{Grishchuk01,Grishchuk05} and references there in). Relic
gravitational waves and primordial density perturbations are
generated by strong variable gravitational field of the early FRW
universe. The generating mechanism is the parametric amplification
of the zero-point quantum oscillations. These generated fields have
specific statistical properties of squeezed vacuum quantum states.
Cosmological perturbations at the early inflation stage
\cite{DolgovZS-book} also continue to be studied (see, e.g., recent
papers \cite{Starobinsky1} - \cite{Starobinsky3} and, e.g., reviews
\cite{Lukash-lect} - \cite{Linde2} and references there in).
Separately, the great interest to AdS spaces has been initiated by
the discovery of the present (not at the earlier inflation stage)
accelerated cosmological expansion (see review \cite{Chernin}). To
explain the acceleration new cosmological solutions are searched
\cite{Chernin2}. On backgrounds of such solutions perturbations are
also to be considered.

Several  black hole solutions \cite{Chandrasekhar}, which could
represent the neighborhoods of relativistic astrophysics objects,
also play a role of backgrounds for evolution  of different kinds of
perturbations. Amplification and dispersion of metric perturbations
(including gravitational waves), electromagnetic field and massless
and mass scalar fields are studied on these backgrounds (see, e.g.,
resent works \cite{BlackHole0} - \cite{BlackHole6}).

The rapid development of the detecting technique stimulates a
development of the gravita\-tional-wave physics (see the recent
detailed review \cite{obzor} and references there in). Thus the
theoretical study of propagation and interaction of gravitational
waves becomes especially important. In the works by Alekseev (see
\cite{Alekseev1,Alekseev2} and references there in), the monodromy
transform approach was developed for constructing exact solutions of
the Einstein equations with spacetime isometries. This method has
been applied for exact solving the characteristic initial value
problem of the collision and subsequent nonlinear interaction of
plane gravitational or gravitational and electromagnetic waves with
distinct wave fronts in a Minkowski space \cite{Alekseev3} -
\cite{Alekseev5}.

Currently, following an extraordinary interest with brane models,
$D$-dimensional metric theories of gravitation have been examined
more and more intensively. Fundamental works on brane worlds have
appeared two decades ago (as we know the works by Rubakov and
Shaposhnikov \cite{RubakovSh} and by Akama \cite{Akama} are the
first in this direction). Later, especially after the works by
Randall and Sundrum \cite{RS1,RS2}, the interest to these models has
risen significantly (see, e.g., a review \cite{Rubakov}).
Perturbations, including gravitational waves, in the framework of
$D$-dimensional metric theories and brane models are also studied
intensively (see, e.g., \cite{branaPert1} - \cite{branaPert6} and
references there in).

The list of directions, where perturbations on curved or flat
backgrounds are studied, could be continued. Impressing results were
obtained in this direction. However, we would like to note the
problems associated with methods of such investigations, rather than
the results. As a rule, these methods are restricted since they are
constructed for examination of particular tasks only. Thus:
 \bit

\item Although the modern cosmic experimental and observable
data require more detailed theoretical results, frequently studies
are carried out in the linear approximation only, without taking
into account the ``back reaction''.

\item Applying particular methods, one generally uses many additional
assumptions. Thus. it is not clear: what results are more general
and what results change under a change of these assumptions.

\item It is difficult to understand: could an approach developed
for a one concrete background be applied to other backgrounds.
Frequently only simplified backgrounds are used; {\it etc}.

\item In each particular case, considering perturbations, one needs to fix
gauge freedoms. A concrete fixation is connected with a concrete
mapping of a perturbed spacetime onto a background spacetime. It
turns out, that it is not so simple to understand what gauge is
``better'', or how to find a connection between different gauges,
{\it etc}.

\eit

\subsection{Conservation laws and their properties}
\m{Einstein-papers}

Very important characteristics for studying perturbations are such
quantities as energy-momentum, angular momentum, their densities,
fluxes, {\it etc.} However, as is well known, the definition of
energy and other conserved quantities in GR has principal problems,
which are well described in many textbooks (see, e.g., \cite{[12]}).
We repeat some related issues to stress importance of these notions
in a theoretical development of gravitational theory also.

It is useful to reconsider works by Einstein who paid special
attention to conservation laws for ``energy components''
$t_\mu{}^\nu$ of the gravitational field. From the early stage, when
the theory was not yet presented in a satisfactory form (its
equations were $R_{\mu\nu} = \k T_{\mu\nu}$), he examined the
conservation law $\di_\nu (t_\mu{}^\nu + T_\mu{}^\nu)= 0$
\cite{Einstein15a,Einstein15b}. Then, the  final form $R_{\mu\nu} =
\k\l(T_{\mu\nu} - \half g_{\mu\nu} T_{\alf}{}^{\alf}\r)$ of the GR
equations was given in the work \cite{Einstein15c}. Einstein himself
explains this change \cite{Einstein15c} by saying that only the
additional term $- \half g_{\mu\nu} T_{\alf}{}^{\alf}$ leads to a
situation, when energy complexes both for the gravitational field,
$t_\mu{}^\nu$, and for matter, $T_\mu{}^\nu$, enter the field
equations in the {\it same} manner. Thus, historically the analysis
of conserved quantities and conservation laws was {\it crucial} for
constructing GR.

In the work \cite{Einstein16}, Einstein finally suggested the
canonical energy-momentum complex for the  gravitational field
$t_\mu{}^\nu$. Later it was called as the Einstein pseudotensor.  As
an application, Einstein used $t_\mu{}^\nu$ to outline gravitational
waves \cite{Einstein16a} - \cite{Einstein18a}. From  the beginning
he stressed that $t_\mu{}^\nu$ is a tensor  {\it only} under linear
transformations.  Under the general coordinate transformations,
$t_\mu{}^\nu$ can change and even be equal to zero. Einstein
interpreted this as a ``non-localization'' of gravitational energy,
which is a special property of the gravitational field, and not a
defect of the theory. This was a reason for numerous criticism  and
discussions. Protecting his theory, Einstein himself was the first
who  gave physically reasonable arguments (see, e.g.,
\cite{Einstein18b}).

The criticism far from killing the theory, was a reason for the
intensive study of properties of gravitational field in GR and its
further development. Following Einstein, in next decades a great
number of methods were suggested for defining conserved quantities
in GR. As a result of these efforts, important theoretical tests,
which restrict an ambiguity in the definition of conserved
quantities, were elaborated. Thus, mathematical expressions have to
give acceptable quantities for black hole masses, for angular
momentum in the Kerr solution, for fluxes of energy and momentum in
the Bondi solution; also, a positive energy density for weak
gravitational waves. However, due to nontrivial peculiar properties
of conserved quantities in GR, up to now:
 \bit
 \item Frequently,  it is very difficult to find a connection between
different definitions; sometimes definitions even contradict one
another; sometimes definitions (especially earlier) are
not-covariant; {\it etc}.
 \item Sometimes definitions are not connected with perturbations.
 \eit

Now, let us present the modern point of view on the non-localization
problem. Considering the physical foundation of the theory it is
clear that the non-localization is directly connected with the
equivalence principle (see, e.g., \cite{[12]}). On the other hand,
the situation  can be also explained by the fact that GR is a
geometric theory where spacetime, in which all physical fields
propagate, itself is a dynamical object. (Of course, the
non-localization problem is related to all the metric theories of
gravity, not only to GR.) Due to these objective reasons, sometimes
the problem of conserved quantities in metric theories is presented
as ill-defined. However, only the fact of non-localization cannot
imply that these notions are meaningless. Without  a doubt
gravitational interaction contributes to the {\it total}
energy-momentum of gravitating system \cite{[12]}. Indeed,
describing a binary star system one needs to consider gravitational
energy as a binding energy; considering gravitational waves in an
empty domain of space one finds a positive energy of this domain as
a whole; {\it etc.} All of these are related to non-local
characteristics. This conclusion is supported by the mathematical
content of GR. Szabados \cite{Szabados04} clearly expresses it as
follows. ``... the Christoffel symbols are not tensorial, but they
do have geometric, and hence physical content, namely the linear
connection. Indeed, the connection is a {\it non-local} geometric
object, connecting the fibers of vector bundle over {\it different}
points of the base manifold.  Hence any expression of the connection
coefficients, in particular the gravitational energy-momentum or
angular momentum, must also be non-local. In fact, although the
connection coefficients at a given point can be taken zero by
appropriate coordinate/gauge transformation, they cannot be
transformed to zero {\it on an open domain} unless the connection is
flat.''

Thus the non-localization is natural and inevitable. However, up to
now there is no simple and clear description of it. Therefore:
 \bit
\item It is important to give mathematical expressions, which
constructively present the non-localization of conserved quantities
in metric theories.
 \eit
The non-localization has to be connected with gauge properties in
the description of perturbations. Thus, a solution of this problem
has to give a possibility to find reasonable assumptions for a gauge
fixation, which allow to describe certain quantities as localized
ones.

It is natural that due to the aforementioned peculiar properties of
gravitational field much attention has been paid just to non-local
characteristics. Thus the {\it total} energy-momentum and angular
momentum of a gravitating system in {\it whole} spacetime were
studied intensively. Such quantities are frequently called as global
ones. In this context asymptotically flat spacetimes are considered
in details (see, for example, earlier reviews
\cite{York80,Ashtekar80}, also recent papers and reviews
\cite{AshtekarBicakSchmidt} - \cite{CJK-book} and numerous
references there in). One of the great achievements was the proof of
the positivity of the total energy for an isolated system
\cite{SchoenYau} - \cite{Nester81} (see also the review
\cite{Faddeev}). The global conserved quantities for asymptotically
curved backgrounds (like AdS space and some others) are also studied
intensively (see, e.g., \cite{AbbottDeser82} - \cite{Olea2}).

The aforementioned development has initiated a more intensive
examination of the energy problem in GR. Conserved quantities became
to be associated with finite spacetime domains. Such quantities are
called as {\it quasilocal} ones and can give a more detailed
information than the global quantities. In the last two-three
decades the quasilocal approach has became very popular. It is not
our goal to present it here, moreover, recently a nice review by
Szabados \cite{Szabados04} has appeared. Nevetherless, below we
shortly outline some of important quasilocal methods.

The Brown and York approach \cite{BY93} is based on the generalized
Hamilton-Jacobi analysis. It considers a spatially restricted
gravitating system on 3-dimensional spacelike section $\Sigma$. A
history of the boundary is a 3-dimensional timelike surface $S$
(cylinder). It is assumed that a 3-metric $\gamma_{ij}$ on $S$ is
fixed and plays a role of a time interval in the usual
non-relativistic mechanics, which defines initial and final
configurations. They define the energy-momentum tensor $\tau^{ij}$
on $S$, as a functional derivative of an action with respect to
$\gamma_{ij}$. An intersection of $\Sig$ with $S$ is a 2-sphere $B$
which is just a spatial boundary of the system. Normal and
tangential projections of $\tau^{ij}$ onto $B$ give surface
densities of energy, momentum and space tensions on $B$ which are
quasi-local expressions. It is crucial to determine a reference flat
space, which is {\it uniquely} defined by the isometric embedding
$B$ (with a positive inner curvature) into a flat space. The
Brown-York method received a significant development in the works by
Brown, Lau and York \cite{Lau93} - \cite{Lau99} and in works of
other authors (see review \cite{Szabados04}). The recent work
\cite{BLY02} could be considered as a mathematical textbook on this
approach.

At the first stages of constructing the Hamiltonian dynamics of GR
by Arnowitt, Deser and Misner \cite{ADM} (ADM), surface integrals
were neglected {\it a priori} and reappeared only after disregarding
non-physical degrees of freedom. One of the way of developing the
standard Hamiltonian description is the symplectic approach by
Kijowski and Tulczyiew \cite{Tulczyiew2}, where it is noted that
surface integrals are not less important than the volume ones.
Jezierski and Kijowski developed this approach in GR \cite{Jacek1} -
\cite{Jacek2}. They use an ``affine formulation'', where  the
connection coefficients $\Gamma^\lam_{\mu\nu}$ are used rather, than
the metric ones. The gravitational field is considered inside  a
closed tube, at a boundary of which some conditions are fixed to
construct a closed Hamiltonian system. The Hamiltonian describes the
full energy inside the boundary and has a quasilocal sense
\cite{Kijowski2}. In the linear gravity, the requirement of
positiveness of the Hamiltonian \cite{Jacek1} leads to a
``localization'' of gravitational energy with {\it unique} boundary
conditions. As an application, gravitational waves on the background
of the Schwarzschild geometry were studied \cite{Jacek2}.

Based on the symplectic method Nester with co-authors \cite{Nester8}
- \cite{ChenNesterTung2005} developed a so-called 4-covariant
Hamiltonian formulation both for GR and for generalized geometrical
gravitational theories. The Hamiltonian on-shell is a surface
integral, which defines a quasilocal conserving quantity inside a
closed volume. For this approach a displacement 4-vector constructed
from the lapse and shift, and a flat space defined at the boundary
of the volume are necessary. For the recent development and
achievements of this fruitful approach see the review paper
\cite{Nester2004}.

Returning to the discussion of the previous subsection, we recall
again on a necessity to operate with local quantities in
cosmological and astrophysical applications. Therefore, to conclude
the subsection let us formulate also the next task:
 \bit
 \item It is important to connect local conserved characteristics
with non-local quantities,  which appear in the theoretical
considerations.
 \eit

\subsection{Goals of the review and plan of the presentation}
\m{Goal}

Analyzing the problems accented in the previous subsections
\ref{Cosmology-and-RA} and \ref{Einstein-papers} a necessity  in a
generalized and universal approach for describing perturbations,
both in GR and in generalized metric gravitational theories, becomes
evident. A description, where perturbations in a geometrical theory
are considered on a curved or flat background, in fact, converts
this theory into the rank of a field theory, like electrodynamics in
a fixed spacetime. A set of all the perturbations acquires  the
sense of the dynamic field configuration. Then,  it is desirable to
represent the perturbed gravitational metric theory in the {\em
field-theoretical} form (or simply, {\em field} form) with all the
properties of a field theory. We formulate these properties as the
following requirements:
 \bit \m{requirements}
\item[(a)] The field-theoretical formulation has to be covariant.
\item[(b)] One can use an arbitrary curved background spacetimes
(solutions to GR or another metric theory).
\item[(c)] The perturbed system has to be represented as a dynamic
field configuration, which is associated with Lagrangian and
corresponding action.
\item[(d)] The
field equations (perturbation equations) have to be derivable from
the action principle.
\item[(e)] The conserved quantities and conservation laws also
have to be derivable using the variational principle.
\item[(f)] The field-theoretical formulation has to have gauge
freedoms. Gauge transformations and their properties have to be
connected with the action.
\item[(g)]  In order not to have restrictions in the use of orders of
perturbations, it is required to have an exact formulation for
perturbed equations, conservation laws and  gauge transformations.
\item[(h)] Lastly, it is desirable
to have a simple and explicit form convenient for applications.
 \eit
Only such a derivation will permit the required universality and
give a full description of perturbations. Of course, the
field-theoretical formulation has to be equivalent to the
geometrical one, without changing the physical content of the
theory.

Thus, the goal of the paper is to suggest an approach, which gives a
possibility to present a perturbed metric (geometrical)
gravitational theory in a field-theoretical form, which satisfies
the above requirements (a) - (h). In last two decades, all the
necessary elements of such an approach were developed in works by
the author together with his co-authors, and in other works all of
which will be cited later. Therefore, from one point of view, the
present paper is a review of these works. On the other hand, the
paper just unites these works into a generalized and universal
approach. Our task is to give an outline of mathematical development
of the approach with necessary mathematical expressions, to
demonstrate possibilities of the approach and its advantages, and to
outline some of its applications. Therefore we hope that the present
work could be interesting both to the experts in gravitational
physics, e.g., in conservation laws in GR, and to cosmologists and
astrophysicists studying the evolution of perturbations on curved
backgrounds. This paper is not a review of all the numerous
perturbation approaches and methods developed during the history of
GR. Therefore we apologize to the authors whose works are not
referred here.

\begin{figure*}[t]
    \centering
    \includegraphics*%
         [width=7cm]%
    {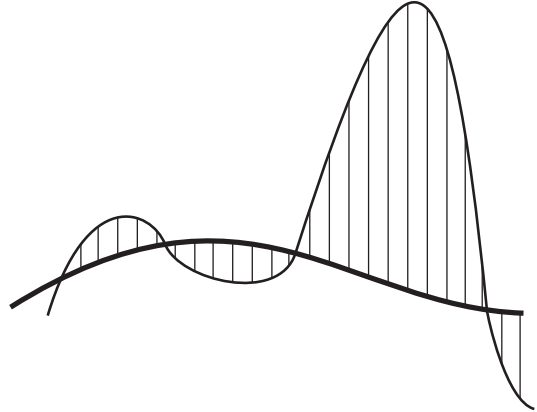}\\
    \vspace*{.0in}
    \caption{A symbolical connection between solutions of the same
    metric theory in the geometrical and field-theoretical forms.}
     \label{F2+}
    \end{figure*}

%
%
Let us discuss some of important points. First,  a possibility to
use an arbitrary curved background means that any solution of the
initial metric theory can be considered as a background. Thus, the
background can be flat, curved vacuum, or even curved including
background matter, i.e. it can be arbitrary. Second, as was
accented, the field and geometrical formulations of the theory have
to be equivalent. This means that a solution of the field
formulation united together with a background solution have to be
transformable into a solution of the geometric formulation (initial
metric theory). Symbolically this situation is explained on the
figure \ref{F2+}. Let the slightly sloping curve be related to a
background and let the oscillating curve mean a solution in the
geometrical form, then the difference between them symbolizes the
solution in the field-theoretical form. Third, usually it is assumed
that a perturbation of a quantity is less than a quantity itself. We
do not impose this restriction here. A realization of the
requirements (a) - (h) means that the field-theoretical formulation
can be thought of as an independent exact field theory. Then, of
course, the ``amplitude'' of an exact solution of the exact theory
can be more than ``amplitude'' of the background (see, e.g., the
right side of the figure \ref{F2+}).

At the earlier stage, Einstein was trying to construct a
gravitational theory as a field theory in Minkowski space, in the
framework of special relativity. However, step by step he had
concluded that one needs to operate with curved spacetime only,
Minkowski background space had disappeared from the consideration
all together. The construction of the field-theoretical formulation
in the framework of the geometrical theory is, in a definite sense,
the revival of the special relativity view. Moreover, as we remarked
above, the field-theoretical formulation really is an independent
field theory and can be constructed by independent ways (see below
the discussion in subsections \ref{revieworiginal} and
\ref{principles}). However, there is no contradiction here. The
geometrical and field-theoretical formulations are two different
formulations of the same theory with the same physical content. The
background spacetime turns out non-observable and has a sense of
only an auxiliary structure.

Up to now the Einstein theory retains its position as the most
popular theory of gravity, leading in all the applications. Thus, in
this paper we pay a significant attention to constructing the
field-theoretical formulation for GR which is presented in detail.
On the other hand, due to the rising precision of experiments and
observables in cosmos and due to the great interest to the brane
models, other metric theories generalizing GR become  more and more
popular and necessary. One of the main properties of the approach
presented here is its universality. We use this advantage and
develop the field theoretical approach  applied to an arbitrary
metric theory.

The paper is organized as follows. The next section
\ref{decompositions} is devoted to the detail description of the
field-theoretical formulation of GR on an arbitrary curved
background, which have all the properties of a self-dependent field
theory. The essential attention is paid to an invariance with
respect to exact gauge transformations. The last do not effect both
coordinates and background quantities in the field-theoretical
formulation and are connected with the general covariance of GR.

In section \ref{CL}, we construct conservation laws for
perturbations in GR in the framework of the field-theoretical
approach. Conserved currents and corresponding superpotentials are
presented. As an important instruments we use the canonical
N{\oe}ther method and the Belinfante symmetrization prescription.
The conserved quantities and conservation laws are used for
examination of asymptotically flat spacetime at spatial infinity,
the closed Friedmann model, the Schwarzschild solution and linear
perturbations on FRW backgrounds.

In section \ref{D-dimensions},  we develop the field-theoretical
approach to describe perturbations in an arbitrary $D$-dimensional
metric theory of gravity on a fixed background. We construct {\em
generalized} conserved currents and corresponding superpotentials
with again the essential use of the N{\oe}ther and Belinfante
methods. The conserved quantities are tested in the
Einstein-Gauss-Bonnet gravity for calculating the mass of the
Schwarzschild-anti-de Sitter black hole.

\subsection{Notations}
\m{Notations}

Here, we present notations, which will appear more frequently and
which are more important. In the text, together with these
notations, numerous other notations will be used, they will be
outlined currently.

\bit

\item Greek indexes numerate 4-dimensional spacetime coordinates
as well as $D$-dimensional spacetime ones. Usually $x^0$ means a
time coordinate, whereas small Latin indexes from the middle of
alphabet $i,\,j,\,k,\,\ldots$ mean 3-dimensional space coordinates
or $(D-1)$-dimensional hyperspace coordinates;

\item Large Latin indexes $A,\,B,\,C,\,\ldots$ are used as generalized
ones for an arbitrary set of tensor densities, for example, $Q^A =
\{\sqrt{-g}g^{\mu\nu},\, \phi,\, T_{\alf\beta} \}$;

\item A dynamic metric of a metric theory is
$g_{\mu\nu}$ ($g = \det g_{\mu\nu}$);

\item Bar means that a quantity ``$\Bar Q^A$'' is a background one;

\item Thus, $\Bar g_{\mu\nu}$ ($\Bar g = \det \Bar g_{\mu\nu}$) is a
background metric. Indexes of all the quantities of a {\it
perturbed} system are raised and lowered by the background metric;

\item Many expressions are presented as densities of the weight +1.
The reasons are as follows. First, all the N{\oe}ther identities,
which are explored intensively,  are such densities in the initial
derivation. Second, conservation laws have to be covariant, however
partial derivatives are crucial for application of the Gauss
theorem. Divergences of both vector densities (conserved currents)
and antisymmetric tensor densities (superpotentials) have this
duality property. To accent these expressions we use ``hats'', as
more economical notations in this situation.  Thus, a quantity
``$\hat Q^A$'' is such a density, it could be a tensor multiplied by
$\sqrt{-g}$ or $\sqrt{-\Bar g}$ (for example, $\hat
Q^{\alf\beta}{}_{\gamma} =\sqrt{-\Bar g} Q^{\alf\beta}{}_{\gamma}$),
or $\hat Q^A$ could be be independent on these determinants, a
situation will be clarified  from the context;

\item $\eta_{\mu\nu}$ is a Minkowskian metric in the Lorentzian
coordinates; sometimes $\sqrt{-\eta}$ is used explicitly instead of
$1$ to stress that a quantity, say $\hat
Q^{\alf\beta}{}_{\gamma}=\sqrt{-\eta}Q^{\alf\beta}{}_{\gamma}$,  is
a density of the weight $+1$;

\item $\hat l^{\mu\nu} = \hat g^{\mu\nu} - \Bar {\hat g}^{\mu\nu} =
\sqrt{- g} g^{\mu\nu} - \sqrt{-\Bar g}\, \Bar {g}^{\mu\nu}$ is a
more important form of the metric perturbations;

\item Partial derivatives are denoted by
 $({}_{,i}),\,({}_{,\alf}),$ or
$\di_i,\,\di_\alf,$;

\item $D_\alf$ and $\BD_\alf$ are covariant derivatives with respect to
 $g_{\mu\nu}$  and $\Bar g_{\mu\nu}$ with the Chistoffel symbols
$\Gamma^\alf_{\beta\gamma}$ and $\Bar \Gamma^\alf_{\beta\gamma}$,
respectively;

\item $\Delta^\alf_{\beta\gamma} = \Gamma^\alf_{\beta\gamma}-
\Bar \Gamma^\alf_{\beta\gamma}$ is the tensor actively used in the
paper;

\item $\xi^\alf$ is an arbitrary displacement vector, whereas
$\lam^\alf$ is a Killing vector of a background;

\item The Lie derivative is defined as
 $$
 \pounds_\xi Q^A =-\xi^\alf \BD_\alf Q^A + \l. Q^A
\r|^\alf_\beta \BD_\alf \xi^\beta\, ,
 $$
note the opposite sign to the usual one, $\l. Q^A \r|^\alf_\beta$ is
defined by the transformation properties of $Q^A$;

\item $\zeta_{\mu\nu}=-\half \pounds_\xi \Bar g_{\mu\nu}$  is the tensor
actively used in the paper;

\item The Lagrangian derivative is defined as usual:
 $$
\frac{\delta{Q^A(q^B,\,q^B_{,\alf},\,q^B_{,\alf\beta},\,\ldots)}}{\delta
q^C} = \frac{\di Q^A}{\di q^C} - \di_{\alf}\l(\frac{\di Q^A}{\di
q^C_{,\alf}} \r) + \di_{\alf\beta}\l(\frac{\di Q^A}{\di
q^C_{,\alf\beta}} \r) - \ldots
 $$
\item $R^\alf{}_{\mu\beta\nu}$, $R_{\mu\nu}$, $G_{\mu\nu}$,
$T_{\mu\nu}$, $R$ and $\Bar R^\alf{}_{\mu\beta\nu}$, $\Bar
R_{\mu\nu}$, $\Bar G_{\mu\nu}$, $\Bar T_{\mu\nu}$,$\Bar R$ are the
Riemannian, Ricci, Einstein, matter energy-momentum tensors and the
curvature scalar for the physical and background  spacetimes.

\item Usually index ``L'' means a linearization, for example,
$G^L_{\mu\nu}$ and $\Phi^L_{\mu\nu}$ mean linearized pure
gravitational and matter parts of the gravitational field equations;

\item The conserved currents are defined in the framework of
different approaches as follows. In the case of GR: $\hat
J^\mu_{(c)}$ is defined with the use of the canonical N{\oe}ther
procedure; $\hat J^\mu_{(s)}$ is defined with the use of the
field-theoretical prescription, based on the symmetrical
energy-momentum tensor; $\hat J^\mu_{(B)}$ is the canonical current
corrected with the use of the Belinfante method. The correspondent
superpotentials in GR are $\hat J^{\mu\nu}_{(c)}$, $\hat
J^{\mu\nu}_{(s)}$ and $\hat J^{\mu\nu}_{(B)}$;

\item Analogous currents in superpotentials in an arbitrary $D$-dimensional metric
theory are respectively $\hat {\cal I}^\mu_{(c)}$, $\hat {\cal
I}^\mu_{(s)}$, $\hat {\cal I}^\mu_{(B)}$ and $\hat {\cal
I}^{\mu\nu}_{(c)}$, $\hat {\cal I}^{\mu\nu}_{(s)}$, $\hat {\cal
I}^{\mu\nu}_{(B)}$;

\item $(\alf\beta)$, $(ik)$ and $[\alf\beta]$, [ik] mean symmetrization
and antisymmetrization;

\item $\k$ --- the ``Einstein'' constant both in GR and in an arbitrary
metric theory.

 \eit

\section{The exact field-theoretical  formulation of GR}
\m{decompositions} \setcounter{equation}{0}

\subsection{Development of the field approach}
\m{revieworiginal}

The study of perturbations in GR, in fact, was begun by Einstein
himself. However, as a separate field, the history of the
field-theoretical approach in GR began in 40's --- 50's of XX
century. Perturbed Einstein equations are rewritten as follows.
Define the metric perturbations on a flat background in the
Lorentzian coordinates as $\hat l^{\mu\nu} = \hat g^{\mu\nu} -
\sqrt{-\eta}\eta^{\mu\nu}$; $l^{\mu\nu} =(\sqrt{-\eta})^{-1} \hat
l^{\mu\nu}$. The terms linear in metric perturbations are placed on
the left hand side of the Einstein equations, whereas all the
nonlinear terms are transported to the right hand side, and together
with a matter energy-momentum tensor are treated as a total
(effective) energy-momentum tensor $t^{(tot)}_{\mu\nu}$. Then
Einstein equations are rewritten in the equivalent perturbed form as
 \be
G^L_{\mu\nu} = \k t^{(tot)}_{\mu\nu}\,
 \m{FullPerturbEqs}
 \ee
where, raising the indexes by $\eta^{\alf\beta}$, one has the left
hand side in the form:
 \be
 G_L^{\mu\nu} \equiv
\half(l^{\mu\nu,\alf}{}_{,\alf} + \eta^{\mu\nu}
l^{\alf\beta}{}_{,\alf\beta} -l^{\alf\mu,\nu}{}_{,\alf}
-l^{\alf\nu,\mu}{}_{,\alf})\,.
 \m{LinearEqs}
 \ee
Its divergence identically is equal to zero: $\di_\nu
G_L^{\mu\nu}\equiv 0$. Then,  one obtains directly the differential
conservation law
 \be
\di_\nu t_{(tot)}^{\mu\nu} = 0\, .
 \m{CLfirst}
 \ee

This picture was developed in a form of a Lagrangian based field
theory with self-interaction in a fixed background spacetime, where
$t^{(tot)}_{\mu\nu}$ is obtained by variation of an action with
respect to a background metric. Following the introduction in the
Deser work \cite{[11]}, below, we shall present the main steps in
this derivation, the corresponding bibliography can also be found in
\cite{[11]}. Assume that a field theory of gravity in a Minkowski
space is constructed. By known observable tests (see, e.g., textbook
\cite{[12]}), the most preferable type of the gravitational field is
the tensor field, say $l^{\mu\nu}$. The linear (approximate)
equations have to have the form $G^L_{\mu\nu} = 0$ and are defined
by the quadratic Lagrangian $\lag_{(2)}^{g}$. Keeping a symmetrical
energy-momentum tensor of matter fields $\phi^A$ as a source of
$G^L_{\mu\nu}$ one obtains
 \be
G^L_{\mu\nu}(l) = \kappa T_{\mu\nu}(\phi,\eta)\, .
 \m{(4.2)}
 \ee
Identically $\di_\nu G^{L\nu}_\mu \equiv 0$, therefore $\di_\nu
T^{\nu}_\mu =0$. However, there is a contradiction between the
conservation law $\di_\nu T^{\nu}_\mu =0$  and equations of motion
for {\it interacting} fields $\phi^A$. How does one avoid this? The
right hand side of Eq. (\ref{(4.2)}) is to be obtained by variation
(conventionally) both with respect to $\eta^{\mu\nu}$ and
$l^{\mu\nu}$. Therefore one needs to make an exchange $\l\{
\phi^A,\eta^{\mu\nu} \r\}\goto \l\{ \phi^A,\eta^{\mu\nu} +
l^{\mu\nu} \r\}$ both in the matter Lagrangian and  in the matter
energy-momentum tensor. This just means the universality of
gravitational interaction. Next, one has to include the
gravitational self-interaction. Therefore one adds the symmetrical
energy-momentum tensor of the gravitational field
$t^{(2)g}_{\mu\nu}(l)$, corresponding to $\lag_{(2)}^{g}$, to the
right hand side of (\ref{(4.2)}) together with $T_{\mu\nu}(\phi,\eta
+ l) $. But the equations that include $t^{(2)g}_{\mu\nu}(l)$ can be
obtained if a cubic Lagrangian is added, $\lag_{(2)}^{g} +
\lag_{(3)}^{g}$. After this one needs to consider the next level,
and so on. In the result, one obtains the final variant of the
gravitational equations:
 \be G^L_{\mu\nu}(l) = \kappa
\left[T_{\mu\nu}(\phi, l + \eta) + \sum_{n=2}^\infty
t^{(n)g}_{\mu\nu}(l) \right]\, .
 \m{(4.4)}
 \ee
It turns out that the equations (\ref{(4.4)}) are exactly the
Einstein equations, i.e. equations (\ref{FullPerturbEqs}). After the
identification $\sqrt{-\eta}\eta^{\mu\nu} +\hat l^{\mu\nu} \equiv
\sqrt{-g}g^{\mu\nu} $ one has only the dynamical metric
$g^{\mu\nu}$, whereas the background metric $\eta^{\mu\nu}$ and the
field $l^{\mu\nu}$ completely disappear from the consideration.

Deser himself \cite{[11]}, unlike (\ref{(4.4)}), has suggested the
field formulation of GR without expansions. As dynamical variables
he used the two independent tensor fields $l^{\mu\nu}$ and
$\Del^\alf_{\mu\nu}$ of the 1-st order formalism. After variation of
the corresponding action he derives the equations in the form:
 \be
G^L_{\mu\nu}(l) = \kappa \left[ t^{g}_{\mu\nu}(l,\Del) +
t^{m}_{\mu\nu}\right]\, ,
 \m{(4.6)}
 \ee
instead of (\ref{(4.4)}). After identifications
$\sqrt{-\eta}\eta^{\mu\nu} +\hat l^{\mu\nu} \equiv
\sqrt{-g}g^{\mu\nu} $ and $\Del^\alf_{\mu\nu} \equiv
\Gamma^\alf_{\mu\nu} $ the equivalence with the Einstein equations
is confirmed, but only in the Palatini form, where $g^{\mu\nu}$ and
$\Gamma^\alf_{\mu\nu}$ are used as independent variables.

In the work \cite{GPP} we have generalized the Deser approach
\cite{[11]}. Instead of the background Minkowski space with the
Lorenzian coordinates we consider an arbitrary curved background
spacetime with a given metric $\Bg_{\mu\nu}$ and given matter fields
$\Bar \Phi^A$ satisfying the background Einstein gravitational and
matter equations. We also use the 1-st order formalism. The
gravitational equations get the generalized form:
 \be
 \hat G^L_{\mu\nu}(l) + \hat
\Phi^L_{\mu\nu}(l,\phi) = \kappa \left[ \hat t^{g}_{\mu\nu}(l,\Del)
+ \hat t^{m}_{\mu\nu}\right]\,
 \m{(4.10)}
 \ee
where the left hand side is linear in $\hat l^{\mu\nu}$ and $\phi^A$
and defined later in Eqs. (\ref{G-L}) - (\ref{(a2.19)}). The term
$\hat \Phi^L_{\mu\nu}$ appears due to $\Bar \Phi^A$. Equivalence
with GR in the ordinary derivation is stated after identifications:
$ \Bar {\hat g}^{\mu\nu} +\hat l^{\mu\nu} \equiv \hat g^{\mu\nu}$,
$\Bar \Ga^\alf_{\mu\nu} + \Del^\alf_{\mu\nu} \equiv
\Ga^\alf_{\mu\nu}$ and $\Bar \Phi^A +\phi^A \equiv \Phi^A$. In the
following years we have developed the principles of the work
\cite{GPP} and have used this approach in many applications. Our
results are presented in the papers \cite{GP86} -
\cite{Petrov2005b}, on the basis of which, in a more part, the
present review was written.

Elements of the field approach in gravity are also actively
developing nowadays in other approaches. Thus, in
\cite{BabakGrishchuk}, a requirement only of the first derivatives
of metrical perturbations in the total symmetrical energy-momentum
tensor has led to a {\it new} field formulation of GR in Minkowski
spacetime, which is different from the formulation in \cite{GPP}.
The new total energy-momentum tensor is the source for the
non-linear left hand side. On the basis of this new field
formulation an interesting variant of the gravitational theory with
non-zero masses of gravitons was developed \cite{BabakGrishchuk1}. A
comparison and a connection of the works
\cite{GPP,BabakGrishchuk,BabakGrishchuk1} are discussed in
\cite{Petrov2004} in detail. In \cite{PintoNetoSilva}, the work
\cite{GPP} was developed to construct the total energies and angular
momenta for $d+1$-dimensional asymptotically anti-de Sitter
spacetime.  The properties of the field approach \cite{GPP} appear
independently in many concrete problems. For example, in
\cite{KRMS,RK} a consideration of linear perturbations on FRW
backgrounds leads to the linear approximation of the exact field
formulation of GR \cite{GPP}. In \cite{PittsSchive2001a}, as a
development of the field approach, a class of so-called ``slightly
bimetric'' gravitation theories was constructed. In
\cite{PittsSchive2001b,Pitts2004}, a behaviour of light cones in
Minkowski space and effectively curved spacetimes was examined.
Then, based on the causality principle, a special criterium was
stated. In \cite{Pitts2003} this criterium was used to show that if
spatially flat  FRW big bang model is considered as a configuration
on a flat background, then the cosmological singularity is banished
to past infinity in Minkowski space. The references to earlier works
and the theoretical foundation for the field approach can be found
in the works \cite{ZG} - \cite{Grishchuk92}.  To the best of our
knowledge up to date bibliography related to the field approach in
gravity can be found in \cite{PittsSchive2001a} - \cite{Pitts2003}.

\subsection{Various directions in the construction} \m{principles}

There are various possibilities to approach the field-theoretical
formulation of GR, which are based on different foundations. Here,
we shall discuss the well known and important ones. The principle
used by Deser \cite{[11]} could be formulated as follows:
 \bit
\item
The source of the linear massless field of spin two (of
gravitational field) in Minkowski space is to be the total
symmetrical (metric) energy-momentum tensor of all the dynamical
fields, including the gravitational field itself.
 \eit
Using this principle Deser has constructed a corresponding
Lagrangian and field equations and energy-momentum tensor following
from it. We have generalized the Deser approach on arbitrary curved
backgrounds \cite{GPP}. His principle has been reformulated in a way
that the linear left hand side of perturbed gravitational equations
has to be of the form in Eq. (\ref{(4.10)}), i.e. together with
$\hat G^L_{\mu\nu}$ one has to include the, linear in matter
perturbations, part $\hat \Phi^L_{\mu\nu}$. The analogous principle
was suggested for perturbed matter equations.

The next known method was most clearly presented by Grishchuk
\cite{Grishchuk92} and shortly can be formulated as:
 \bit
\item {A transformation from gravistatic (Newton law) to
gravidynamics, i.e. to a relativistic theory of gravitational field
(general relativity), equations of which (Einstein equations)
describe  gravitational waves}.
 \eit Following this direction, one has to
transform the Newton law $\Delta \phi = -4\pi G\rho$ into a special
relativity description. To satisfy the relativistic requirement a)
the mass density $\rho$ has to be generalized to 10 components of
the matter stress-energy tensor $T_{\mu\nu}$; b) the single
component $\phi$ should also be replaced by  10  gravitational
potentials $l^{\mu\nu}$; c) the Laplace operator should be replaced
by the d'Alembert operator; d) the gravitational field has to be
nonlinear and, thus, has to be a source for itself. Following this
reformations one obtains the generalized equations: $\half
l^{~~~,\alpha}_{\mu\nu~~,\alpha} = \kappa(t^{g}_{\mu\nu} +
T_{\mu\nu})$. They imply that the gauge condition
$l^{\mu\nu}_{~~~,\nu} = 0$ is already chosen. e) To reconstruct the
gauge invariance properties it is necessary to add to the left hand
side the terms $ \half\l(
\eta_{\mu\nu}l^{\alpha\beta}_{~~~,\alpha,\beta} -
l^\alpha_{~\mu,\nu,\alpha}
 - l^\alpha_{~\nu,\mu,\alpha}\r)$.
As a result, one obtains Eq. (\ref{FullPerturbEqs}), which is just
the Einstein equations.

The field formulation of GR can be also constructed based on the
gauge properties (see our work \cite{[16]}). This direction is
analogous to the method in the gauge theories of the Yang-Mills
type, which are constructed by localizing parameters of a gauge
group. However, unlike usual, we postulate a non-standard way of
localization, namely:
 \bit
\item
{ A ``localization'' of Killing vectors of the background
spacetime.}
 \eit
This assumes the existence of a fixed background spacetime with
symmetries presented by a Killing vector field $\lam^\alf$, in which
initial dynamic fields $\phi^A$ are propagated. It is noted that an
action for the initial fields is invariant, up to a surface term,
under the transformation $\phi^A \goto \pounds_\lam \phi^A$. Next,
the Killing vector is changed for an arbitrary vector $\xi^\alf$,
that is localized. Then the invariance is destroyed. To restore it
the compensating (gauge) field has to be included. In doing so the
coordinates and the background metric do not change. The
requirements to have the gauge field as an universal field and to
have the simplest sought-for action for the free gauge field lead
just to the field formulation of GR, developed in \cite{GPP}.

As a rule, a fixed background spacetime, in which perturbations are
studied, is determined by the problem under consideration (see
subsections \ref{Cosmology-and-RA} and \ref{Einstein-papers} in
Introduction). Thus the background could be assumed as a known
solution to the Einstein equations. Then one has to study the
perturbed (with respect to this background) Einstein equations. As
already was noted above, this picture can be developed as a
Lagrangian based field theory, where the first step is:
 \bit
\item { The decomposition of dynamical variables of GR
into background variables and dynamic perturbations.}
 \eit
This method is evident in itself and has the explicit connection
with the ordinary geometrical formulation of GR. Namely this method
is easily adopted for constructing the field-theoretical formulation
in the framework of an arbitrary metric theory.  In the next
subsections, basing on the works \cite{GPP,PP87,[15]}, we present it
in detail. However, although the construction was developed for both
in the 1-st and in the 2-nd order formalisms, here, we use the 2-nd
order formalism only since it is more convenient and suitable. To
the best of our knowledge, Barnebey \cite{Barnebey} was the first
who suggested to use the 2-nd order formalism for an exact (without
expansions) description of perturbations in GR.

\subsection{A dynamical Lagrangian}
\m{DynamicLagrangian}

Let us consider the Einstein theory and write out its action
 \be S = {1\over c} \int d^4x \hat{\cal L}^E \equiv
 -{1 \over {2\kappa c}} \int d^4x \hat R(g_{\mu\nu})
+ {1 \over c} \int d^4x \hat{\cal L}^M (\Phi^A,~g_{\mu\nu})\, ,
\m{(a2.1)}
 \ee
$\hat{\cal L}^M$ depends on $\Phi^A$ and their derivatives up to a
finite order. Then, the gravitational and matter equations
corresponding to the Lagrangain (\ref{(a2.1)}) are
 \bea
 {{\del {\lag}^E} \over {\del \hat g^{\mu\nu}}} =
 -{1 \over {2\k }}
 {{\del \hat R} \over {\del \hat g^{\mu\nu}}} +
 {{\del {\lag}^M} \over {\del \hat g^{\mu\nu}}}& =& 0\, ,
 \m{(a2.2)}\\
  {{\del {\lag}^E} \over {\del \Phi^A}} =
 {{\del \lag^M} \over {\del \Phi^A}} &=& 0\, .
\m{(a2.3)}
 \eea
The form of Eq. (\ref{(a2.2)}) corresponds to the form of the
Einstein equations
 \be
 R_{\mu\nu} - \k\l(T_{\mu\nu} - \half g_{\mu\nu} T^\alf_\alf\r)=0\, ,
 \m{(a2.2+)}
 \ee
whereas a more customary form is
 \be
 \sqrt{-g}\l({G}_{\mu\nu} - \k {T}_{\mu\nu}\r) \equiv
 {\hat G}_{\mu\nu} - \k {\hat T}_{\mu\nu} = 0\,,
 \m{EinsteinEquations}
 \ee
which is obtained by variation with respect to $g^{\mu\nu}$.

Next, let us define the metric and matter perturbations with the use
of the decompositions:
 \be \hat g^{\mu\nu}
\equiv \Bar {\hat g}^{\mu\nu} + \hat l^{\mu\nu}\,,
 \m{(a2.4)}
 \ee
 \be \Phi^A  \equiv \Bar {\Phi}^A + \phi^A\, .
  \m{(a2.5)}
  \ee
It is assumed that the background quantities $\Bar{\hat g}^{\mu\nu}$
and $\Bar{\Phi}^A$ are known and satisfy the background (given)
system, which is defined as follows. Its action is
 \be
\Bar S = {1 \over c} \int{ d^4x \Bar{{\lag}^E}} \equiv
 -{1 \over {2\k c}} \int {d^4x \Bar{\hat R}}
+ {1 \over c} \int {d^4x \Bar{{\lag}^M}}\, .\m{(a2.8)}
 \ee
The corresponding to the Lagrangian in (\ref{(a2.8)}) background
gravitational and matter equations have the form of the barred
equations (\ref{(a2.2)}) and (\ref{(a2.3)}):
 \bea
-{1 \over {2\k }}  {{\del \Bar{\hat R}} \over {\del \Bar{\hat
g}^{\mu\nu}}} +
 {{\del \Bar{{\lag}^M}} \over {\del \Bar{\hat g}^{\mu\nu}}}& =& 0\, ,
 \m{(a2.6)}\\
 {{\del \Bar{{\lag}^M}} \over {\del \Bar{\Phi}^A}}& =& 0\, .
 \m{(a2.6+)}
 \eea
Frequently we use a Ricci-flat background with the background
equations
 \be
 \Bar{\hat G}_{\mu\nu}= \Bar{\hat R}_{\mu\nu}  = 0.
 \m{BackRicciFlat}
 \ee

Now let us classify the perturbations $\hat l^{\mu\nu}$ and $\phi^A$
as {\it  independent dynamic} variables, which present a field
configuration on the background of the system (\ref{(a2.6)}) and
(\ref{(a2.6+)}). To describe this dynamical configuration we
construct a corresponding Lagrangian called as a {\it dynamical} one
\cite{[15]}. After substituting the decompositions (\ref{(a2.4)})
and (\ref{(a2.5)}) into the Lagrangian $\lag^E$ of the action
(\ref{(a2.1)}) and subtracting zero's and linear in $\hat
l^{\mu\nu}$ and  $\phi^A$ terms of the functional expansion of the
Lagrangian $\lag^E$ one has
 \be
 {\lag}^{dyn}  =
 {\lag}^E(\Bg+l,\,\Bar \Phi+\phi) -
 \hat l^{\mu\nu}
 {{\del \Bar{\lag^E}} \over {\del \Bar{\hat g}^{\mu\nu}}} -
\phi^A {{\delta \Bar {\lag^E}}\over{\delta \Bar\Phi^A}} -
 \Bar{\lag^E} - {1 \over {2\k }}\di_\alf \hat k^\alf
= -{1\over{2\k}}\lag^g + \lag^m\, .
 \m{(2.10)}
 \ee
As is seen, zero's order term is the background Lagrangian, whereas
the linear term is proportional to the left hand sides of the
background equations (\ref{(a2.6)}) and (\ref{(a2.6+)}).

In Eq. (\ref{(2.10)}) a vector density $\hat k^\alf$ is not
concreted. However, consider $\hat k^\alf$ defined as
 \be
\hat k^\alf \equiv \hat g^{\alpha\nu}\Del^\mu_{\mu\nu} - \hat
g^{\mu\nu} \Del^\alpha_{\mu\nu}\,
 \m{k-KBL}
 \ee
with $\Delta^\rho_{\mu\nu}$ presented as the perturbations of the
Cristoffel symbols
 \be
\Del^\alpha_{\mu\nu} \equiv \Gamma^\alpha_{\mu\nu} - \Bar
{\Gamma}^\alpha_{\mu\nu} = \half g^{\alf\rho}\l( \BD_\mu g_{\rho\nu}
+ \BD_\nu g_{\rho\mu} - \BD_\rho g_{\mu\nu}\r)\, , \m{DeltaDef}
 \ee
and depending on $\hat l^{\mu\nu}$ through the decomposition
(\ref{(a2.4)}). Then a pure gravitational part in the Lagrangian
(\ref{(2.10)}) is presented in the form:
 \bea \lag^{g}& =& \hat R (\Bar {\hat
g}^{\mu\nu} + \hat l^{\mu\nu}) - \hat l^{\mu\nu} \Bar{R}_{\mu\nu} -
\Bar{\hat g}^{\mu\nu}\Bar R_{\mu\nu} +\di_\mu \hat k^\mu\,
 \nonumber \\&=&
-(\Delta^\rho_{\mu\nu} - \Delta^\sig_{\mu\sig}\delta^\rho_\nu)\Bar
D_\rho \hat l^{\mu\nu} + (\Bar{{\hat g}}^{\mu\nu} + \hat l^{\mu\nu})
\l(\Delta^\rho_{\mu\nu}\Delta^\sig_{\rho\sig}
-\Delta^\rho_{\mu\sig}\Delta^\sig_{\rho\nu}\r)\,.
 \m {(a2.16)}
 \eea
It depends only on the first derivatives of the gravitational
variables $\hat l^{\mu\nu}$. In the case of a flat background the
Lagrangian (\ref{(a2.16)}) transfers to the covariant Lagrangian
suggested by Rosen \cite{[2]}, which has been rediscovered in
\cite{Katz85} and \cite{GP86}. The matter part of (\ref{(2.10)}) is
\be
 \lag^m  =
{\lag}^M\l(\Bg+ l, \,\Bar {\Phi} + \phi\r) \nonumber - \hat
l^{\mu\nu} {{\delta \Bar {\lag^M}}\over{\delta \Bar {\hat
g}^{\mu\nu}}}- \phi^A {{\delta \Bar {\lag^M}}\over{\delta
\Bar\Phi^A}} - \Bar{\lag^M}\, . \m {(a2.15)} \ee

\subsection{The Einstein equations in the field formulation}
\m{generalequations}

The variation of the action with the Lagrangian (\ref{(2.10)}) with
respect to $\hat l^{\alf\beta}$ and the contraction with
$\sqrt{-\Bar g} (\delta^\alf_\mu\delta^\beta_\nu - \half \Bar
g_{\mu\nu}\Bar g^{\alf\beta})$ give the field equations in the form:
 \be \hat
G^L_{\mu\nu} + \hat \Phi^L_{\mu\nu} = \k\l({\hat t}^g_{\mu\nu} +
{\hat t}^m_{\mu\nu}\r) \equiv \k{\hat t}^{(tot)}_{\mu\nu}\, .
 \m{(a2.17)}
 \ee
They coincide with the form (\ref{(4.10)}), only now in the 2-nd
order formalism. The left hand side of Eq. (\ref{(a2.17)}) is linear
in $\hat l^{\mu\nu}$ and $\phi^A$.  It consists of the pure
gravitational part
 \bea &{}& \hat G^L_{\mu\nu}(\hat l) \equiv
{\delta \over {\delta\Bar{g}^{\mu\nu}}} \hat l^{\rho\sig}
{{\delta\Bar{\hat R}}\over{\delta \Bar{\hat g}^{\rho\sig}}}
\label{G-L} \\ &\equiv & \half \l({\Bar D_\rho}{\Bar D^\rho}\hat
l_{\mu\nu} + {\Bar g_{\mu\nu}}{\Bar D_\rho}{\Bar D_\sig}\hat
l^{\rho\sig} - {\Bar D_\rho}{\Bar D_\nu}\hat l_{\mu}^{~\rho} - {\Bar
D_\rho}{\Bar D_\mu}\hat l_{\nu}^{~\rho}\r) ,
 \m {(a2.18)}
 \eea
which is the covariantized expression (\ref{LinearEqs}), and of the
matter part
 \be
\hat \Phi^L_{\mu\nu}(\hat l, \phi) \equiv -2\k {\delta \over
{\delta\Bar{g}^{\mu\nu}}} \l(\hat l^{\rho\sig}
{{\delta\Bar{\lag^M}}\over{\delta \Bar{\hat g}^{\rho\sig}}} + \phi^A
{{\delta {\Bar{\lag^M}}}\over{\delta {\Bar\Phi^A}}}\r)\, .
\m{(a2.19)}
 \ee
It disappears for Ricci-flat backgrounds (\ref{BackRicciFlat}), and
then Eq. (\ref{(a2.17)}) acquires the form of Eq.
(\ref{FullPerturbEqs}). The right hand side of Eq. (\ref{(a2.17)})
is the symmetrical energy-momentum tensor density
 \be
{\hat t}^{(tot)}_{\mu\nu} \equiv 2{{\delta{\lag}^{dyn}}\over{\delta
\Bar g^{\mu\nu}}} \equiv 2{{\delta}\over {\delta \Bar
g^{\mu\nu}}}\l(-{1\over{2\k}}\lag^g + \lag^m  \r) \equiv {\hat
t}^g_{\mu\nu} +  {\hat t}^m_{\mu\nu}. \m{(2.20)}
 \ee
The explicit form of the gravitational part is
 \be
{\hat t}^g_{\mu\nu} = {1 \over \k} \l[\sqrt{-\Bar g}
\l(-\del^\rho_\mu \del^\sig_\nu +\half \Bar g_{\mu\nu} \Bar
g^{\rho\sig}\r)\l(\Del^\alf_{\rho\sig}\Del^\beta_{\alf\beta} -
\Del^\alf_{\rho\beta}\Del^\beta_{\alf\sig}\r) + \Bar D_\tau \hat
Q^\tau_{\mu\nu}\r]\, ;\m{(2.20')}
 \ee
 \bea 2\hat Q^\tau_{\mu\nu} &\equiv & -\Bar g_{\mu\nu} \hat
l^{\alf\beta}\Del^\tau_{\alf\beta}+ \hat l_{\mu\nu}
\Del^\tau_{\alf\beta}\Bar g^{\alf\beta}- \hat l^\tau_{\mu}
\Del^\alf_{\nu\alf}- \hat l^\tau_{\nu} \Del^\alf_{\mu\alf} +\hat
l^{\beta\tau}\l( \Del^\alf_{\mu\beta}\Bar g_{\alf\nu} +
 \Del^\alf_{\nu\beta}\Bar g_{\alf\mu}\r)
\nonumber \\ &+ & \hat l^{\beta}_{\mu}\l( \Del^\tau_{\nu\beta}-
 \Del^\alf_{\beta\rho}\Bar g^{\rho\tau}\Bar g_{\alf\nu}\r)
+\hat l^{\beta}_{\nu}\l( \Del^\tau_{\mu\beta}-
 \Del^\alf_{\beta\rho}\Bar g^{\rho\tau}\Bar g_{\alf\mu}\r)\, .
\m{(2.20'')}
 \eea
The matter part is expressed through the usual matter
energy-momentum tensor density $\hat T_{\mu\nu}$ of the Einstein
theory as
 \bea
 \hat t^m_{\mu\nu}& = &\hat T_{\mu\nu} -
{\half}g_{\mu\nu}\hat T_{\alpha\beta}g^{\alpha\beta} -
{\half}\Bg_{\mu\nu}\Bg^{\alpha\beta}\l(\hat T_{\alpha\beta}
- {\half}g_{\alpha\beta}\hat T_{\pi\rho}g^{\pi\rho}\r)\nonumber \\
&{}& -\,2{\delta \over {\delta\Bar g^{\mu\nu}}} \l(\hat l^{\rho\sig}
 {{\delta\Bar{\lag^M}}\over{\delta \Bar{\hat g}^{\rho\sig}}} + \phi^A
{{\delta {\Bar{\lag^M}}}\over{\delta {\Bar\Phi^A}}}\r) - \Bar{\hat
T}_{\mu\nu}\, . \m{(2.20+)}
 \eea

Taking into account the definitions (\ref{(a2.19)}), (\ref{(2.20)})
and (\ref{(2.20+)}) in the field equations (\ref{(a2.17)}) one can
rewrite them in the form:
 \be \hat G^L_{\mu\nu} = \k\l( {\hat
t}^g_{\mu\nu} +  \delta{\hat t}^M_{\mu\nu} \r) = \k\hat
t^{(eff)}_{\mu\nu}\,;
 \m{B30}
 \ee
 \be
\delta{\hat t}^M_{\mu\nu}\equiv {\hat t}^M_{\mu\nu} - \Bar {\hat
t^M}_{\mu\nu} = \hat T_{\mu\nu}
 -\half g_{\mu\nu}g^{\rho\sig}\hat T_{\rho\sig} - \half \Bar
 g_{\mu\nu}\Bar g^{\rho\sig} \l(\hat T_{\rho\sig} -\half
 g_{\rho\sig}g^{\lam\tau}\hat T_{\lam\tau}\r) - \Bar{\hat T}_{\mu\nu}\, .
 \m{B31}
 \ee
The equation (\ref{B30}) has the form of Eq. (\ref{FullPerturbEqs})
even on {\it arbitrary} backgrounds. The price is that the effective
source $\hat t^{(eff)}_{\mu\nu}$, including the matter part as
$\delta{\hat t}^M_{\mu\nu}$, does not follow from the Lagrangian
(\ref{(2.10)}) directly. However, this matter part could be
classified as a perturbation of
 \be
{\hat t}^M_{\mu\nu} \equiv \frac{\delta \lag^M\l(\Bar\Phi^A +
\phi^A;~\Bar{\hat g}^{\mu\nu} + \hat l^{\mu\nu} \r)}{\delta \Bar{
g}^{\mu\nu}}\, .
 \m{HAT-tM}
 \ee
Return to the equations (\ref{(a2.17)}) in the whole. Transfer the
energy-momentum tensor density to the left hand side  and use the
definitions (\ref{G-L}), (\ref{(a2.19)}) and (\ref{(2.20)}) with
(\ref{(2.10)}):
 \bea
 &{}& \hat G^L_{\mu\nu} + \hat \Phi^L_{\mu\nu}
- \k{\hat t}^{(tot)}_{\mu\nu} \equiv \nonumber\\
&{}& - 2\k {{\di \Bar {\hat g}^{\rho\sig}}\over {\di \Bar
g^{\mu\nu}}} {{\delta}\over {\delta {\hat l^{\rho\sig}}}} \l[ -{1
\over {2\k }} \hat R\l(\Bar{\hat g}^{\alf\beta} + \hat
l^{\alf\beta}\r) + {\lag}^M \l(\Bar\Phi^A +
\phi^A;~\Bar{\hat g}^{\mu\nu} + \hat l^{\mu\nu}\r)\r] \nonumber \\
&{}&+~ 2\k{{\delta}\over {\delta \Bar g^{\mu\nu}}} \l( -{1 \over
{2\k }} \Bar{\hat R} +  \Bar{\lag}^M \r) = 0\,. \m {(2.17')}
 \eea
Because the second line contains the operator of the background
Einstein equations in (\ref{(a2.6)}) one can assert that Eq.
(\ref{(a2.17)}) is equivalent to the Einstein equations
(\ref{(a2.2)}).

By the same way the perturbed matter equations are constructed. They
have the form:
 \be
 \Phi^L_A = t^m_A
 \m{MatterPert}
 \ee
where
  \bea
\Phi^L_A(\hat l, \phi) &\equiv & -{\delta \over {\delta\Bar \Phi^A}}
\l(\hat l^{\rho\sig} {{\delta\Bar{\lag^M}}\over{\delta \Bar{\hat
g}^{\rho\sig}}} + \phi^B {{\delta {\Bar{\lag^M}}}\over{\delta
{\Bar\Phi^B}}}\r)\, , \m{(a2.19+)}\\
 t^m_A & \equiv & {{\delta \lag^m}\over {\delta\Bar \Phi^A}}\, .
 \m{MatterCurrent}
 \eea

\subsection{Expansions}
\m{Expansions}

The methods of the exact field-theoretical formulation give a
possibility to construct an approximate scheme easily and in clear
expressions up to an arbitrary order in perturbations. Let us show
this. Assuming enough smooth functions, expand the Lagrangian
$\lag^E(\Bar g + l,\, \Bar \Phi + \phi)$ as
 \bea
{\lag}^E & = &\Bar {\lag^E} + \hat l^{\rho\sig}
{{\delta\Bar{\lag^E}}\over{\delta \Bar{\hat g}^{\rho\sig}}} + \phi^B
{{\delta {\Bar{\lag^E}}}\over{\delta {\Bar\Phi^B}}} + \nonumber\\
&+& \frac{1}{2!} \hat l^{\alf\beta} {{\delta} \over{\delta \Bar
{\hat g}^{\alf\beta}}} \hat l^{\rho\sig} {{\delta  {{{\Bar{\lag^E}
}}}}\over{\delta \Bar { \hat g}^{\rho\sig}}}+ \hat l^{\rho\sig}
{{\delta }\over{\delta \Bar { \hat g}^{\rho\sig}}}\phi^A {{\delta
{{\Bar{\lag^E}}}}\over{\delta \Bar\Phi^A}} + \frac{1}{2!}\phi^B
{{\delta }\over{\delta \Bar\Phi^B}}\phi^A {{\delta {{\Bar{\lag^E}}}}
\over{\delta \Bar\Phi^A}}+ \nonumber\\
&+& \frac{1}{3!}\hat l^{\mu\nu} {{\delta} \over{\delta \Bar {\hat
g}^{\mu\nu}}} \hat l^{\alf\beta} {{\delta} \over{\delta \Bar {\hat
g}^{\alf\beta}}} \hat l^{\rho\sig} {{\delta
{{\Bar{\lag^E}}}}\over{\delta \Bar { \hat g}^{\rho\sig}}}+    \ldots
+ div\, .
 \m{LE-perturb}
 \eea
Then the dynamical Lagrangian (\ref{(2.10)})  acquires the form:
 \be
{\lag}^{dyn}  = \frac{1}{2!} \hat l^{\alf\beta} {{\delta}
\over{\delta \Bar {\hat g}^{\alf\beta}}} \hat l^{\rho\sig} {{\delta
{{{\Bar{\lag^E}}}}}\over{\delta \Bar { \hat g}^{\rho\sig}}}+ \hat
l^{\rho\sig} {{\delta }\over{\delta \Bar { \hat
g}^{\rho\sig}}}\phi^A {{\delta {{\Bar{\lag^E}}}}\over{\delta
\Bar\Phi^A}} + \frac{1}{2!}\phi^B {{\delta }\over{\delta
\Bar\Phi^B}}\phi^A {{\delta  {{\Bar{\lag^E}}}}\over{\delta
\Bar\Phi^A}}+   \ldots  + div\, .
 \m{Ldyn-perturb}
 \ee
First, note that a divergence in (\ref{Ldyn-perturb}) is not so
important because divergences vanish under the Lagrange derivative.
Second, now we can explain why in the linear terms of the Lagrangian
(\ref{(2.10)}) the background equations are not taken into account
before variation. Indeed, the Lagrangian $\lag^E(\Bar g + l,\, \Bar
\Phi + \phi)$ in (\ref{(2.10)}) contains the same linear terms with
only the opposite sign not explicitly (the formula
(\ref{LE-perturb}) shows this). Therefore, in fact, these linear
terms are compensated, and the real lowest order in the expansion of
$\lag^{dyn}$ is a quadratic one (\ref{Ldyn-perturb}). In this
relation, recall that in the usual geometrical description of GR the
attempt to define the energy-momentum tensor through $\delta
\lag^E/\delta g^{\mu\nu}$ leads to zero on the solutions of the
Einstein equations. In contrast, ${\hat t}^{(tot)}_{\mu\nu}$ in
(\ref{(2.20)}) does not vanish on the field equations. The reason is
just in the presence in the Lagrangian (\ref{(2.10)}) of these
linear terms.

Third, under necessary assumptions the series (\ref{Ldyn-perturb})
can be interrupted at a corresponding order. Thus the approximate
Lagrangian for the perturbed system can be obtained. Its variation
gives both approximate field equations and energy-momentum tensor.
For example, the quadratic approximation of (\ref{Ldyn-perturb})
gives a possibility a) to construct the linear equations
 \bea
-\frac{1}{2\k}\l(\hat G^L_{\mu\nu}(\hat l, \phi)+\hat
\Phi^L_{\mu\nu}(\hat l, \phi)\r) &\equiv & {\delta \over {\delta\Bar
g^{\mu\nu}}} \l(\hat l^{\rho\sig} {{\delta\Bar{\lag^E}}\over{\delta
\Bar{\hat g}^{\rho\sig}}} + \phi^B {{\delta
{\Bar{\lag^E}}}\over{\delta
{\Bar\Phi^B}}}\r)= 0\, , \m{linear-l}\\
-\Phi^L_A(\hat l, \phi) &\equiv & {\delta \over {\delta\Bar \Phi^A}}
\l(\hat l^{\rho\sig} {{\delta\Bar{\lag^E}}\over{\delta \Bar{\hat
g}^{\rho\sig}}} + \phi^B {{\delta {\Bar{\lag^E}}}\over{\delta
{\Bar\Phi^B}}}\r)= 0\, , \m{linear-phi}
 \eea
b) to construct the quadratic energy-momentum tensor:
 \be
 t^{(tot)}_{\mu\nu}  = \frac{2}{\sqrt{\Bar g}}\frac{\delta}{\delta
\Bar g^{\mu\nu}}\l(\frac{1}{2!} \hat l^{\alf\beta} {{\delta}
\over{\delta \Bar {\hat g}^{\alf\beta}}} \hat l^{\rho\sig} {{\delta
{{{\Bar{\lag^E} }}}}\over{\delta \Bar { \hat g}^{\rho\sig}}}+ \hat
l^{\rho\sig} {{\delta }\over{\delta \Bar { \hat
g}^{\rho\sig}}}\phi^A {{\delta {{\Bar{\lag^E}}}}\over{\delta
\Bar\Phi^A}} + \frac{1}{2!}\phi^B {{\delta }\over{\delta
\Bar\Phi^B}}\phi^A {{\delta  {{\Bar{\lag^E}}}}\over{\delta
\Bar\Phi^A}}\r) \, .
 \m{t-tot-perturb}
 \ee
The cubic approximation of (\ref{Ldyn-perturb}) gives a possibility
a) to construct the field equations including quadratic terms (which
are related to the energy-momentum tensor), and  b) to construct the
energy-momentum tensor, including quadratic and cubic parts, {\em
etc. }

Fourth, the expansions, like (\ref{LE-perturb}) -
(\ref{t-tot-perturb}), are used in quantum field theories
\cite{DeWitt-book} and called as functional expansions. As is seen,
in the framework of the classic theory the functional expansions
(\ref{Ldyn-perturb}) - (\ref{linear-phi}) give, in fact, the
algorithm for constructing the approximate systems, thus an each
order can be obtained automatically.

\subsection{Gauge transformations and their properties}
\m{fieldgauge}

The important property of the field formulation is the gauge
invariance. Usually gauge transformations in GR and other metric
theories are connected with mapping a spacetime onto itself that is
connected with differentiable coordinate transformations
 \be
 {x^\prime}^\alpha =
f^\alpha(x^\beta)\, .
 \m{x'=fx}
 \ee
These transformations can be connected with the smooth vector field
$\xi^\alf$:
 \be
{x^\prime}^\alpha = x^\alpha + \xi^\alpha (x) + {1\over{2!}}~
\xi^\beta
 \xi^\alpha_{~~,\beta} + {1\over{3!}}~ \xi^\pi (\xi^\beta
\xi^\alpha_{~~,\beta})_{,\pi} + {...}\,.  \m {(5.1)}
 \ee
To map the spacetime onto itself one has to follow the standard way
\cite{Eisenkh} - \cite{Mitzk}. After the coordinate transformations
(\ref{x'=fx}) (or (\ref{(5.1)})) the metric density, for example, is
transformed as ${\hat g}^{\mu\nu}(x) \goto {\hat g}'^{\mu\nu}(x')$.
Then return to the points with quantities $x$ within a new frame
$\l\{ x'\r\}$. After that one has to compare geometrical objects of
the initial spacetime and of the mapped spacetime in the points with
quantities $x$. The comparison can be carried out both without
$\xi^\alf$ in the terms of (\ref{x'=fx}) and with  $\xi^\alf$
included in  (\ref{(5.1)}):
 \bea
{\hat g}'^{\mu\nu}(x) &=& \hat g^{\mu\nu}(x) + \delta_f \hat
g^{\mu\nu}= \hat g^{\mu\nu}(x) + \sum^{\infty}_{k = 1}{1\over{k!}}~
\hbox{$\pounds$}_\xi^k  {\hat g^{\mu\nu}}(x)\,,\m{(5.6)}\\
{\Phi}'^A(x) &=& \Phi^A(x) + \delta_f \Phi^A(x)=\Phi^A(x) +
\sum^\infty_{k = 1}{1\over{k!}}~\hbox{$\pounds$}_\xi^k
{\Phi^A}(x)\,.
 \m{(5.6+)}
 \eea

The next property is very useful. Assume that geometrical objects
$\Psi^{B}$ are differentiable functions of other geometrical objects
$\psi^{A}$ and their derivatives, but are not explicit functions of
coordinates. Then it is clear that a simple substitution gives
$\Psi^{B}(\psi'^{A}) = \Psi'^{B}$, and one has
 \be
\Psi^{B}(\psi'^A(x)) = \Psi^{B}(\psi^A(x)) + \delta_f \Psi^{B} =
\Psi^{B}(\psi^A(x)) + \sum^\infty_{k=1} {1\over k!} {\Lix}^k
\Psi^{B}\, .
 \m{(5.3)}
 \ee

Now let us define the gauge transformations for the dynamical
variables in the framework of the field formulation of GR:
 \bea
{\hat l}'^{\mu\nu} &= &\hat l^{\mu\nu} + \delta_f \l(\Bar {\hat
g}^{\mu\nu} + \hat l^{\mu\nu}\r)=\hat l^{\mu\nu} + \sum^{\infty}_{k
= 1}{1\over{k!}}~ \hbox{$\pounds$}_\xi^k \l(\Bar {\hat g}^{\mu\nu} +
\hat l^{\mu\nu}\r),
\m{(5.9)} \\
{\phi}'^A &=& \phi^A + \delta_f\l(\Bar\Phi^A+\phi^A\r) =\phi^A +
\sum^\infty_{k = 1}{1\over{k!}}~\hbox{$\pounds$}_\xi^k
\l(\Bar\Phi^A+\phi^A\r). \m{(5.9+)}
 \eea
The assumptions above, in fact, state that  the operators $\delta_f$
and $\sum^\infty_{k = 1}{\textstyle {1\over{k!}}}~\hbox
{$\pounds$}_\xi^k $ are equivalent. However, the operator $\delta_f$
could present a wider class of transformations, which cannot be
expressed through infinite series. We conserve a possibility to
consider such transformations,  however more frequently we will use
the sums, keeping in mind that they can be changed by $\delta_f$
also.

Now, let us substitute (\ref{(5.9)}) and (\ref{(5.9+)}) into the
dynamical Lagrangian (\ref{(2.10)}). One finds that this
substitution into ${\lag}^{E}(\Bar{\hat g}+\hat l, \Bar{\Phi}+\phi)$
just permits to use the property (\ref{(5.3)}). Then finally one
obtains
 \bea
{\lag}'^{dyn} &=&{\lag}^{dyn} + \sum^\infty_{k=1} {1\over k!}
{\Lix}^k {\lag}^{E}(\Bar{\hat g}+\hat l, \Bar{\Phi}+\phi) -
{1 \over {2\k }}\di_\alf \l(\hat k'^\alf - \hat k^\alf\r)\nonumber \\
&{} &- \l(\hat l'^{\mu\nu}  - \hat l^{\mu\nu}\r)
 {{\del {\Bar{\lag^E}}} \over {\del \Bar{\hat g}^{\mu\nu}}}  -
\l(\phi'^A - \phi^A\r) {{\delta  {\Bar{\lag^M}}}\over{\delta
\Bar\Phi^A}}\, . \m{lag-dyn'}
 \eea
 Because ${\lag}^{E}$ is the scalar density of the weight
$+1$ all the terms under the sum in (\ref{lag-dyn'}) are
divergences. Thus ${\lag}^{dyn}$ is gauge invariant up to a
divergence if the background equations (\ref{(a2.6)}) and
(\ref{(a2.6+)}) hold.

Considering the gauge invariance properties of the field equations
we use their form (\ref{(2.17')}). The substitution of (\ref{(5.9)})
and (\ref{(5.9+)}) with the  use of the property (\ref{(5.3)}) gives
 \bea
&{}& \l[\hat G^L_{\mu\nu} + \hat \Phi^L_{\mu\nu} - \k{\hat
t}^{(tot)}_{\mu\nu} \r]' =
 \l[\hat G^L_{\mu\nu} + \hat \Phi^L_{\mu\nu}
- \k{\hat t}^{(tot)}_{\mu\nu} \r]
\nonumber \\
&+& {{\di \Bar {\hat g}^{\rho\sig}}\over {\di \Bar g^{\mu\nu}}}
\sum^\infty_{k=1} {1\over k!} {\Lix}^k\l[ {{\di \Bar
g^{\delta\pi}}\over {\di \Bar{\hat g}^{\rho\sig}}}
 \l(\hat G^L_{\delta\pi} + \hat \Phi^L_{\delta\pi}
- \k{\hat t}^{(tot)}_{\delta\pi} \r) -2\k {{\delta{\Bar {\lag^E}}
}\over {\delta \Bar{\hat g}^{\rho\sig}}} \r]\, . \m {(2.17-gauge)}
 \eea
Thus the field equations are gauge invariant on solutions of
themselves and with using the background equations (\ref{(a2.6)})
and (\ref{(a2.6+)}). Analogous transformations could be presented
for the matter equations (\ref{MatterPert}).

Concerning the energy-momentum tensor, it is not gauge invariant.
Even on the dynamical equations, as it follows from
(\ref{(2.17-gauge)}), under the transformations (\ref{(5.9)}) and
(\ref{(5.9+)}) one has
 \be
 \k{{\hat t}}^{\prime(tot)}_{\mu\nu} =  \k{\hat t}^{(tot)}_{\mu\nu}
  + \hat G^L_{\mu\nu}(l'-l) + \hat \Phi^L_{\mu\nu}(l'-l;\,\phi' -\phi)\, .
  \m{tei-gauge}
  \ee
The mathematical reason is that the background equations in the
Lagrangian (\ref{lag-dyn'}) cannot be taken into account before
variation. In the case of the Ricci-flat backgrounds
(\ref{BackRicciFlat}) one has $\hat \Phi^L_{\mu\nu}=0$, therefore
the energy-momentum  ${\hat t}^{(tot)}_{\mu\nu}$ is gauge invariant
up to $\hat G^L_{\mu\nu}$
--- covariant divergence (see (\ref{(a2.18)})). Let us turn also to
the equivalent form (\ref{B30}) of the field equations with the
effective source ${\hat t}^{(eff)}_{\mu\nu}$. For the operator of
the field equations $\hat G^L_{\mu\nu}  - \k{\hat
t}^{(eff)}_{\mu\nu} $ the form of the transformations
(\ref{(2.17-gauge)}) can be used without changing. Then on the field
equations one has
 \be
\k{{\hat t}}^{\prime(eff)}_{\mu\nu} =  \k{\hat t}^{(eff)}_{\mu\nu} +
\hat G^L_{\mu\nu}(l'-l)\, ,
 \m{eff-gauge}
 \ee
i.e. for all the kinds of backgrounds ${\hat t}^{(eff)}_{\mu\nu}$ is
gauge invariant up to a covariant divergence. It is not surprising
that both the energy-momentum tensors are not gauge invariant. It is
a
\begin{figure*}[t]
    \centering
    \includegraphics*%
    [width=7cm]%
    {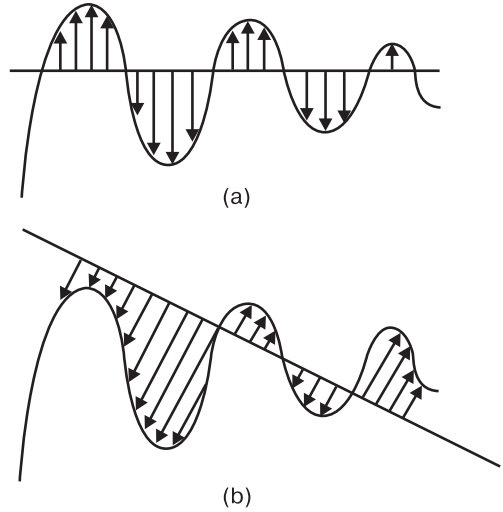}\\
    \caption{The perturbations (a) and (b) connected by the
gauge transformations.}
    \label{F1}%
    \end{figure*}
waited result. Indeed, this reflects the fact that energy and other
conserved quantities in GR are not localized. Moreover, the formulae
(\ref{tei-gauge}) and  (\ref{eff-gauge}) are very useful because
they give a {\it quantitative} and {\it constructive} description of
the non-localization. See the discussion in subsection
\ref{Einstein-papers}.

The transformations (\ref{(5.9)}) and (\ref{(5.9+)}) are directly
connected with a way of mapping a perturbed spacetime onto a given
background spacetime (a perturbed solution onto a given solution).
In the other words, they are connected with a definition of
perturbations with respect to a given background. Let us consider a
solution in the geometrical form $\hat g^{\mu\nu}(x)$. Next let us
map a spacetime onto itself following the prescription at the
beginning this subsection. Then we obtain $\hat g'^{\mu\nu}(x)$.
After that we decompose both of the solutions into dynamic and
background parts:
 \bea
\hat g^{\mu\nu}(x) &=& \Bar {\hat g}^{\mu\nu}(x) + \hat
l^{\mu\nu}(x)\, , \m{(5.10)}\\
 (\hat g'^{\mu\nu})(x)& =&
\Bar {\hat g}^{\mu\nu}(x) + \hat l'^{\mu\nu}(x)\,. \m{(5.12)}
 \eea
The main property of (\ref{(5.10)}) and (\ref{(5.12)}) is that the
background metric is the same. Then it turns out  that $\hat
l^{\mu\nu}(x)$ and $\hat l'^{\mu\nu}(x)$ are connected by the
transformations in (\ref{(5.9)}). This situation is interpreted as
follows. The same background in (\ref{(5.10)}) and (\ref{(5.12)}) is
chosen by different ways that symbolically is illustrated on the
figure \ref{F1}. The curves in both the cases (a) and (b) symbolize
the same solution of GR in the geometrical form, whereas the
straight lines present a background, say, a Minkowski space.  The
perturbations in the cases (a) and (b) are different, but they are
solutions to the equations of the field formulation connected by the
gauge transformations, and, thus, they are the same solution in
different forms. In spite of that the gauge transformations in the
field formulation have the evident geometrical origin, they can be
interpreted as inner gauge transformations, like in standard gauge
theories. Indeed, they act only onto the dynamical variables
(perturbations), whereas the backgrounds variables and coordinates
do not change.

As discussed in Introduction, in many of applications it is
important to consider equations and gauge transformations in linear
and quadratic approximations. Assume that perturbations and their
derivatives are small: $\hat l^{\mu\nu} \ll \Bar{\hat g}^{\mu\nu}$,
$\phi^A \ll \Bar\Phi^A$, $\hat l^{\mu\nu} \approx \di_\alf\hat
l^{\mu\nu}\ldots$ and $\phi^A \approx \di_\alf \phi^A \approx
\ldots~$. Assuming that the background equations (\ref{(a2.6)}) give
a connection $\Bar{\hat g}^{\mu\nu} \approx h(\k)\Bar\Phi^A$ with a
coefficient $h(\k)$ depending on the Einstein constant one has to
set ${\hat l^{\mu\nu}} \approx h(\k){\phi^A}$, {\it etc}. Now,
rewrite the equations (\ref{(a2.17)}) up to the second order in
perturbations:
 \be
\hat G^L_{\mu\nu}(\hat l) + \hat \Phi^L_{\mu\nu}(\hat l,~\phi) -
\k\hat t^{(tot2)}_{\mu\nu}(\hat l\hat l,~\hat l\phi,~\phi\phi) = 0\,
. \m{(2.17-q)}
 \ee
Assuming iterations  the perturbations can be expanded as $\hat
l^{\mu\nu} = \hat l^{\mu\nu}_{1} + \hat l^{\mu\nu}_{2} + ...~$, and
$\phi^{A} = \phi^{A}_{1} + \phi^{A}_{2} + ...~$. Then the linear
equations acquire the form:
 \be \hat G^L_{\alf\beta}(\hat
l_{1}) + \hat \Phi^L_{\alf\beta}(\hat l_{1},~\phi_{1}) = 0\,,
\m{lin}
 \ee
whereas the quadratic equations have the form:
  \be
\hat G^L_{\alf\beta}(\hat l_{2}) + \hat \Phi^L_{\alf\beta}(\hat
l_{2},~\phi_{2}) -
 \k \l(\hat t^{(gr2)}_{\alf\beta}(\hat l_{1}\hat l_{1}) +
\hat t^{(m2)}_{\alf\beta}(\hat l_{1}\hat l_{1}, ~\hat
l_{1}\phi_{1},~\phi_{1}\phi_{1})\r) = 0\, . \m{quadr}
 \ee
Besides, assuming $\xi^{\mu} = \xi_{1}^{\mu} + \xi_{2}^{\mu}+ ...$
with  $ \xi_{1}^{\mu} \approx \di_\alf \xi_{1}^{\mu}\approx \ldots
\approx \hat l^{\mu\nu}_{1}\approx h(\k) \phi^A_{1}$  and
$\xi_{2}^{\mu} \approx \di_\alf \xi_{2}^{\mu} \approx \ldots \approx
\hat l^{\mu\nu}_{2}\approx h(\k) \phi^A_{2}$, one has a linear
version of (\ref{(5.9)}) in the simple form:
 \bea \hat l'^{\mu\nu}_{1} &=&
\hat l^{\mu\nu}_{1} + {{\hbox{$\pounds$}}}_{\xi_{1}} \Bar{\hat
g}^{\mu\nu} = \hat l^{\mu\nu}_{1}+ \Bar D^\mu \hat \xi_{1}^\nu +
\Bar D^\nu \hat \xi_{1}^\mu - \Bar {\hat g}^{\mu\nu}~\Bar D_\rho
\xi_{1}^\rho\, ,
\m{linear-trans} \\
\phi'^{A}_{1} &=& \phi^{A}_{1} +{{\hbox{$\pounds$}}}_{\xi_{1}}
\Bar{\Phi}^{A}. \m{linear-trans+}
 \eea
Under these transformations the equations (\ref{lin}) are
transformed as
 \bea
&{}&\l[\hat G^L_{\mu\nu} + \hat \Phi^L_{\mu\nu}\r]' = \l[\hat
G^L_{\mu\nu} + \hat \Phi^L_{\mu\nu}\r] \nonumber \\&+&
\l(\sqrt{-\Bar g}\del^\rho_\mu\del^\sig_\mu - \half \Bar
g_{\mu\nu}\Bar{\hat g}^{\rho\sig}\r) {{\hbox{$\pounds$}}}_{\xi_{1}}
\l[ \Bar R_{\rho\sig} -\k\l(\Bar T_{\rho\sig} - \half \Bar
g_{\rho\sig}\Bar T\r) \r] \m{(2.17-gauge-1)}
 \eea
and are, thus, gauge invariant on the background equations. In the
simple case of the Ricci-flat background (\ref{BackRicciFlat}) the
linear transformations have only the form (\ref{linear-trans}),
without (\ref{linear-trans+}). Then the formula
(\ref{(2.17-gauge-1)}) transfers to the formula $\hat G'^L_{\mu\nu}
= \hat G^L_{\mu\nu}$, which expresses the gauge invariance of the
linear field of spin 2. It is the well known gauge transformations
in the linear gravity \cite{LL,[12]}.

The quadratic order of the gauge transformations has a form;
 \bea
\hat l'^{\mu\nu}_{2} &=& \hat l^{\mu\nu}_{2} +
{{\hbox{$\pounds$}}}_{\xi_{2}} \Bar{\hat g}^{\mu\nu}+
{1\over{2!}}{{\hbox{$\pounds$}}}^2_{\xi_{1}} \Bar{\hat g}^{\mu\nu} +
{{\hbox{$\pounds$}}}_{\xi_{1}} \hat l^{\mu\nu}_{1}
\nonumber \\
\phi'^{A}_{2} &=& \phi^{A}_{2} + {{\hbox{$\pounds$}}}_{\xi_{2}}
\Bar{\Phi}^{A}+ {1\over{2!}}{{\hbox{$\pounds$}}}^2_{\xi_{1}}
\Bar{\Phi}^{\mu\nu} + {\hbox{$\pounds$}}_{\xi_{1}} \phi^{A}_{1}\, .
\m{trans-(2)}
 \eea
Substitution of Eqs. (\ref{linear-trans}) and  (\ref{trans-(2)})
into (\ref{quadr}) give
 \bea
&{}& \l[\hat G^L_{\mu\nu}(\hat l_{2}) + \hat \Phi^L_{\mu\nu}(\hat
l_{2},~\phi_{2})
 - \k \hat t^{(tot2)}_{\mu\nu}\r]' =
\l[\hat G^L_{\mu\nu}(\hat l_{2}) + \hat \Phi^L_{\mu\nu}(\hat
l_{2},~\phi_{2})
 - \k \hat t^{(tot2)}_{\mu\nu}\r]
 \nonumber\\
 &+&
{{\di \Bar {\hat g}^{\rho\sig}}\over {\di \Bar g^{\mu\nu}}}
\l({\hbox{$\pounds$}}_{\xi_{2}}  + {1\over
2!}{\hbox{$\pounds$}}^2_{\xi_{1}}\r) \l[ \Bar R_{\rho\sig}
-\k\l(\Bar T_{\rho\sig} - \half \Bar g_{\rho\sig}\Bar T\r) \r] +
\nonumber\\
&+& {{\di \Bar {\hat g}^{\rho\sig}}\over {\di \Bar g^{\mu\nu}}}
{\hbox{$\pounds$}}_{\xi_{1}}\l[ {{\di \Bar g^{\delta\pi}}\over {\di
\Bar{\hat g}^{\rho\sig}}} \l[\hat G^L_{\delta\pi}(\hat l_{1}) + \hat
\Phi^L_{\delta\pi} (\hat l_{1},~\phi_{1})\r]\r]. \m{(2.17-gauge-2')}
 \eea
Thus, equations (\ref{quadr})  are gauge invariant on the background
equations (\ref{(a2.6)}) and on the linear equations (\ref{lin}). Of
course, the procedure can be continued in the next orders.

In this subsection, we were  based on the {\it exact} theory of
gauge transformations developed in our works \cite{GPP,[15]}.
Together with this, for the presentation we have used some of
details given in \cite{GP86,Petrov95,Petrov97}. An arising interest
to cosmological perturbations stimulates their more detailed study
including a second order approximation \cite{Bartolo}. Thus it turns
out necessary to examine the gauge transformations up to a second
order also. Such studies (independently on ours, but repeating them
in main properties) were carried out, for examples, in
\cite{Bruni,Abramo}.

\subsection{Differential conservation laws on special backgrounds}
\m{conservationlaws}

The energy-momentum tensor is the one of important objects of a
field theory in Minkowski space. Its differential conservation
together with symmetries of the Minkowski space permit to construct
integral conserved quantities (see subsection \ref{UsingPseudo}).
The field formulation of GR has also the conserved energy-momentum
in Minkowski space with the same properties (\ref{CLfirst}). In this
short subsection we demonstrate that the conservation law, like
(\ref{CLfirst}), also has a place on curved backgrounds, which are
important in many applications. Although conservation laws and
conserved quantities are constructed and studied in the next section
\ref{CL} in detail, by the logic of the presentation we include this
subsection here.

Firstly, consider Ricci-flat (including flat) backgrounds
(\ref{BackRicciFlat}), which have an independent meaning. This means
that $\Bar\Phi^A \equiv 0$ and $\Bar {\lag^M} \equiv 0$. Then the
Lagrangian (\ref{(2.10)}) is simplified to
 \be
 {\lag}^{dyn}  =
 -{1 \over {2\k }}  \lag^g  + \lag^m =
 -{1 \over {2\k }}  \lag^g
+ {\lag}^M \l(\phi^A;~\Bar{\hat g}^{\mu\nu} + \hat l^{\mu\nu}\r).
\m{(2.22)}
 \ee
The field equations (\ref{(a2.17)}) transform into the form
 \be \hat
G^L_{\mu\nu} = \k\l({\hat t}^g_{\mu\nu} + {\hat t}^m_{\mu\nu}\r)
\equiv \k{\hat t}^{(tot)}_{\mu\nu}\, . \m {(a2.23)}
 \ee
For  (\ref{BackRicciFlat})  one has identically $\Bar D_\nu{\hat
G}^{L\nu}_{\mu} \equiv 0$ and, thus, a divergence of Eq.
(\ref{(a2.23)}) leads to the differential conservation law:
 \be \Bar D_\nu{\hat t}^{(tot)\nu}_{\mu} = 0\, . \m{(2.23')} \ee

Now consider backgrounds presented by Einstein spaces in Petrov's
definition \cite{Petrov}, then the background equations are
 \be
\Bar R_{\mu\nu} = \Lambda \Bg_{\mu\nu}\,
 \m{R=Lg}
 \ee
where $\Lambda$ is a constant. Ricci-flat and AdS backgrounds are
particular cases of Einstein's spaces. The Lagrangian of the
background system has the form:
 \be \Bar {\lag^E} = -{1\over{2\k}}\Bar
{\hat R} + \Bar {\lag^M} = -{1\over{2\k}}\l(\Bar {\hat R} - 2\Lambda
\sqrt{-\bar g}\r)\, . \m {(a2.25)}
 \ee
Here, the constant $\Lambda$ is rather interpreted as `degenerated'
matter. Then, the dynamical Lagrangian (\ref{(2.10)}) transforms
into
 \be
 {\lag}^{dyn}  =
 -{1 \over {2\k }}  \lag^g  + \lag^m =
 -{1 \over {2\k }}  \lag^g
+  \l[{\lag}^M \l(\phi^A;~\Bar{\hat g}^{\mu\nu} + \hat l^{\mu\nu}\r)
- \hat l^{\mu\nu} {{\delta\Bar{\lag^M}}\over{\delta \Bar {\hat
g}^{\mu\nu}}} - {\Bar{\lag^M}}\r] \m{(a2.26)}
 \ee
and leads to the field equations (\ref{(a2.17)}) in the simple form:
 \be
\hat G^L_{\mu\nu} +  \Lambda\hat l^{\mu\nu} = \k\l( {\hat
t}^g_{\mu\nu} + {\hat t}^m_{\mu\nu}\r) \equiv \k {\hat
t}^{(tot)}_{\mu\nu}\, . \m {(a2.27)}
 \ee
Taking into account the background equations (\ref{R=Lg}) one has
identically
 \be
\Bar D_\nu\l(\hat G^{L\nu}_{\mu} + \Lambda\hat l_{\mu}^{\nu}\r)
\equiv 0.
 \m{DGLl=0}
 \ee
Thus, the differentiation of Eq. (\ref{(a2.27)}) gives the same
conservation law (\ref{(2.23')}). In heuristic form the differential
conservation law on AdS and de Sitter (dS) backgrounds was used in
\cite{AbbottDeser82}; in the Lagrangian description it was shortly
noted in \cite{GPP}; and, in the paper \cite{Deser87}, it was
studied in more detail.

\subsection{Different definitions of metric perturbations}
\m{arbitrarydecompositions}

In GR, the metric perturbations can be defined by different
decompositions:
 \bea
 g_{\mu\nu} & =& \Bg_{\mu\nu} + h_{\mu\nu},\nonumber\\
\hg^{\mu\nu} & =& \Bar {\hg}^{\mu\nu} + \hat l^{\mu\nu},\nonumber\\
g^{\mu\nu} &=& \Bg^{\mu\nu} + r^{\mu\nu},\nonumber\\
\ldots &=&
 \ldots + \ldots\, ,
 \m{B38+}
 \eea
not only by (\ref{(a2.4)}). Denoting components of metrical
densities in the united way
 \be
 g^{a} = \l\{g^{\mu \nu },~g_{\mu
\nu },~\sqrt{-g}g^{\mu \nu },~\sqrt{-g}g_{\mu \nu },~(-g)g^{\mu \nu
},~\ldots\r\} \m{(1)}
 \ee
we rewrite the action of GR as
 \be S={1\over c}\int
d^{4}x\lag^{E(a)}\equiv -{1\over {2\k c}}\int
d^{4}x\hat{R}(g^a)+{1\over c}\int d^{4}x\lag^{M}(\Phi ^{A},~g^a)\, .
\m{(2)}
 \ee
After its variation the Einstein equations take the generalized
form:
 \be
-{1\over 2\k}{{\delta \hat{R}}\over{\delta g^{a}}}+ {{\delta
\lag^M}\over {\delta g^{a}}} = 0\, .\m{(3)}
 \ee
The background action and equations have the corresponding to
(\ref{(2)}) and (\ref{(3)}) barred form. After that we present the
united form for  decompositions (\ref{B38+})
 \be
 g^a = \Bg^a + h^a\,,
 \m{B38}
 \ee
and, following the rules of (\ref{(2.10)}), construct the
generalized dynamical Lagrangian:
 \bea
&{}& {\lag}^{dyn}_{(a)}  =
 -{1 \over {2\k }}  \hat R\l(\Bar g^a + h^a\r)
+  {\lag^M} \l(\Bar\Phi^A + \phi^A;~\Bar g^a + h^a\r) \nonumber
 \\ &-&   h^a\l(
 -{1 \over {2\k }}  {{\del \Bar{\hat R}} \over {\del \Bar g^a}} +
 {{\del {\Bar{\lag^M}}} \over {\del \Bar g^a}} \r) -
\phi^A {{\delta {\Bar {\lag^M}}}\over{\delta \Bar\Phi^A}} - \l( -{1
\over {2\k }} \Bar{\hat R} +  {\Bar{\lag^M}} \r)
 -{1 \over {2\k }}\di_\nu \hat k^\nu .
\m{(2.10-a)}
 \eea
The total symmetrical energy-momentum tensor density is defined as
usual:
 \be
  {\hat t}^{(tot~a)}_{\mu\nu} \equiv
 2{{\delta\lag^{dyn}_{(a)}}\over{\delta
 \Bar g^{\mu\nu}}}\, .
 \m{B43}
 \ee
After substituting the expression (\ref{(2.10-a)}) into this
definition (identity)
 and taking into account Eq. (\ref{(3)}) and the barred Eq. (\ref{(3)})
 we obtain the Einstein equations
in the form (\ref{(a2.17)}):
 \be \hat G^{L(a)}_{\mu\nu} + \hat
\Phi^{L(a)}_{\mu\nu} = \k{\hat t}^{(tot~a)}_{\mu\nu}\, .\m{B39}
 \ee
Here, the linear left hand side is defined by the same operators
(\ref{G-L}) - (\ref{(a2.19)}), only now with independent variables
 \be
 \hat l^{\mu\nu}_{(a)} \equiv
 h^a {{ \di \Bar {\hat g}^{\mu\nu}} \over {\di \Bar g^a}}\, .
 \m{B40}
 \ee

However, a choice of two different arbitrary decompositions as $
g^a_1 = \Bar g^a_1 + h^a_1$ and  $g^a_2 = \Bar g^a_2 + h^a_2$, gives
the difference
 \be \hat l^{\mu\nu}_{(a2)} - \hat
 l^{\mu\nu}_{(a1)} = \hat \beta^{\mu\nu}_{(a)12}\, ,
  \m{beta}
  \ee
which is not less than quadratic in perturbations. Because
differences enter the linear expressions of equations (\ref{B39})
the energy-momentum tensor densities  ${\hat t}^{(tot~1a)}_{\mu\nu}$
and ${\hat t}^{(tot~2a)}_{\mu\nu}$ have the same differences too.
For the case of flat backgrounds this ambiguity was noted by
Boulware and Deser \cite{B-Deser}. Later we \cite{[15]} have
examined it for arbitrary curved backgrounds and arbitrary metric
theories. However, only in our works \cite{PK,PK2003b} this
ambiguity has been resolved, and we present this solution in
subsection \ref{BelOnArbitrary}.

\section{Conservation laws in GR}
\m{CL} \setcounter{equation}{0}

\subsection{Classical pseudotensors and superpotentials}
 \m{ClassPseudo}

During many decades after constructing GR pseudotensors and
superpotentials were main objects in constructing conservation laws
and conserved quantities. In the framework of the field-theoretical
approach these notions and quantities, in a definite sense, are
developed and generalized. Therefore in this subsection we give a
short review describing classical pseudotensors and superpotentials,
only which are necessary in our own presentation. On their examples
we outline the general properties of such objects, their problems
and some of modern applications.

Let us present the general way for constructing pseudotensors and
superpotentials. Using the metric and its derivations construct an
{\it arbitrary} quantity $\hat {\cal U}^{\mu\alf}_\nu$, satisfying
the condition
 \be
 \di_{\mu\alf}\hat {\cal U}^{\mu\alf}_\nu \equiv 0\, .
 \m{Us-l}
 \ee
Next, define the complex
 \be \hat \theta^{\mu}_\nu \equiv
\di_{\alf}\hat {\cal U}^{\mu\alf}_\nu -\frac{1}{\k} \hat
G^{\mu}_\nu\,, \m{(1.13)}
 \ee
which usually is called as an energy-momentum pseudotensor of
gravitational field. Using the Einstein equations one obtains
 \be \hat T_\nu^{\mu} + \hat
\theta_\nu^{\mu} = \di_\alf \hat {\cal U}^{\mu\alf}_\nu\, \m{(1.14)}
 \ee
where $\hat {\cal U}^{\mu\alf}_\nu$ plays the role of a
superpotential.  Thus Eq. (\ref{(1.14)}) is another form of the
Einstein equations. Due to (\ref{Us-l}) one has a differential
conservation law
 \be \di_\mu \l(\hat
T_\nu^{\mu} + \hat \theta_\nu^{\mu}\r) =0\, . \m{(1.15)}
 \ee
Concerning the physical sense, it is the 4-dimensional continuity
equation, which is the differential conservation law  derived
directly from the field equations. As is seen, the above
construction has problems. First, it is an ambiguity in a definition
of $\hat \theta_\nu^{\mu}$ and $\hat {\cal U}^{\mu\alf}_\nu$,
second, these expressions are not covariant in general.

The presented picture, as a generalization,  has been formulated
after prolonged history of constructing pseudotensors and
superpotentials in GR (see a review \cite{Trautman62}). Below we
recall constructing some of well known expressions. In the work
\cite{Einstein16}, following the standard rules Einstein had
suggested his famous pseudotensor. Firstly, a non-covariant
Lagrangian
 \be \lag^{cut} = - \frac {1}{2\k}
\hat g^{\mu\nu}\l( \Gamma^\sig_{\mu\rho} \Gamma^\rho_{\sig\nu} -
\Gamma^\rho_{\mu\nu} \Gamma^\sig_{\rho\sig}\r) \m{(1.1)}
 \ee
had been suggested. It differs from the covariant Hilbert Lagrangian
 \be \lag^H = -
\frac {1}{2\k} \hat R \m{HilbertL}
 \ee
by a divergence and leads to the same field equations
(\ref{EinsteinEquations}). Next, the corresponding to (\ref{(1.1)})
canonical complex had been constructed:
 \be \hat t_\nu^{E\mu} =  {\frac{\di
\lag^{cut}} {\di(\di_\mu g_{\alf\beta})}} \di_\nu g_{\alf\beta}
-\delta^\mu_\nu \lag^{cut}\, , \m{(1.2)}
 \ee
which is just the Einstein pseudotensor.  Both the Lagrangian
(\ref{(1.1)}) and the pseudotensor (\ref{(1.2)}) depend on the
metric  $g_{\mu\nu}$ and only its first derivatives. Combining
(\ref{(1.2)}) and the field equations Einstein had obtained the
conservation law
 \be \di_\mu
\l(\hat T_\nu^{\mu} + \hat t_\nu^{E\mu}\r) =0\, , \m{(1.10)}
 \ee
which is a variant of (\ref{(1.15)}).

Later Tolman \cite{Tolman-book} had found the quantity $\hat {\cal
T}^{\mu\alf}_\nu$ with a property (\ref{Us-l}), and for which, as a
particular case of (\ref{(1.14)}), one has $ \hat T_\nu^{\mu} + \hat
t_\nu^{E\mu} \equiv \di_\alf \hat {\cal T}^{\mu\alf}_\nu$. From here
the conservation law (\ref{(1.10)}) follows directly. However, $\hat
{\cal T}^{\mu\alf}_\nu$ is not antisymmetrical in $\alf$ and
$\beta$, whereas the use of antisymmetrical superpotentials ${\cal
U}^{\mu\alf}_\nu = -{\cal U}^{\alf\mu}_\nu$ is more reasonable
because it makes explicit the identity (\ref{Us-l}) and presents
less difficulties under covariantization, {\it etc}. Assuming an
antisymmetry Freud \cite{Freud39} had found out the superpotential
 \be \hat
{\cal F}^{\mu\alf}_\nu \equiv
 \frac{1}{2\k}\frac{g_{\nu\rho}}{ \sqrt{-g}}\di_\lam
\l[(-g)\l(g^{\mu\rho}g^{\alf\lam} - g^{\mu\lam}g^{\alf\rho}\r)\r]\,
, \m{(1.12)} \ee such that \be \hat T_\nu^{\mu} + \hat t_\nu^{E\mu}
= \di_\alf \hat {\cal F}^{\mu\alf}_\nu, \m{(1.11)} \ee and connected
with the Tolman one  by $ \hat {\cal F}^{\mu\alf}_\nu \equiv
 \hat {\cal T}^{\mu\alf}_\nu +
\di_\beta \hat {\Phi}^{\mu\alf\beta}_\nu$; $ \di_{\alf\beta} \hat
{\Phi}^{\mu\alf\beta}_\nu \equiv 0. $

Many authors (see, e.g., more known publications \cite{Bergman49} -
\cite{Moller58} and the review \cite{Trautman62}) considered the
problem of an ambiguity in definition of conserved quantities
described in (\ref{Us-l}) - (\ref{(1.15)}). It has turned out that
some of pseudotensors and superpotentials are directly connected
with a Lagrangian through the N{\oe}ther procedure. This  restricts
some ambiguities in their definitions. Naturally, the Eienstein
pseudotensor is defined by the Lagrangian (\ref{(1.1)}) which, being
generally non-covariant, is covariant with respect to linear
coordinate transformations. Then for translations $\lam^\mu_{(\alf)}
= \delta^\mu_{\alf}$ one can write out the N{\oe}ther identity:
 \be {\pounds}_{\lam} \lag^{cut} + \di_\mu
\l(\lam^\mu_{(\alf)} \lag^{cut}\r) \equiv 0\, . \m{(1.16)}
 \ee
Direct calculations lead to $ \di_\mu \l({\k}^{-1} \hat G_\nu^{\mu}
+ \hat t_\nu^{E\mu}\r) \equiv 0 $ with the definition (\ref{(1.2)})
that after using Eq. (\ref{EinsteinEquations}) gives the
conservation law (\ref{(1.10)}). The general conclusion is
formulated as follows.
 \bit
\item {In the sense of the N{\oe}ther procedure
the pseudotensor {\rm (\ref{(1.2)})} is uniquely defined by the
Lagrangian {\rm (\ref{(1.1)})}; other Lagrangians $\lag = \lag^{cut}
+ div$ lead to other pseudotensors.}
 \eit
Thus the identity (\ref{(1.16)}) with the Hilbert Lagrangian
(\ref{HilbertL}) leads to
 \be
\hat T_\nu^{\mu} + \hat {\cal M}_\nu^{\mu} = \di_\alf \hat
{\chi}^{\mu\alf}_\nu\,, \m{(1.19)}
 \ee
instead of (\ref{(1.11)}), where the M{\o}ller \cite{Moller58}
superpotential
 \be
\hat {\chi}^{\mu\alf}_\nu \equiv {1\over 2\k} \sqrt{-g}
g^{\mu\beta}g^{\alf\rho} \l(\di_\beta g_{\nu\rho} -\di_\rho
g_{\nu\beta}\r) \m{(1.20)}
 \ee
is used. The components $\hat {\cal M}^0_\nu(x^\alf)$ of the
pseudotensor in Eq. (\ref{(1.19)}) are transformed as a
4-dimensional vector density under transformations $ x'^k =
f^k(x^l)$, $ x'^0 = x^0 + g(x^l) $ \cite{Moller58}. In contrary, the
corresponding components of the Einstein pseudotensor  (\ref{(1.2)})
do not. This property of $\hat {\cal M}^0_\nu(x^\alf)$, in a part,
improves the general non-covariant picture.

The other problem is in a definition of angular momentum of a
system. Both Einsten's and M{\oe}ller's pseudotensors themselves, as
canonical N{\oe}ther complexes, cannot  define it. One needs to
modify these expressions or to construct new ones (see the review
\cite{Trautman62}). One of methods to construct angular momentum,
using {\it unique} energy-momentum complex, is to construct
symmetrical pseudotensors. Thus, Landau and Lifshitz have suggested
a such expression \cite{LL}. Their famous symmetrical pseudotensor
${{t}}_{LL}^{\mu\nu}$, like $\hat t_{\mu}^{E\nu}$, has only the
first derivatives, and is expressed through Einstein's (\ref{(1.2)})
and the Freud's (\ref{(1.12)}) expressions: ${}{{t}}_{LL}^{\mu\nu} =
\hat g^{\mu\sig}_{~~,\rho} \hat {\cal F}^{[\nu\rho]}_{\sig} + \hat
g^{\mu\rho} {}\hat t_{\rho}^{E\nu}$. However, as is seen, it has an
anomalous weight $+2$, unlike expressions in (\ref{(1.11)}) and
(\ref{(1.19)}) with the weight $+1$. Goldberg \cite{Goldberg58}
generalized the Landau-Lifshitz approach, suggesting a symmetrical
pseudotensor of an arbitrary weight.

Concerning symmetrical expressions, return also to the the equation
(\ref{FullPerturbEqs}) and the expression (\ref{LinearEqs}). The
last can be rewritten as
 \bea
\hat G_L^{\mu\nu} &\equiv & \k\di_{\beta}\hat{\cal
P}^{\mu[\nu\beta]}\,,
 \m{HlinLlin}\\
 \hat{\cal P}^{\mu[\nu\beta]}& \equiv &\frac{1}{\k}\di_{\alf}\l(\hat
l^{\nu[\mu}\eta^{\alf]\beta} - \hat
l^{\beta[\mu}\eta^{\alf]\nu}\r)\,.
 \m{Papapetrou}
 \eea
One can find that $ \hat{\cal P}^{\mu[\nu\beta]}$ is the well known
superpotential by Papapetrou \cite{Papapetrou48}; in the linear
approximation it is Weinberg's superpotential \cite{Weinberg-book};
and, up to factor $\sqrt{-\eta}$, it coincides also with the linear
version of the Landau-Lifshitz superpotential \cite{LL}. Then the
equation (\ref{FullPerturbEqs}) can be rewritten as
 \be
\hat t_{(tot)}^{\mu\nu} =\di_{\beta}\hat{\cal P}^{\mu[\nu\beta]}\,.
 \m{GL-Sup}
 \ee
This means that the conservation law (\ref{CLfirst}) can be
interpreted by the same way as the conservation law (\ref{(1.15)}),
i.e. in the framework of the superpotential approach.

It was a problem  to relate {\it modified} superpotentials to
Lagrangians, like it was done for Freud's (\ref{(1.12)}) and
M{\o}ller's (\ref{(1.20)}) ones. Only recently with the use of  a
covariant Hamiltonian formalism \cite{ChenNester} (see subsection
\ref{Einstein-papers}) Chang, Nester and Chen \cite{ChangNesterChen}
closed this gap. They
 describe GR by Hamiltonians with different surface terms
corresponding to different boundary conditions under variation. For
each of known (covariantized) superpotentials one can find its own
Hamiltonian. Thus, the Dirichlet conditions correspond to the Freud
superpotential (\ref{(1.12)}), Neuman's ones --- to M{\o}ller's
superpotential (\ref{(1.20)}). All the other superpotentials require
more complicated boundary conditions. Recently Babak and Grishchuk
\cite{BabakGrishchuk}  suggested a special Lagrangian including
additional terms with Lagrangian multipliers, which leads to a
covariantized Landau-Lifshitz pseudotensor of normal weight $+1$.

Of course, a non-covariance of pseudotensors and superpotentials
lead to the evident problems described in all the textbooks. In this
relation, return to the Hilbert Lagrangian (\ref{HilbertL})
recalling that it is generally covariant. Thus, for constructing
conservation laws one can use arbitrary diplacement vectors
$\xi^\mu$, not just the linear translations $\lambda^\mu_{(\alf)}$,
like in (\ref{(1.16)}). Hence there has to exist a generally
covariant energy-momentum complex and superpotential. Such
quantities were found by Komar \cite{Komar59}, his superpotential is
 \be \hat {\cal K}^{[\mu\alf]}=
{{\sqrt{-g}}\over 2\k}\l(D^\mu \xi^{\alf} - D^\alf  \xi^{\mu}\r)=
{{\sqrt{-g}}\over \k}D^{[\mu} \xi^{\alf]}= {{1}\over \k}D^{[\mu}
\hat \xi^{\alf]}\, . \m{K.2}
 \ee
With $\xi^\mu = \const$ it goes to the M{\o}ller superpotential
(\ref{(1.20)}). However, in spite of the advantage with covariance,
there are problems. The superpotential (\ref{K.2}) does not make a
correct ratio of angular momentum to mass in the Kerr solution (see,
e.g., \cite{Katz85,Mitzk}).

The mentioned above problems could be neglected if resonable
assumptions are made. Then many classical pseudotensors and
superpotentials could be useful for a description of differerent
physical systems. In this sense it is useful to return to Einstein's
arguments. First, in spite of that the local conservation laws
(\ref{(1.10)}) are not covariant they have the same form in every
coordinate system. Being the continuity equations they give a {\it
balance} between  a loss and a gain of a density of {\it
particularly} defined physical quantities, and hence they state the
{\it equilibrium } of a system. The second remarkable Einstein's
example \cite{Einstein18b} is  related to masses connected by a
rigid rod. Tensions in the road (which exist because $\hat
T^\mu_\nu$ is present in (\ref{(1.10)})) can be interpreted {\it
only} as a compensation to tensions of gravitational field included
in $t_\mu{}^{E\nu}$. Third, Einstein had suggested also a simple
method for the ``localization'' \cite{Einstein18a}, when an isolated
system is considered as placed into ``Galileian space'', and
``Galileian coordinates'' are used for description. Finally, he had
used his energy-momentum complex to describe weak gravitational
waves as metric perturbations with respect to a Mikowski metric
\cite{Einstein16a} - \cite{Einstein18a}.

At reasonable assumptions, pseudotensors and superpotentials are
left useful nowadays too. Thus, usually quantities of  global
energy-momentum and angular momentum for asymptotically flat
spacetimes are calculated. For example, Bergmann-Tomson's and
Landau-Lifshitz's complexes give the same expressions for global
angular momentum in an isolated system \cite{Garecki01}. In
\cite{Vibhadra1,Vibhadra1+}, with the use of the Einstein, Tolman
and Landau-Lifshitz complexes reasonable energy and momentum
densities of cylindrical gravitational waves were obtained.
Recently, in the work \cite{Favata01} with a careful set of
assumptions it was shown that a tidal work in GR calculated for
different non-covariant energy-momentum complexes is the same and
unambiguous. Authors frequently use pseudotensors and
superpotentials to describe many popular exact solutions of GR (see
\cite{Vibhadra2} - \cite{Xulu2} and references there in).
Particulary, it was shown that in many cases {\em different}
pseudotensors give {\em the same} energy distribution  for the same
solution.

However, in spite of some successes in applying pseudotensors and
superpotentials, it is clear that more universal and  unambiguous
quantities are more preferable. The next subsection is devoted to
possible ways in improving properties of the classical complexes.

\subsection{Requirements for constructing conservation laws on curved backgrounds}
\m{UsingPseudo}

From the beginning we  recall how conservation laws are constructed
in a field theory in a Minkowski space. These principles  are useful
both to understand how classical pseudotensors and superpotentials
could be modified to avoid their problems and  to better describe
and explain conservation laws in the field-theoretical formulation.

Consider an arbitrary theory of fields $\psi^A$ in the Minkowski
space in curved coordinates with the Lagrangian
 \be
\lag_{(\psi)} = \lag_{(\psi)} \left(\psi^A, \BD_\alpha \psi^A, \Bar
g_{\mu\nu} \right)\, . \m{(2.9)}
 \ee
For the sake of simplicity assume that it contains the first
derivatives only. The symmetrical (metric) energy-momentum tensor
density is defined as usual:
 \be
\hat t_{(s)\alf\beta} \equiv
 2{{\delta \lag_{(\psi)}} \over {\delta \Bar g^{\alf\beta}}}\,.
\m{(2.12)}
 \ee
The density of the canonical energy-momentum tensor has the form:
 \be \hat t^\alf_{(c)\sig} \equiv {{\di
\lag_{(\psi)}} \over {\di \left(\Bar D_\alpha \psi^A\right)}} \Bar
D_\sigma \psi^A - \lag_{(\psi)} \delta^\alf_\sig\,.
 \m{(2.13)}
 \ee
Both of them are differentially conserved
 \bea
  \Bar D_\alf \hat t^\alf_{(s)\sig} &=& 0\, ,
 \m{Spsi-t}\\
  \Bar D_\alf \hat t^\alf_{(c)\sig} &=& 0\, ,
 \m{Cpsi-t}
 \eea
on the field equations. To construct integral conserved quantities
it is necessary to use Killing vectors of the Minkowski space.
Contracting $\hat t^\alf_{(s)\sig}$ with an each of the 10 Killing
vectors $\lambda^\nu$ one obtains the currents
 \be
 \hat{I}^\mu_{(s)}(\lambda)\equiv \hat t^\mu_{(s)\nu}\lambda^\nu\, ,
 \m{Spsi-current}
 \ee
which are conserved differentially
 \be
\Bar D_\alf \hat {I}^\alf_{(s)}(\lambda)\equiv \di_\alf \hat {
I}^\alf_{(s)}(\lambda)=0\,.
 \m{dipsi-current}
 \ee

The other situation with the canonical energy-momentum $\hat
t^\alf_{(c)\sig}$. If one tries to contract it with the {\em
4-rotation} Killing vectors, then the correspondent currents, like
(\ref{Spsi-current}), are not conserved. The reason is that $\hat
t^{\alf\sig}_{(c)}$ is not symmetrical in general. A one of the ways
to construct differentially conserved currents is
 \be
\hat {I}^\alf_{(c)}(\lambda)\equiv \hat
t^\alf_{(c)\sigma}\lambda^\sigma + \hat
\sigma^{\alpha\beta}_{(\psi)}{}_{\sigma}\Bar D_\beta\lambda^\sigma\,
 \m{C1psi-current}
 \ee
where
 \be
\hat \sigma^{\alpha\beta}_{(\psi)}{}_{\sigma} \equiv - {{\di
\lag_{(\psi)}} \over {\di \left(\Bar D_\alpha \psi^A\right)}}
\left.\psi^A \right|^\beta_\sigma\,  \m{(2.14)}
 \ee
is a spin density. Another way follows the Belinfante symmetrization
\cite{Belinfante}, who has introduced the antisymmetric in $\alf$
and $\beta$ expression
 \be \hat S^{\alf\beta\ga}_{(\psi)} = \hat
\sigma^{\ga[\alpha\beta]}_{(\psi)}+ \hat
\sigma^{\alf[\ga\beta]}_{(\psi)} - \hat
\sigma^{\beta[\ga\alf]}_{(\psi)}\,  \m{(2.17)}
 \ee
and has defined the new (symmetrized) energy-momentum tensor density
as follows
 \be
\hat t^{\alpha}_{(B)\sigma}   = \hat t^\alpha_{(c)\sigma}  + \Bar
D_\beta \hat S^{\alpha\beta}_{(\psi)\sigma}. \m{(2.19)}
 \ee
It is also conserved, $\Bar D_\alf \hat t^{\alf}_{(B)\sig} = 0$, on
the  equations of motion. Besides, it is symmetrical on the
equations of motion. Then the symmetrized current
 \be
 \hat {I}^\alf_{(B)}(\lambda)\equiv \hat t^{\alf}_{(B)\sig}\lambda^\sig\,
 \m{Bpsi-current}
 \ee
is also  conserved for all the Killing vectors, i.e. one can use the
{\it unique} object $\hat t^{\alpha}_{(B)\sigma}$ without additional
quantities, like the spin tensor in (\ref{C1psi-current}).

For the Minkowski background it is easy to show
 \be \hat
t_{(s)\sig}^\alpha = \hat t^\alpha_{(c)\sigma}  + \Bar D_\beta \hat
S^{\alpha\beta}_{(\psi)\sigma}= \hat t^{\alpha}_{(B)\sigma}\,
\m{T=t(psi)}
 \ee
on the field equations. Thus the {\it metric} energy-momentum
(\ref{(2.12)}) is equivalent to the Belinfante {\it symmetrized}
quantity (\ref{(2.19)}) (in this relation see also
\cite{Szabados91}), i.e. $\hat {I}^\mu_{(s)}= \hat {I}^\mu_{(B)}$.
Note also that $\hat {I}^\mu_{(s)}$ differs from $\hat
{I}^\mu_{(c)}$ by a divergence. One can see that the Belinfante
procedure suppresses the spin term in the current
(\ref{C1psi-current}). Classical electrodynamics has just the
Lagrangian of the type (\ref{(2.9)}), and is a good illustration of
the above.

The first example where the Belinfante method was used in GR is the
paper by Papapertrou \cite{Papapetrou48}. He has symmetrized the
Einstein pseudotensor and obtained his remarkable superpotential. By
the original definition \cite{Belinfante} it is necessary to use a
background metric. Thus, Papapetrou used an auxiliary Minkowski
metric. Later Berezin \cite{Berezin92}, in the framework of the
field approach in GR on a flat background, has shown that effective
energy-momentum tensor can be constructed by the Belinfante method.
Applications of the Belinfante method in GR to the pseudotensors
without using an auxiliary background
\begin{figure*}[t]
    \centering
    \includegraphics*%
    [width=7cm]%
     {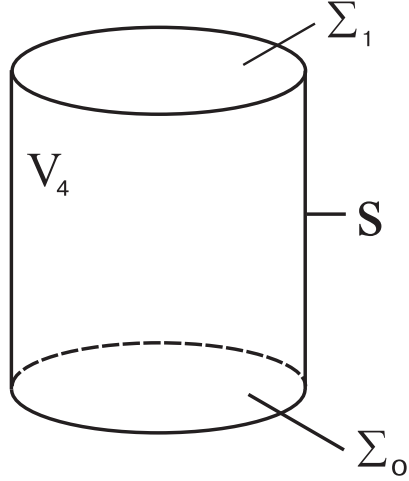}\\
     \caption{4-dimensional volume $V_4$ restricted by a
     truncated cylinder.}
     \label{F2}%
     \end{figure*}
metric \cite{Szabados91,Szabados92} lead uniquely to the Einstein
tensor. It is important and interesting result. However, this cannot
be useful for description of energy characteristics, e.g.,  in
vacuum case. Recently Borokhov \cite{Borokhov} generalized the
Belinfante method for an arbitrary field theory on arbitrary
backgrounds, with arbitrary Killing vectors and with non-minimal
coupling. With simplification to minimal coupling his model goes to
the standard description \cite{Szabados04,Szabados91,Szabados92}.

Denoting $\hat {I}^\mu_{(\psi)}= \l\{\hat {I}^\mu_{(s)},\,\hat {
I}^\mu_{(c)}\r\}$  consider 4-dimensional volume $V_4$ in the
Mikowski space, the boundary of which consists of timelike ``wall''
$S$ (cylinder) and two spacelike sections: $\Sigma_0 :=  t_0 =
\const$ and $\Sigma_1 := t_1 = \const$ (see figure \ref{F2}).
Because the conservation laws, like (\ref{dipsi-current}), are
presented by the scalar densities, they can be integrated through
the 4-volume: $ \int_{V_4} \di_\mu \hat {I}^\mu_{(\psi)}(\lam) d^4 x
= 0\, . $ The generalized Gauss theorem gives
 \be \int_{\Sigma_1}  \hat {I}^0_{(\psi)}(\lam) d^3 x -
\int_{\Sigma_0}  \hat {I}^0_{(\psi)}(\lam) d^3 x +\oint_{S} \hat
{I}^\mu_{(\psi)}(\lam) dS_\mu = 0\, \m{UseGauss}
 \ee
where $dS_\mu$ is the element of integration on $S$. If in Eq.
(\ref{UseGauss})
 \be \oint_{S}  \hat {I}^\mu_{(\psi)}(\lam) dS_\mu = 0\,,
 \m{BoundaryS}
  \ee
then the quantity
 \be {\cal P}_{(\psi)}(\lam) = \int_{\Sigma} \hat
{I}^0_{(\psi)}(\lam)d^3 x\, \m{(1.26)}
 \ee
is conserved on $\Sigma$ restricted by $\di\Sigma$ --- intersection
of $\Sigma$ with $S$. If the equality (\ref{BoundaryS}) does not
hold, then Eq. (\ref{UseGauss}) describes changing the quantity
(\ref{(1.26)}), i.e. its flux trough $\di\Sigma$. Due to the
difference between $\hat {I}^\mu_{(s)}$ and $\hat { I}^\mu_{(c)}$
the quantities ${\cal P}_{(s)}(\lam)$ and ${\cal P}_{(c)}(\lam)$
differ one from another by surface integrals. However, as a rule,
boundary conditions suppress these surface terms, and then ${\cal
P}_{(s)}(\lam) = {\cal P}_{(c)}(\lam)$.

Now let us return to pseudotenors and superpotentials. To avoid some
problems they have to be covariantized. Usually a covariantization
is carried out by including an auxiliary metric of a Minkowski
space. After that partial derivatives are related to Lorentzian
coordinates of this flat spacetime. In arbitrary coordinates,
partial derivatives naturally transform into covariant ones. Thus,
one can keep in mind that in usual expressions, like (\ref{(1.14)})
in general form, already there exist the Minkowski metric
$\eta_{\mu\nu}$ and its determinant $\eta \equiv \det{\eta_{\mu\nu}}
= -1$. For constructing integral conservation laws based on
covariantized pseudotensors it is also natural to use the Killing
vectors of the Minkowski space.

Let us present the program of the covariantization in detail.
Consider the generalized conservation law  (\ref{(1.15)}) and
rewrite it in the equivalent form:
 \be
 \di_\mu \l[\sqrt{-\eta} \l(({\hat \theta^\mu_\rho}+
 {\hat T^\mu_\rho})/ {\sqrt{-\eta}} \r)
\lam^\rho_{(\nu)}\r] = 0\,.
 \m{pseudo-covariant}
 \ee
Here, the left hand side is a scalar density. The translation
Killing vector $\lam^\rho_{(\nu)}$ of the background Minkowski space
is expressed in the Lorentzian coordinates, therefore one has
$\lam^\rho_{(\nu)} = \delta^\rho_{\nu}$ (the lower index is not
coordinate one). Thus, in (\ref{pseudo-covariant}) the
differentially conserved vector density (current)
 \bea \hat {\cal J}^\mu(\lam)
&\equiv &\sqrt{-\eta} \l(({\hat \theta^\mu_\rho} +
 {\hat T^\mu_\rho})/ {\sqrt{-\eta}} \r)
\lam^\rho_{(\nu)}\, ,\m{(1.25)}\\
\Bar D_\mu \hat {\cal J}^\mu(\lam)&\equiv & \di_\mu \hat {\cal
J}^\mu(\lam) = 0\, \m{(1.25+)}
 \eea
is presented. Here, one has to accent that if $\hat \theta^{\mu\nu}$
is symmetrical, then the conservation law (\ref{(1.25+)}) could be
extended for 4-rotation Killing vectors. However, if $\hat
\theta^{\mu\nu}$ is not symmetrical, then the simple and direct
covariantization is restricted to using only translation Killing
vectors.

Next, as is seen, the programm of constructing integral quantities
(\ref{UseGauss}) - (\ref{(1.26)}) can be repeated for the
covariantized current (\ref{(1.25)}). The superpotential in Eq.
(\ref{(1.14)}) can also be rewritten in the covariantized form:
 \be \hat {\cal J}^{\mu\alf}(\lam) =
\sqrt{-\eta} \l({\hat {\cal U}^{\mu\alf}_\rho} / {\sqrt{-\eta}} \r)
\lam^\rho_{(\nu)}\,. \m{CovarSuper}
 \ee
Thus the expression (\ref{(1.14)}) can be rewritten in a fully
covariant form:
 \be \hat {\cal J}^\mu(\lam) = \di_\alf \hat {\cal
J}^{\mu\alf}(\lam) \equiv \Bar D_\alf \hat {\cal
J}^{\mu\alf}(\lam)\m{(1.14+)}
 \ee
that has a sense of the conservation law because $\hat {\cal
J}^{\mu\alf}(\lam)$ is an antisymmetric tensor density. Due to
antisymmetry of the superpotential (\ref{CovarSuper}), the
4-momentum presented in the volume form, like (\ref{(1.26)}), is
expressed as a surface integral
 \be {\cal P}(\lam)
= \oint_{\di\Sigma}  \hat {\cal J}^{0k}(\lam) ds_k \m{int-surface+}
\ee where $ds_k$ is the element of integration on $\di\Sigma$.

Up to now in constructing conservation laws we considered Killing
vectors of a background only. However, other displacement vectors
can be used also. Let us give two examples. First, for study of
perturbations on FRW backgrounds a so-called notion of an {\it
integral constraint}, which connects a volume integral of only
matter perturbations with a surface integral of only metric
perturbations, frequently is very important \cite{Traschen}. With
the use of such constraints, e.g., measurable effects of the cosmic
background radiation were analyzed \cite{TraschenEardley}. In the
definition of integral constraints {\it integral constraint
vectors}, not necessarily Killing vectors, play a crucial role
\cite{Traschen}. Second, in \cite{Stebbin}, a new conserved
energy-momentum pseudotensor was found and used in an effort to
integrate Einstein's equations with scalar perturbations and
topological defects on FRW backgrounds for a sign of spatial
curvature $k=0$. In \cite{UzanDerTurok} it was realized that these
conservation laws might be associated with the {\it conformal}
Killing vector of time translation, and can be also generalized for
$k=\pm 1$.

Taking into account the above we formulate the generalized
requirements for constructing conservation laws for perturbations on
curved backgrounds as follows.

\bit

\item { Expressions have to be covariant on a chosen
background and valid for arbitrary curved backgrounds.}

\item
{ One has to construct  differentially conserved currents (vector
densities) $\hat {\cal J}^\mu(\xi)$, such that $\di_\mu \hat {\cal
J}^\mu (\xi) = 0 $, and which are defined by the canonical
N{\oe}ther procedure applied to a Lagrangian of a perturbed system.}

\item  {The currents have to be expressed
through corresponding superpotentials $\hat {\cal J}^{\mu\nu}$
(antisymmetric tensor densities), such that $\di_{\mu\nu} \hat {\cal
J}^{\mu\nu} (\xi) \equiv 0 $,  by the way $\hat {\cal J}^\mu(\xi) =
\di_\nu \hat {\cal J}^{\mu\nu} (\xi)$.}

\item  {There has to be a possibility to use
arbitrary displacement vectors $\xi^\mu$, not only Killing vectors
of the background.}

\item  {Applications of suggested conserved quantities and
conservation laws have to satisfy the known tests (see discussion in
subsection \ref{Einstein-papers}).}

\eit

\subsection{The Katz, Bi\v c\'ak and Lynden-Bell
conservation laws}
 \m{CL-KBL}

Katz, Bi\v c\'ak and Lynden-Bell \cite{KBL} (later we call as KBL)
have  satisfied the requirements at the end of previous subsection.
They describe a perturbed system by the Lagrangian:
 \be
 \lag_{KBL} = \hat{\cal L}^E - \Bar{\lag^E} -{1\over
2\k} \di_\alf \hat k^\alf\, , \m{KBLLagrangian}
 \ee see
notations in (\ref{(a2.1)}) and (\ref{(a2.8)}). In fact they use
bimetric formalism, where $g_{\mu\nu}$ and $\Bar g_{\mu\nu}$ are
thought as independent metric coefficients. The
 gravitational part of (\ref{KBLLagrangian}) is
 \be \lag_G = -{1\over 2\k} \l(\hat R - \Bar {\hat R} + \di_\alf \hat
k^\alf \r)\, . \m{(1.27)}
 \ee
Here, for the physical scalar curvature density $\hat R$ an external
background metric $\Bar g_{\mu\nu}$ is introduced with the use of
the presentation for the curvature tensor:
 \be
 R^\lam{}_{\tau\rho\sig}  =
\BD_\rho \Delta^\lam_{\tau\sig} -  \BD_\sig\Delta^\lam_{\tau\rho} +
 \Delta^\lam_{\rho\eta} \Delta^\eta_{\tau\sig} -
 \Delta^\lam_{\sig\eta} \Delta^\eta_{\tau\rho}
 + \Bar R^\lam{}_{\tau\rho\sig}\,.
 \m{(17-Dm)}
 \ee
It is also very important to note that $\hat k^\alf$ is defined in
(\ref{k-KBL}).

For the Lagrangian (\ref{(1.27)}), as for a scalar density, one has
the identity:
 \be
{\Lix} \lag_G + \di_\mu \l(\xi^\mu \lag_G\r) \equiv 0\, .
\m{Lix-xi-arbitrary}
 \ee
It is a generalization of Eq. (\ref{(1.16)}) for arbitrary
displacement vectors $\xi^\mu$. Consequent identical transformations
give:
 \bea
\di_\mu \hat \jmath^\mu &\equiv & 0\, ,\m{B16}\\
\hat \jmath^\mu &\equiv & \l[ {{\di \lag_G} \over {\di \left(\Bar
D_\mu g_{\rho\sig}\right)}} \Bar D_\nu g_{\rho\sig} - \lag_G
\delta^\mu_\nu\r]\xi^\nu - {{\di \lag_G} \over {\di \left(\Bar D_\mu
 g_{\rho\sig}\right)}} \l.g_{\rho\sig}\r|^\rho_\lam \bar
 g^{\lam\sig}\BD_{\rho}\xi_{\sig}  \nonumber\\ &-&  {{\delta
{\lag_G}} \over {\delta g_{\rho\sig}}} \left.{ g_{\rho\sig}}
\right|^\mu_\nu \xi^\nu - {{\delta {\lag_G}} \over {\delta \Bar
g_{\rho\sig}}} \left.{ \Bar g_{\rho\sig}} \right|^\mu_\nu \xi^\nu +
\hat Z^\mu_{(c)} \equiv  0\,.
 \m{B16+}
 \eea
Note that here
 \be - {{\delta
{\lag_G}} \over {\delta g_{\rho\sig}}} \left.{
 g_{\rho\sig}} \right|^\mu_\nu \xi^\nu \equiv
 {1 \over {\k}}{\hat G^\mu_\nu }\xi^\nu\, ,\qquad
- {{\delta {\lag_G}} \over {\delta \Bar g_{\rho\sig}}} \left.{ \Bar
g_{\rho\sig}}
 \right|^\mu_\nu \xi^\nu \equiv - {1 \over
 {\k}}{\Bar{\hat G}^\mu_\nu }\xi^\nu\, .
 \m{B17}
 \ee
Then substituting the Einstein equations (\ref{EinsteinEquations})
and their barred version into Eq. (\ref{B16}) one obtains the
differential conservation law
 \bea
\di_\mu \hat J^\mu_{(c)}(\xi) &=& 0\, , \m{(1.32+)}\\
\hat J^\mu_{(c)}(\xi) & \equiv &\hat \Theta^\mu_{(c)\nu} \xi^\nu +
\hat \sig^{\mu\rho\sig}\Bar D_{\rho}\xi_{\sig}+ \hat
Z^\mu_{(c)}(\xi)\, . \m{(1.32)}
 \eea

The first term in (\ref{(1.32)}), the generalized total
energy-momentum tensor density, is
 \be
\hat \Theta^\mu_{(c)\nu} =\hat t^\mu_\nu +  \l(\hat T^\mu_\nu -
\Bar{\hat T}^\mu_\nu\r) + {1\over 2\k}\hat l^{\rho\sig} \Bar
R_{\rho\sig}\delta^\mu_\nu\, , \m{(1.33)}
 \ee
with $\hat l^{\rho\sig} =  \hat g^{\rho\sig}  - \Bar {\hat
g}^{\rho\sig}$. The first term in Eq. (\ref{(1.33)}) is the
canonical energy-momentum tensor density of the gravitational field:
 \bea \hspace*{-1.0cm}2\k \hat t^\mu_\nu & = & \hat g^{\rho\sig}
\l(\Delta^\lam_{\rho\lam} \Delta^\mu_{\sig\nu}  +
\Delta^\mu_{\rho\sig} \Delta^\lam_{\lam\nu} -
2\Delta^\mu_{\rho\lam} \Delta^\lam_{\sig\nu}\r) \nonumber \\
\hspace*{-1.0cm}& - & \hat g^{\rho\sig} \l(\Delta^\eta_{\rho\sig}
\Delta^\lam_{\eta\lam} - \Delta^\eta_{\rho\lam}
\Delta^\lam_{\eta\sig}\r)\delta^\mu_\nu + \hat
g^{\mu\lam}\l(\Delta^\sig_{\rho\sig} \Delta^\rho_{\lam\nu} -
\Delta^\sig_{\lam\sig} \Delta^\rho_{\rho\nu}\r)\, , \m{(1.34)}
 \eea
the second term in Eq. (\ref{(1.33)}) is a perturbation of the
matter energy-momentum, the third term describes an interaction with
the background. The second term in (\ref{(1.32)}) is the spin tensor
density:
 \bea
2\k \hat \sig^{\mu\rho\sig}\equiv &-& 2\k{{\di \lag_G} \over {\di
\left(\Bar D_\mu
 g_{\rho\sig}\right)}} \l.g_{\rho\sig}\r|^\rho_\lam \bar
 g^{\lam\sig}  = (\hat g^{\mu\rho} \Bar g^{\sig\nu}
+\Bar g^{\mu\sig} \hat g^{\rho\nu} - \hat g^{\mu\nu} \Bar
g^{\rho\sig}) \Del^\lam_{\nu\lam} \nonumber \\ & -& (\hat
g^{\nu\rho} \Bar g^{\sig\lam} +\Bar g^{\nu\sig} \hat g^{\rho\lam} -
\hat g^{\nu\lam} \Bar g^{\rho\sig}) \Del^\mu_{\nu\lam}\, .
\m{(1.35)}
 \eea
If one simplifies to a Minkowski background then it is the quantity
presented by Papapetrou \cite{Papapetrou48} to construct angular
momentum with the use of the Einstein pseudotensor. The last term in
(\ref{(1.32)}) has the form:
 \be
{\hat Z^\mu_{(c)}(\xi)} ={1\over 2\k}\l [
 \hat l^{\mu\lam} \di_\lam \zeta^\rho_\rho + \hat l^{\rho\sig}\l(
\Bar D^\mu \zeta_{\rho\sig} - 2\Bar D_\rho \zeta^\mu_\sig \r)\r ],
\m{Zmu}
 \ee
with $2\zeta_{\rho\sigma} = - {\pounds}_\xi \Bg_{\rho\sigma} =
2\BD_{(\rho}\xi_{\sigma)}$. It,  thus, disappears if $\xi^\mu =
\lam^\mu$ is a Killing vector of the background spacetime. Besides,
in this case the part of the second term in (\ref{(1.32)}): $\hat
\sig^{\mu\rho\sig} \Bar D_{(\rho}\lam_{\sig)}$ disappears also.

Because  Eq. (\ref{B16}) is the identity then the quantity $\hat
\jmath^\mu(\xi)$  must be presented through a superpotential $\hat
J^{\mu\nu}_{(c)}(\xi)$, thus indeed
 \bea
 \hat \jmath^\mu(\xi) &\equiv& \di_\nu \hat J^{\mu\nu}_{(c)} (\xi)\,,
 \m{(1.29)}\\
 {\hat J^{\mu\nu}_{(c)}(\xi)}&\equiv &{1\over \k}\big({D^{[\mu}\hat
\xi^{\nu]}}- \Bar {D^{[\mu}\hat \xi^{\nu]}}\big)+ {1\over \k}\hat
\xi^{[\mu} k^{\nu]}\, . \m{(1.31)}
 \eea
This superpotential has been found earlier independently  by Katz
for flat backgrounds \cite{Katz85} and by Chru\'sciel for Ricci-flat
backgrounds \cite{Chrusciel85}. However, as it turns out, it has the
same form (\ref{(1.31)}) for generally curved backgrounds. Note that
the expression (\ref{(1.31)}) modifies the Komar superpotential
(\ref{K.2}). With the use of the Einstein equations, the identity
(\ref{(1.29)}) transforms into
 \be \hat J^\mu_{(c)}(\xi) = \di_\nu \hat J^{\mu\nu}_{(c)}
(\xi)\, , \m{(1.29+)}
 \ee
that is another form of the conservation law (\ref{(1.32+)}).

Now, let us compare the KBL expressions with the Einstein
prescription. For the simplification $\bar g_{\mu\nu} \goto
\eta_{\mu\nu} $ the KBL Lagrangian $\lag_G$ transforms into the
Einstein Lagrangian (\ref{(1.1)}), and the KBL gravitational
energy-momentum $\hat t^\mu_\nu$ transforms into the Einstein
pseudotensor (\ref{(1.2)}).  Next, if $\Bar g_{\mu\nu} \goto
\eta_{\mu\nu}$ and $\xi^{\nu} \goto \lam^\nu = \delta^\nu_{(\mu)}$,
then the KBL current ${\hat J^{\mu}_{(c)}(\xi)}$ goes to $\hat
t^{E\mu}_\nu + \hat T^{\mu}_\nu$ in (\ref{(1.10)}), the KBL
superpotential ${\hat J^{\mu\nu}_{(c)}(\xi)}$ transforms into the
Freud superpotential (\ref{(1.12)}). That is generally  Eq.
(\ref{(1.29+)}) transforms into Eq. (\ref{(1.11)}). However, both
the current and superpotential in Eq. (\ref{(1.29+)}) {\it are not a
simple direct} covariantization of the quantities (\ref{(1.2)}) and
(\ref{(1.12)}). Indeed, first, the KBL expressions hold on arbitrary
curved backgrounds, not only on flat backgrounds in curved
coordinates. Second, $\hat J^\mu_{(c)}$ includes the spin term
(\ref{(1.35)}) analogous to the quantity (\ref{(2.14)}). It plays
its crucial role because then rotation Killing vectors can be used
giving a reasonable definition of angular momentum. Thus the KBL
current gives the right ratio of mass to angular momentum for
rotating black hole solutions.

The KBL approach received a significant development in applications.
Thus, in \cite{Katz-LB-Israel,KatzOri} the problem of localization
was considered. In \cite{KatzLerer}, the conserved quantities and
their fluxes at null infinity for an isolated sysytem  were studied
in detail. In the works \cite{UzanDerTurok,DerKatzUzan}
perturbations on FRW backgrounds and conservation laws for them were
examined. At last, recently \cite{DerKatzOgushi} the approach was
developed for constructing the conserved charges in Gauss-Bonnet
5-dimensional cosmology.

The KBL quantities were also checked from the point of view of the
problem of uniqueness. Thus, Julia and Silva
\cite{JuliaSilva98,Silva99}, and independently Chen and Nester
\cite{ChenNester}, stated that the KBL quantities are uniquely
defined as associated with the Dirichlet boundary conditions, under
which the action with the Lagrangian (\ref{(1.27)}) is variated.
This gives evident advantages. Both the Lagrangian (\ref{(1.27)})
and the energy-momentum (\ref{(1.34)}) have only the first
derivatives. Thus, the Cauchy problem is stated simply.

Keeping in mind successes presented by the KBL approach it is
necessary to note the next generic problem related to the canonical
derivation. If one chooses different boundary conditions, adding
different divergences to the Lagrangian, different expressions  both
for currents and for superpotentials appear by the N{\oe}ther
procedure.  On the one hand, such a situation could be useful and
interesting in gravitational theory, when analogies, e.g.,  with
thermodynamics are carried out \cite{ChangNesterChen}. On the other
hand, it is evident that {\em unique} expressions independent on
boundary consditions are necessary. As examples, the symmetric
energy-momentum tensor in classical electrodynamics is such a
quantity. The expressions in the field-theoretical formulation of GR
are also independent on boundary conditions (see section
\ref{decompositions}). In the next subsection we just construct
conserved currents and superpotentials in the framework of the field
approach.

\subsection{Conserved quantities in the field-theoretical formulation}
\m{NewCLinFieldGR}

Already the form of Eq. (\ref{GL-Sup}) indicates that there is a
possible to construct currents and superpotentials in the framework
of the field approach. Recall that for backgrounds presented by an
Einstein space the conservation law (\ref{(2.23')}) has a place.
Then if the Einstein space has Killing vectors $\lam^\alf$  one can
construct the conserved currents by the same way as in the usual
field theory (\ref{Spsi-current}). At the same time the left hand
sides both of Eq. (\ref{(a2.23)}) and of Eq. (\ref{(a2.27)})
contracted with $\lam^\mu$ are presented as divergences of
superpotentials \cite{AbbottDeser82,GPP,Deser87}. However, in the
general case of a curved background the conservation law
(\ref{(2.23')}) has no a place. The reason is that there is no an
identical conservation both in Eq. (\ref{(a2.17)}): $ \Bar
D_\nu\l(\hat G^{L\nu}_{\mu} + \hat \Phi_{\mu}^{L\nu}\r) \neq 0$ and
in  Eq. (\ref{B30}):  $ \Bar D_\nu\hat G^{L\nu}_{\mu} \neq 0$. The
physical reason is that the system (\ref{(2.10)}) interacts with a
complicated background geometry defined by the background matter
fields $\Bar\Phi^A$.

To construct conserved quantities in the general case we
\cite{PK2003b} combine the technique of previous subsection with the
results \cite{GPP}. Therefore, in the framework of the field
approach, we are going to the KBL independent variables of the
bimetrical approach by changing $\hat l^{\mu\nu} \goto \hat
g^{\mu\nu} - \Bar {\hat g}^{\mu\nu}$. Then the dynamical Lagrangian
$\lag^g$ (\ref{(a2.16)}) transforms into $2 \k\lag_{G2}$. After that
we connect it with the KBL Lagrangian $\lag_G $ (\ref{(1.27)}):
 \be \lag_{G2}  \equiv
 \lag_G - \lag_{G1}
\equiv {1\over 2\k}{{\hat g}^{\mu\nu}}
\l(\Delta^\rho_{\mu\nu}\Delta^\sig_{\rho\sig}
-\Delta^\rho_{\mu\sig}\Delta^\sig_{\rho\nu}\r)
 \m{B19}
 \ee
where $
 \lag_{G1} = -(2\k)^{-1} (\hat g^{\mu\nu} - \Bar {\hat
 g}^{\mu\nu}) \Bar R_{\mu\nu}$. Returning to the
 comparison with the Rosen Lagrangian $\lag_R$ \cite{[2]}
 we note that $\lag_{G2}$ is a direct generalization
 of $\lag_R$ to arbitrary backgrounds, whereas
 $\lag_G$ is reduced to $\lag_R$ for the Ricci-flat
 backgrounds.

Now, consider the identity
 \be
 {\Lix} \lag_{G2} + \di_\mu (\xi^\mu \lag_{G2})
\equiv {\rm 0}\,
 \m{B21}
 \ee
 for the Lagrangian (\ref{B19}) and transform it identically to
 \bea &{}&\l[{{\di \lag_{G2}} \over {\di \left(\bar D_\mu
g_{\rho\sig}\right)}} \Bar D_\nu g_{\rho\sig} -
\lag_{G2}\delta^\mu_\nu\r] \xi^\nu - {{}^{2}\!}{\hat
{S}}^{~\mu\rho}_\lam \Bar g^{\sig\lam}\BD_{\rho}\xi_{\sig}
\nonumber\\ &{}&- {{\delta \lag_{G2}} \over {\delta g_{\rho\sig}}}
\left.g_{\rho\sig} \right|^\mu_\nu \xi^\nu - {{\delta \lag_{G2}}
\over {\delta \Bar g_{\rho\sig}}} \left.\Bar g_{\rho\sig}
\right|^\mu_\nu \xi^\nu \equiv  \Bar D_\nu \left[- {{}^{2}\!}{\hat
{S}}^{~\mu\nu}_\lam \xi^\lam \right]\, .
 \m{B22}
 \eea
The quantity
 \be
 {{}^{2}\!}\hat {S}^{~\mu\nu}_\lam
 \equiv {{\di
\lag_{G2}} \over {\di\l({\Bar D_\mu} g_{\rho\sig}\r)}} \left.{
g_{\rho\sig}} \right|^\nu_\lam + {{\di \lag_{G2}} \over
{\di\l({\di_\mu}\Bar g_{\rho\sig}\r)}} \left.{\Bar g_{\rho\sig}}
\right|^\nu_\lam\,
 \m{+B26+}
 \ee
is antisymmetric in $\mu$ and $\nu$. Thus on the right hand side of
(\ref{B22})  the quantity $[- {{}^{2}\!}{\hat {S}}^{~\mu\nu}_\lam
\xi^\lam] $ plays the role of a superpotential. Besides, the
expression (\ref{+B26+}) is connected with the spin term
(\ref{(1.35)}) as
 \be
 {{}^{2}\!}\hat {S}^{~\mu\nu}_\lam \Bar g^{\lam\rho} =
 \hat\sig^{\rho[\mu\nu]}+
 \hat\sig^{\mu[\rho\nu]}-\hat\sig^{\nu[\rho\mu]}\, .
 \m{B26+}
 \ee
One has also to note that in (\ref{B22})
 \be
 {{\delta  \lag_{G2}} \over
{\delta \Bg_{\rho\sig}}} \left.{ \Bg}_{\rho\sig} \right|^\mu_\nu
\equiv {1 \over {\k}}\hat G^{L\mu}_{\nu}\, ,
 \m{GL-L2}
 \ee
it is exactly the expression defined in (\ref{G-L}) and
(\ref{(a2.18)}).

Now, substitute $\lag_{G2}  \equiv
 \lag_G - \lag_{G1}$  into Eq. (\ref{B22}) and subtract it from the
identity (\ref{(1.29)}) where $\hat \jmath^\mu$ is defined  in
(\ref{B16+}). Returning to the variables $ \hat l^{\mu\nu}$: $\hat
g^{\mu\nu}-
 \Bar{\hat g}^{\mu\nu} \goto
 \hat l^{\mu\nu}$, one obtains the new identity:
\be
 {1 \over {\k}} \hat G^{L\mu}_{\nu}\xi^\nu + {1 \over
 \k} \hat l^{\mu\lam} \Bar
 R_{\lam\nu}\xi^\nu + \hat Z_{(s)}^\mu \equiv {\Bar D_\nu} \hat
 J^{\mu\nu}_{(s)}\,
 \m{B27'}
 \ee
with the new superpotential connected with the KBL superpotential
$\hat J^{\mu\nu}_{(c)}$ given in (\ref{(1.31)}) as
 \bea
 \hat J^{\mu\nu}_{(s)} & \equiv &
 \hat J^{\mu\nu}_{(c)} +{{}^{2}\!}{\hat
{S}}^{~\mu\nu}_\lam \xi^\lam
 \m{B29}\\
&=& {1 \over \k} \hat l^{\rho[\mu}\Bar D_\rho\xi^{\nu]} + \hat {\cal
P}^{\mu\nu}{_\rho} \xi^\rho
\m{(2.26)}\\
&=& {1 \over \k} \l(\hat l^{\rho[\mu}\Bar D_\rho\xi^{\nu]}+
\xi^{[\mu}\Bar D_\sig \hat l^{\nu]\sig}-\bar D^{[\mu}\hat l^{\nu
]}_\sig \xi^\sig \r). \m{alaAbbottDeser}
 \eea
For $\Bar g_{\mu\nu}\goto\eta_{\mu\nu}$ and $\xi^\rho \goto
\lam^\rho = \delta^\rho_{(\alf)}$ the superpotential $\hat
J^{\mu\nu}_{(s)}$ transforms into the the Papapetrou superpotential
(\ref{Papapetrou}) (see the form (\ref{(2.26)})). The last term  on
the left hand side of Eq. (\ref{B27'}) is connected with the
expression (\ref{Zmu}) as
 \bea
 2\k \hat Z_{(s)}^\mu &\equiv &2\k\l({{}^{2}\!}{\hat
{S}}^{~\mu\nu}_\tau \Bg^{\tau\lam}
  + {{\di \lag_G} \over {\di \left(\bar D_\mu
 g_{\rho\sig}\right)}} \l.g_{\rho\sig}\r|^{\nu}_\tau \Bar
 g^{\lam\tau} \r)
 \zeta_{\nu\lam} + 2\k\hat Z^\mu_{(c)}\nonumber\\
&=& 2\l(\zeta^{\rho\sig}\Bar D_\rho \hat l^\mu_\sig - \hat
l^{\rho\sig}\Bar D_\rho \zeta^\mu_\sig\r) - \l(\zeta_{\rho\sig}\Bar
D^\mu \hat l^{\rho\sig} - \hat l^{\rho\sig}\Bar D^\mu
\zeta_{\rho\sig}\r) \nonumber\\ &+& \l(\hat l^{\mu\nu}\Bar D_\nu
\zeta^\rho_\rho - \zeta^\rho_\rho\Bar D_\nu \hat l^{\mu\nu}\r).
 \m{Zmu-new}
 \eea

The way of constructing (\ref{B19}) - (\ref{B27'}) is natural, but a
cumbersome one. Formally, one concludes that the final identity
(\ref{B27'}) is a result of  a difference between identities
(\ref{Lix-xi-arbitrary}) and (\ref{B21}). Thus, the N{\oe}ther
procedure could be applied directly to the difference $\lag_G -
\lag_{G2} = \lag_{G1}$. Indeed, recalculating the identity
 \be
 {\Lix} \lag_{G1} + \di_\mu (\xi^\mu \lag_{G1})
\equiv {\rm 0}\,
 \m{B21'}
 \ee
one obtains directly the identity (\ref{B27'}). The result is not
degenerated because $\lag_{G1}$ contains derivatives of the
background metric up to the second order.

After substituting the field equations (\ref{(a2.17)}) into the
identity (\ref{B27'}) one obtains the conservation law in the form:
 \be
 \hat
J^{\mu}_{(s)} \equiv \hat \Theta^{\mu}_{(s)\nu} \xi^\nu + \hat
Z_{(s)}^\mu = {\di_\nu} \hat J^{\mu\nu}_{(s)}\,
 \m{B32}
 \ee
where the conserved current $\hat J^{\mu}_{(s)}$ is expressed
through the superpotential $\hat J^{\mu\nu}_{(s)}$. The generalized
total energy-momentum tensor density is
 \be
 \hat
 \Theta^\mu_{(s)\nu} =
 \hat t^{(tot)\mu}_{\nu} +  {1 \over \k} \l(\hat l^{\mu\lam} \Bar
 R_{\lam\nu}-\hat \Phi^{L\mu}_{\nu}\r)\, .
 \m{B34}
 \ee
The role of interaction with the background is played by the term
\noindent $ (\k)^{-1} (\hat l^{\mu\lam} \Bar R_{\lam\nu}-\hat
\Phi^{L\mu}_{\nu})$. Next, if one uses the form of the field
equations (\ref{B30}), then (\ref{B34}) looks as
 \be
 \hat
 \Theta^\mu_{(s)\nu} =
\hat t^{(eff)\mu}_{\nu} +  {1 \over \k} \hat l^{\mu\lam} \Bar
R_{\lam\nu}\, ,
 \m{B34+}
 \ee
and one needs to consider $\hat t^{(eff)\mu}_{\nu} $ as the
energy-momentum of perturbations and  $(\k)^{-1} \hat l^{\mu\lam}
\Bar R_{\lam\nu}$ as the interacting term.  It is interesting to
note that the form of the energy-momentum (\ref{B34+}) is more close
to the form of the KBL energy-momentum (\ref{(1.33)}). The other
point is that the general term $\sim\hat l^{\mu\lam} \Bar
R_{\lam\nu}$ destructs the symmetry of $\hat \Theta^{\mu\nu}_{(s)}$.

Let us present also gauge invariance properties of the presented
current. Substituting  (\ref{(5.9)}) and (\ref{(5.9+)}) into
(\ref{B32}) and using the identity (\ref{B27'}) one obtains for
$\xi^\alf = \lam^\alf$:
 \bea
&{}&\hat J'^{\mu}_{(s)}(\lam) = \nonumber\\
&{}&\hat J^{\mu}_{(s)}(\lam)  + {1 \over \k}\di_\nu \l[\hat
(l'^{\rho[\mu} - l^{\rho[\mu})\Bar D_\rho\lam^{\nu]}+
\lam^{[\mu}\Bar D_\rho (\hat l'^{\nu]\rho}-\hat l^{\nu]\rho}) - \Bar
D^{[\mu}(\hat l'^{\nu]\rho}-\hat l^{\nu]\rho}) \lam_\rho \r]
 \nonumber\\& -&
 {1 \over
\k}\lam^\nu \Bar g^{\mu\lam} {{\di \Bar {\hat g}^{\rho\sig}}\over
{\di \Bar g^{\lam\nu}}} \sum^\infty_{k=1} {1\over k!} {\Lix}^k\l[
{{\di \Bar  g^{\delta\pi}}\over {\di \Bar{\hat g}^{\rho\sig}}}
 \l(\hat G^L_{\delta\pi} + \hat \Phi^L_{\delta\pi}
- \k{\hat t}^{(tot)}_{\delta\pi} \r) -2\k {{\delta\Bar
{\lag^E}}\over {\delta \Bar{\hat g}^{\rho\sig}}} \r]\, ,
\m{GaugeFCurrent}
 \eea
thus $\hat J^{\mu}_{(s)}$ is gauge invariant up to a divergence on
the dynamic and background equations.

Conserved quantities in Eq. (\ref{B32}), like the KBL quantities,
satisfy all the requirements of subsection \ref{UsingPseudo}.
Besides, the field-theoretical derivation has some advantages.
First, the quantities in Eq. (\ref{B32}) do not depend on
divergences in the dynamical Lagrangian. Second, the current in Eq.
(\ref{B32}) is defined by the unique quantity, the generalized
energy-momentum $\Theta^\mu_{(s)\nu}$, without a spin term. Third,
the conservation law (\ref{B32}) explicitly describes perturbations.

However, the currents and superpotentials in the framework of the
field approach bring in themselves the Boulware-Deser ambiguity
defined for  the total energy-momentum tensor in subsection
\ref{arbitrarydecompositions}. Let us outline this problem in
detail. Basing on the Lagrangian (\ref{(2.10-a)}) one can write out
the identity:
 \be {1 \over {\k}} \hat
 G^{L(a)\mu}_{\nu}\xi^\nu + {1 \over \k} \hat l^{\mu\lam}_{(a)}
 \Bar R_{\lam\nu}\xi^\nu + \hat Z^\mu_{(sa)} \equiv {\di_\nu}
 \hat J^{\mu\nu}_{(sa)}\, .
 \m{B50}
 \ee
Substituting the field equations (\ref{B39}) one obtains
 \be
\hat
 J^\mu_{(sa)} = \l[\hat t_{\nu}^{(tot~a)\mu}  +
{\k}^{-1} \l(\hat l^{\mu\lam}_{(a)} \Bar R_{\lam\nu}- \hat
\Phi^{L\mu}_{(a)\nu}\r)\r]\xi^\nu + \hat Z^\mu_{(sa)}
 = {\di_\nu}  \hat J^{\mu\nu}_{(sa)}
 \m{B51}
 \ee
where $\hat Z^\mu_{(sa)}$ is defined in (\ref{Zmu-new}) with
exchanging $\hat l^{\mu\nu}$ by $\hat l^{\mu\nu}_{(a)}$ (see the
definition  (\ref{B40})). The generalized superpotential is
 \be
 \hat {J}^{\mu\nu}_{(sa)} = {1 \over \k} \l(\hat
l^{\rho[\mu}_{(a)}\Bar D_\rho\xi^{\nu]}+ \xi^{[\mu}\Bar D_\sig \hat
l^{\nu]\sig}_{(a)}-\bar D^{[\mu}\hat l^{\nu ]}_{(a)\sig} \xi^\sig
\r).
 \m{B54}
 \ee
Thus, taking into account the difference in perturbations
(\ref{beta}) one obtains an analog of the Boulware-Deser ambiguity
in the definition of the superpotentials:
 \be
 \Delta\hat {J}^{\mu\nu}_{(sa)12} = {1 \over \k} \l(\hat
\beta^{\rho[\mu}_{(a)12}\Bar D_\rho\xi^{\nu]}+ \xi^{[\mu}\Bar D_\sig
\hat \beta^{\nu]\sig}_{(a)12}-\bar D^{[\mu}\hat \beta^{\nu
]\sig}_{(a)12} \xi_\sig \r)\, .
 \m{B55}
\ee In the next subsection we present arguments to resolve this
ambiguity.

\subsection{The Belinfante procedure on curved backgrounds}
 \m{BelOnArbitrary}

Returning to the problems of the canonical approach (see the end of
subsection \ref{CL-KBL}) we remark that the classical Belinfante
method resolves analogous problems in a canonical field theory in
Minkowski space. Indeed, the Belinfante corrected current
(\ref{Bpsi-current}), unlike the current (\ref{C1psi-current}), is
presented without a spin term. At the same time the Belinfante
symmetrized energy-momentum is equal to the symmetrical (metric)
energy momentum (\ref{T=t(psi)}), and thus it does not depend on
divergences in Lagrangian. Of course, a perturbed system on a curved
background in GR is a more complicated case. However, the KBL model
\cite{KBL} looks an appropriate one for an application of the
Belinfante procedure. Indeed, for a Killing vector the current $\hat
J^\mu_{(c)}$ in (\ref{(1.32)}) has the form (\ref{C1psi-current})
with the correspondent energy-momentum $\hat \Theta^\mu_{(c)\nu}$
and spin term $\hat \sig^{\mu\rho\sig}$. Thus, it is anticipated
that the Belinfante procedure could 1) resolve the problems of the
KBL approach, 2) connect it with the field-theoretical approach and,
consequently, 3) resolve problems of the last also. Therefore, in
this subsection, we generalize the Belinfante method for
constructing conservation laws in the perturbed GR on arbitrary
curved backgrounds and following the papers \cite{PK-lett,PK} apply
it to the KBL model.

Thus, for the spin tensor density (\ref{(1.35)}) with the rules
(\ref{(2.17)}) we construct the quantity:
 \be \hat{S}^{\mu\nu\rho}=
- \hat{S}^{\nu\mu\rho}=\hat\sig^{\rho[\mu\nu]}+
\hat\sig^{\mu[\rho\nu]}-\hat\sig^{\nu[\rho\mu]}\, \m{(2.23)}
 \ee and
present the KBL conservation law (\ref{(1.29)}) in the equivalent
form:
 \be \hat J^\mu_{(c)} + \di_\nu\l(\hat{S}^{\mu\nu\rho}\xi_\rho\r) =
\di_\nu\l(\hat J^{\mu\nu}_{(c)}+\hat{S}^{\mu\nu\rho}\xi_\rho\r).
\m{Katz+S}
 \ee
Defining the left hand side as a symmetrized current $\hat
{J}^{\mu}_{(B)}$ and the right hand side as a divergence of a
superpotential $\hat {J}^{\mu\nu}_{(B)}$, we rewrite (\ref{Katz+S})
in the form:
 \be \hat {J}^{\mu}_{(B)} = \hat \Theta^\mu_{(B)\nu}\xi^\nu + \hat {
Z}^\mu_{(B)} = \di_\nu\hat  {J}^{\mu\nu}_{(B)}. \m{(2.24)}
 \ee
Here, the spin term is absent; if $\xi^\mu$ is a Killing vector of
the background  $Z$-term disappears; the symmetrized energy-momentum
tensor density has the form:
 \be
\hat \Theta^\mu_{(B)\nu}= \hat \Theta^\mu_{(c)\nu} + \Bar D_\rho
\hat S^{\mu\rho}_{~~~\nu}\,. \m{(2.25)}
 \ee

Now, consider  properties of Belinfante symmetrized quantities.  The
symmetrized energy-momentum tensor density (\ref{(2.25)}) has the
structure
 \be
\hat \Theta^{\mu\nu}_{(B)} = \hat t^{\mu\nu}_{B} + (\hat
T^{(\mu}_\rho\Bar g^{\nu)\rho} -\Bar {\hat T}^{\mu\nu})+{1\over
2\k}\hat l^{\rho\sig} \Bar R_{\rho\sig} \Bar g^{\mu\nu} + {1\over
\k} {\hat l}^{\lam[\mu} \Bar R^{\nu]}_\lam. \m{(2.28)}
 \ee
Here, the first term is the symmetrical energy-momentum tensor
density for the free gravitational field:
 \bea
\k \hat t^{\mu\nu}_{B} & = & \half \l(\hat l^{\mu\nu}\Bar
g^{\rho\sig} - \Bar g^{\mu\nu}\hat l^{\rho\sig}\r) \Bar D_\sig
\Delta^\lambda_{\rho\lambda} \nonumber \\ & + &
  \l(\hat l^{\rho\sig} \Bar g^{\lambda(\mu} -
\Bar g^{\rho\sig}\hat l^{\lambda(\mu}\r) \Bar D_\sig
\Delta^{\nu)}_{\lambda\rho} \nonumber \\ &+ & \Bar
g^{\rho\sig}\l(\half\hat g^{\mu\nu}\Delta^\lambda_{\rho\lambda}
\Delta^\eta_{\sig\eta}+ \hat
g^{\lambda\eta}\Delta^{(\mu}_{\lambda\rho}\Delta^{\nu)}_{\eta\sig}\r)
\nonumber \\ & + & \Bar
g^{\rho\sig}\l(\Delta^{\lambda}_{\sig\eta}\Delta^{(\mu}_{\lambda\rho}\hat
g^{\nu)\eta} -
2\Delta^{\lambda}_{\sig\lambda}\Delta^{(\mu}_{\eta\rho}\hat
g^{\nu)\eta}\r) \nonumber \\ & + & \half \hat g^{\lambda\eta} \Bar
g^{\mu\nu}\Delta^\sig_{\rho\lambda}\Delta^\rho_{\sig\eta} \nonumber
\\ &+ & \hat g^{\lambda\eta}
\l(\Delta^{\sig}_{\rho\sig}\Delta^{(\mu}_{\lambda\eta} -
\Delta^{\sig}_{\lambda\eta}\Delta^{(\mu}_{\rho\sig} -
\Delta^{\sig}_{\lambda\rho}\Delta^{(\mu}_{\eta\sig}\r)\Bar
g^{\nu)\rho}. \m{(2.29)}
 \eea
The second term in (\ref{(2.28)}) is the perturbation of the matter
energy-momentum, the third and fourth terms describe the interaction
with the background geometry.

Because the fourth term in (\ref{(2.28)}) is antisymmetric, the
energy-momentum is symmetric: $\hat \Theta^{\mu\nu}_{(B)}=\hat
\Theta^{\nu\mu}_{(B)}$ if and only if $\Bar R_{\mu\nu} = \Bar
\Lambda \Bar g_{\mu\nu}$, i.e., for the cases when backgrounds are
Einstein spaces in Petrov's terminology \cite{Petrov}. Thus, the
Belinfante symmetrization is restricted in the sense of constructing
really {\it symmetric} quantities. Similary, the differential
conservation law $\Bar D_\nu \hat \Theta^{\mu\nu}_{(B)}= 0$ has also
a place if and only if backgrounds are Einstein spaces. However,
even on arbitrary curved backgrounds for Killing vectors $\lam^\mu$
the current is defined, like in (\ref{Bpsi-current}): $\hat
{J}^\mu_{(B)} =\hat \Theta^{\mu}_{(B)\nu}\lam^\nu$, and is
conserved: $\di_\mu \l(\hat \Theta^{\mu}_{(B)\nu}\lam^\nu\r) = 0$.
Thus both ``defects'' compensate one another. This could be useful,
e.g., for constructing angular momenta of relativistic astrophysical
objects on a FRW background, which is not an Einstein space, but
which has rotating Killing vectors.

Unlike the KBL quantities, second derivatives of $g_{\mu\nu}$ appear
in the energy-momentum tensor density (\ref{(2.29)}).  This needs
some comments also. The canonical  energy-momentum  $\hat
t^\mu_\nu$, see (\ref{(1.34)}), is quadratic in first order
derivatives, and this is the normal behaviour of a conserved
quantity relating to initial conditions. Consider  the local
quantities $\hat {J}^\mu_{(B)}$ in (\ref{Katz+S}) and
(\ref{(2.24)}). Suppose that initial conditions are defined on a
hypersurface at a given time coordinate $x^0=\const$.  Then it is
important to examine initial conditions for zero's component $\hat
{J}^0_{(B)}= \hat {J}^0_{(c)} + \di_k (\hat S^{0k\sig}\xi_\sig)$
only, see Eq. (\ref{(1.26)}). Such a form is because $\hat
S^{\mu\nu\sig}$  is anti-symmetric in the first two indices, see
(\ref{(2.23)}). Thus since $\hat {J}^0_{(c)}$ and $\hat
S^{0k\sig}\xi_\sig$ contain only first order {\it time} derivatives,
$\hat {J}^0_{(B)}$ itself contains only first order {\it time}
derivatives of the metric and therefore has the normal behaviour
with respect to initial conditions.  Thus, this requirement may be
unnecessarily restrictive.

Now let us discuss the properties of the quantities   in
(\ref{(2.24)}). Comparing the expressions in Eqs. (\ref{B26+}) and
(\ref{(2.23)}) and definitions in Eqs. (\ref{B29}) and
(\ref{Katz+S}) we find out that the superpotentials are identical.
The same conclusion is related to $Z$-terms, thus
 \bea
 \hat {J}^{\mu\nu}_{(B)} &=& \hat {J}^{\mu\nu}_{(s)}\, ,
 \m{JB=Js}\\
 \hat {Z}^\mu_{(B)} &=& \hat Z_{(s)}^\mu\, .
 \m{ZB=Zs}
 \eea
Next, taking into account these relations and  comparing (\ref{B32})
with (\ref{(2.24)}) we conclude that $\hat {J}^{\mu}_{(B)} = \hat
{J}^{\mu}_{(s)}$ and, consequently it has to be
 \be
 \hat \Theta^\mu_{(B)\nu} =  \hat \Theta^\mu_{(s)\nu}\,.
 \m{B35}
 \ee
Although the structure of these generalized energy-momenta are
different, direct calculations with the use of the field equations
also assert the claim (\ref{B35}). Thus, on arbitrary curved
background one cannot use separately the energy-momentum of the free
gravitational field neither $\hat t^B_{\mu\nu}$, nor $\hat
t^g_{\mu\nu}$. One needs to use the {\it total} energy-momentum
tensor density, which is the same, as is seen from in (\ref{B35}).

From all of these one concludes
 \bit
\item
{ The construction of the conservation laws for perturbations on
arbitrary curved backgrounds in GR with the use of the generalized
Belinfante symmetrization transforms the conservation laws of the
canonical system into the ones of the field-theoretical approach.
Thus, the Belinfante method is a ``bridge'' between the canonical
N{\oe}ther and the field-theoretical approaches.}
 \eit
In spite of that Eq. (\ref{B35}) is an analog of the relation
(\ref{T=t(psi)}) in a simple field theory (\ref{(2.9)}), this
conclusion is not so evident. Indeed, we cannot apply the Belinfante
method {\em only} in the framework of the KBL bimetric system
(\ref{KBLLagrangian}) because we cannot define a symmetrical
energy-momentum analogously to (\ref{(2.12)}). Really, the variation
of the Lagrangian (\ref{KBLLagrangian}) with respect to the
background metric gives only the background Einstein equations,
whereas the variation of the Lagrangian (\ref{(1.27)}) gives only
the non-dynamical background Einstein tensor. Only a non-trivial
connection with the field approach turns out fruitful leading to
(\ref{JB=Js}) - (\ref{B35}).

Relations (\ref{JB=Js}) - (\ref{B35}) solve the aforementioned
problems of the canonical construction. Indeed the quantities $\hat
\Theta^\mu_{(B)\nu}$, $\hat {Z}^\mu_{(B)}$ and $\hat {
J}^{\mu\nu}_{(B)}$ are, in fact, the quantities in the
field-theoretical frame and therefore they do not depend on
divergences in the Lagrangian. A direct mathematical substantiation
is given in subsection \ref{NIdentities}. Thus we have arrived a one
of desirable results.

Now, return to the ambiguity of the field approach in currents and
superpotentials noted at the end of subsection \ref{NewCLinFieldGR}.
Remark that the KBL approach and the Belinfante symmetrization do
not depend on a choice of the variables, like $g^{\mu \nu },~g_{\mu
\nu },~\sqrt{-g}g^{\mu \nu },~\ldots$,  and {\it uniquely} lead to
the conservation law (\ref{(2.24)}). On the other hand, it coincides
{\it uniquely} with conservation law (\ref{B32}) defined by the
second decomposition in (\ref{B38+}). Thus it is a theoretical
argument in favor of a choice $\hat l_{(a)}^{\mu \nu }= \hat l^{\mu
\nu }$ in (\ref{B40}). To support this theoretical conclusion below
we present a useful test. Recall that differences between
superpotentials in the family (\ref{B35}) appear in the second
order. Just the second order is crucial, e.g., in calculations of
energy and its flux \cite{BMS} at null infinity. As it was checked
in \cite{PK}, only the superpotential (\ref{alaAbbottDeser}) from
the family (\ref{B55}) gives the standard result \cite{BMS}, whereas
all the others, e.g., the superpotential corresponding to the first
decomposition in (\ref{B38+}) and defined in \cite{AbbottDeser82},
do not.

\subsection{Applications} \m{Applications}

The above presented formalism, its technique and expressions (in
particular, the energy-momentum tensor ${\hat t}^{(tot)}_{\mu\nu}$
and its properties, including gauge invariance ones) were used in
various applications in GR. Shortly we note that ${\hat
t}^{(tot)}_{\mu\nu}$ was used in development of quantum mechanics
with non-classical gravitational self-interaction \cite{PP1,PP1a},
in the framework of which inflation scenario was analyzed
\cite{PP2,PP3}. Some other applications are reviewed below in more
detail.

\subsubsection{The weakest falloff at spatial infinity} \m{AFST}

The proof of the positive-energy theorem in \cite{SchoenYau} -
\cite{Nester81} has renewed a great interest to asymptotically flat
spacetimes in GR. In particular, these investigations showed that
the standard asymptotic conditions at spatial infinity  ($1/r$
falloff in metric) can be significantly relaxed (see, e.g.,
\cite{Soloviev} - \cite{Chrusciel87} and references there in). These
models are very appropriate for applications in the framework of the
field approach. Indeed, asymptotically a background is just chosen
naturally, it is a Minkowski space at spatial infinity.
Gravitational field at spatial infinity is weak metrical
perturbations with respect to this background. We have done such
applications: both in Lagrangian \cite{Petrov95} and in Hamiltonian
\cite{Petrov97} derivation we examine the weakest asymptotic
behaviour at spatial infinity. We consider so-called {\it real}
isolated systems, for which all the physical fields are effectively
concentrated in a confined space at finite time intervals, and a
falloff of which is invariant under asymptotic Poincar\'e
transformations.

As an initial falloff we use the standard one: $1/r$. Then choosing
a background spacetime as the Minkowski space we evaluate a field
configuration in the framework of the field approach. For simplicity
it is assumed that a manifold, which globally supports the physical
metric, supports also the auxiliary flat metric. To search for the
weakest falloff global conserved quantities of the isolated system
(integrals of motion) are examined. We construct them in the form of
the surface integrals (\ref{int-surface+}) with the use of the
superpotential defined in (\ref{alaAbbottDeser}) and the ten Killing
vectors of the Minkowski space.

To find the weakest falloff conditions one uses the gauge invariance
properties of the field formulation of GR in section
\ref{fieldgauge}. Taking into account the general formula
(\ref{GaugeFCurrent}) for the current (\ref{B32}) we conclude that
the integrals of motion are gauge invariant up to surface term on
the solutions of  the field equations. Then we search for the
weakest falloff for the gauge potentials $\xi^\alpha$ and their
derivatives which  ensure vanishing the non-invariant terms. After
that we find out the weakest falloff for the gravitational
potentials:
 \be {l}^{\mu\nu} =
O^{+}(r^{-\varepsilon}) + O^{-}(r^{-\delta})\,;\qquad \varepsilon +
\delta > 2\,,~~1 \ge \varepsilon > {1\over 2}\,,~~\delta
> 1\,   \m{-(3.18)}
 \ee
where $(+)$ and $(-)$ mean even and odd functions with respect to
changing sign of the 3-vector ${\vec n} = {x^k}/r $. To obtain the
simple conclusion that the falloff can be weakened up to $r^{-1/2}$
one needs to examine energy and momentum only (see, e.g.,
\cite{Bartnic,OMurchadha86}).  To obtain the more detailed falloff
(\ref{-(3.18)}) one has to study all the ten integrals of motion. In
this relation we remark also the works \cite{Soloviev,Chrusciel87}.
In \cite{Soloviev} the algebra of asymptotic Poincar\'e generators
is studied, and in \cite{Chrusciel87}  the finiteness of the ADM
definition of angular momentum  is examined. The result
(\ref{-(3.18)}) based on the gauge invariance requirement coincides
with the results in \cite{Soloviev} corrected not significantly, and
with the results in \cite{Chrusciel87}, if the last are adopted by
some natural requirements (see \cite{Petrov95,Petrov97}). Thus the
three different approaches give the same weakest falloff.

Historically, among the ten integrals of motion, not enough
attention was payed to the Lorentzian momentum. In this relation the
paper by Regge and Teitelboim \cite{ReggeTeitelboim} who included
the falloff of odd and even parity, and the paper by  Beig and
\'OMurchadha \cite{BOM} who improved the results of the work
\cite{ReggeTeitelboim} were the most important earlier works.
Currently this gap is closing \cite{BLP03}, \cite{Szabados03} -
\cite{NesterMengChen}. In \cite{BLP03} we demonstrate that the
quasi-local Brown-York \cite{BY93} center of mass integral
asymptotically agrees with the one given in \cite{BOM} and with the
one expressed by the curvature integrals, e.g., in
\cite{AshtekarHansen}.  In \cite{Szabados03} the falloff of the
gravitational potentials is analyzed in $n+1$ dimensions with
special attention to the Lorenzian momentum and matter variables. In
\cite{BLP03} and \cite{Szabados03} (for $n=3$) the falloff exploited
by Beig and \'OMurchadha \cite{BOM} is used. It is $r^{-1}$ for the
even parity first term and $r^{-\delta},~\delta>1$, for the next of
arbitrary parity term. Comparing with the spectrum of the conditions
(\ref{-(3.18)}) one finds that this condition corresponds to
$\varepsilon = 1$ and $\delta
>1$ in  the full spectrum (\ref{-(3.18)}). As an example, the other
end of the spectrum is $\varepsilon > 1/2$ and $\delta \ge 3/2$. In
\cite{NesterHoChen} and \cite{NesterMengChen}, in the framework of
the covariant Hamiltonian approach \cite{ChenNester} the {\it
quasi-local} center of mass integral in the teleparallel gravity and
GR is considered. This approach gives {\it quasi-local} 4-momentum,
angular momentum and center of mass integral in the framework of an
unique relativistic invariant (Poincar\'e invariant) description.

\subsubsection{Closed worlds as gravitational fields} \m{Friedmann}

In the paper \cite{GP86},  we show how a closed-world geometry can
be viewed as a gravitational field configuration, superposed on a
topologically trivial geometrically flat spacetime. As seems, it is
not very natural. However, we do this without contradictions. The
space components of this configuration  corresponds  to a so-called
stereographic projection of 3-sphere $S^3$ onto a flat 3-space
$E^3$, when  the ``south'' pole corresponds to the origin of the
coordinate system, whereas the ``north'' pole is identified with all
the points at infinity of $E^3$. Thus, to make the full
identification the ``north'' pole is ``knocked out'' of the sphere
$S^3$ to simplify its topology.

The background Minkowski space has an auxiliary character, as it has
to be. Formally, the field configuration is specified in an infinite
volume. However, with the use of the physically reasonable
measurements examining light signals in gravitational fields one
finds the standard volume for $S^3$ (see \cite{LL}). By considering
the metric relations established through the physical measurements
an observer will infer that homogeneity, an equivalence of observers
at all points of space, has a place.

A picture when topological and geometrical properties are replaced
by the properties of an effective field propagated in a trivial
spacetime could be interesting and useful. Thus, Rubakov and
Shaposhnikov \cite{RubakovSh} have shown that not too energetic
scalar particles could become effectively trapped in a potential
well even in a topologically trivial universe, although nontrivial
classical solutions would have to be present to play a role of an
external field.

The field configuration presenting the closed world is static and
has an asymptotic behaviour $1/r^2$. This ensures zero energy,
momentum and angular momentum of the system, which then could be
treated as a microuniverse. Indeed, such characteristics are true
for a Minkowski vacuum where classical fields and particles are
absent, and thus an idea of quantum birth of the universe could be
supported.

\subsubsection{A point particle in GR} \m{PointParticle}

The Schwarzschild solution is the one of the most popular models in
GR. Frequently, in spite of a complicated geometry, it is treated as
a point mass solution in GR \cite{LL}; currently the interest to
this viewpoint is renewed \cite{JVWitten} - \cite{GolubevKelner}.
However, if one considers GR in the usual geometrical description,
then this interpretation meets conceptual difficulties (see, for
example, \cite{Narlikar2} and also \cite{PetrovNarlikar1}).

Except a pure theoretical interest the point particle description
could be interesting and useful for experimental gravity problems.
Gravitational wave detectors such as LIGO and VIRGO will definitely
discover gravitational waves from coalescing binary systems
comprising of compact relativistic objects. Therefore it is
necessary to derive equations of motion of such components, e.g.,
two black holes. As a rule, at an {\em initial} step the black holes
are modeled by point-like particles presented by Dirac's
$\delta$-function. Then consequent post-Newtonian approximations are
used (see, for example, \cite{Damour-1,Damour-4} and references
therein).

The aforementioned difficulties \cite{Narlikar2}, at least, do not
appear in Newtonian gravity. To describe a point particle one has to
assume that the mass distribution has the form $\rho({\bf r}) =
m\delta({\bf r})$ where $\delta$-function satisfies the ordinary
Poisson equation, which in spherical coordinates is
 \be
 \nabla^2 \left({1\over r}\right) \equiv \l({{d^2}\over {dr^2}} +
 {2\over r}\,{d \over {dr}}\r){1\over r} = - 4\pi\delta({\bf r})\, .
 \m{Poisson}
 \ee
Then, the Newtonian potential $\phi=m/r$ satisfies the Newtonian
equation in the whole space including the point $r = 0$. Besides,
the whole mass of the system is defined by the {\it unique}
expression $\int_\Sigma dx^3 \rho({\bf r})$ where both the usual
regular distribution and the point particle density can be
integrated providing the standard mass $m$.

In \cite{PetrovNarlikar1}) we show that, analogously to the
Newtonian prescription, the point mass in GR can be described in a
non-contradictory manner in the framework of the field formulation.
The Schwarzschild solution was presented as a gravitational field
configuration in a background Minkowski space described in the
standard spherical Schwarzschild ({\it static}) coordinates. The
concept of Minkowski space was extended from spatial infinity (frame
of reference of a distant observer) up to the horizon $r = r_g$, and
even under the horizon including the worldline $r = 0$ of the true
singularity. The configuration satisfies the Einstein equations at
all the points of the Minkowski space, including $r = 0$, if the
operations with the generalized functions \cite{GSh} are valid. The
energy-momentum tensor was constructed. Its components are presented
by  expressions proportional to $\del({\bf r})$ and by free
gravitational field outside $r = 0$. The picture is clearly
interpreted as a point particle distribution in GR. Indeed, the
configuration is essentially presented by $\delta$-function, one can
use the volume integration over the {\it whole} Minkowski space and
obtain the total energy $mc^2$ in the natural way. However at $r=
r_g$ the energy density and other characteristics have
discontinuities. A``visible'' boundary between the regions outside
and inside the horizon exists and does not allow to consider an
evolution of events continuously.

However, it is not a real singularity. In the field formulation this
situation is interpreted as a ``bad'' fixing of  gauge freedom,
which can be improved. That is the break at $r= r_g$ can be
countered with the use of an appropriate choice of a flat
background, which is determined by related coordinates for the
Schwarzschild solution. At least, the use of the coordinates without
singularities at the horizon could resolve the problem locally at
neighborhood of $r= r_g$. Next, it would be natural  to describe the
true singularity by the world line $r=0$ of the chosen polar
coordinates. Besides, it is desirable to have appropriate
coordinates, in which the Schwarzschild solution conserves the form
of an asymptotically flat spacetime.

Recently Pitts and Schieve \cite{Pitts2004} defined and studied
properties of  a so-called ``$\eta$-causality''. Its fulfilment
means that  the physical light cone is inside the flat light cone at
all the points of the Minkowski space. It is necessary to avoid
interpretation difficulties under the field-theoretical presentation
of GR. By this requirement all the causally connected events in the
physical spacetime are described by the right causal structure of
the Minkowski space. A related position of the light cones is not
gauge invariant. We consider this requirement {\em only} to
construct a more convenient in applications and interpretation field
configuration for the Schwarzschild solution. The requirement of the
$\eta$-causality can be strengthened   by the requirement of a
``stable $\eta$-causality'' \cite{Pitts2004}. The last means that
the physical light cone has to be {\it strictly} inside the flat
light cone, and  this is important when quantization problems are
under consideration. Indeed, in the case of tangency a field is on
the verge of $\eta$-causality violation. Returning to the
presentation in the Schwarzschild coordinates in
\cite{PetrovNarlikar1} we note that it does not satisfy the
$\eta$-causality requirement.

In the paper \cite{Petrov2005a}, taking into account the above
requirements we  have found a desirable description. A more
appropriate gauge fixing corresponds to the {\it stationary} (not
static) coordinates presented in \cite{Petrov-H2,Petrov-H3} and
recently improved in \cite{Pitts2004}. We consider also the
contracting Eddingtom-Finkenstein coordinates in stationary form
\cite{[12]}. These two coordinate systems belong to a parameterized
family where all of systems satisfies all the above requirements.
The transformation
 \be
ct' = ct + r_g \ln \l|\l(\frac{r}{r_g}-
1\r)\l(\frac{r_g}{r}\r)^\alf\r|
 \m{AlphaTrans}
 \ee
gives just the parameterized by $\alf \in [0,2]$ family of metrics.
The cases  $\alf = 1$ and $\alf = 0$ correspond to the first and to
the second examples above. The {\em stable} $\eta$-causality {\em is
not} satisfied with $\alf = 0$ at $0 \le r \le \infty$. Thus, all
the configurations $\alf \in (0,2]$ are appropriate for the study
both classical and quantum problems, whereas the case $\alf = 0$
could not be useful for the study quantized fields. Properties of
field configurations corresponding to  $\alf \in (0,2]$
qualitatively are the same as for $\alf = 1$. In the terms of the
field approach \cite{GPP}, all the field configurations for $\alf
\in [0,2]$ are connected by gauge transformations and are physically
equivalent.

 \begin{figure*}[t]
     \centering
     \includegraphics*%
     [width=9cm]%
      {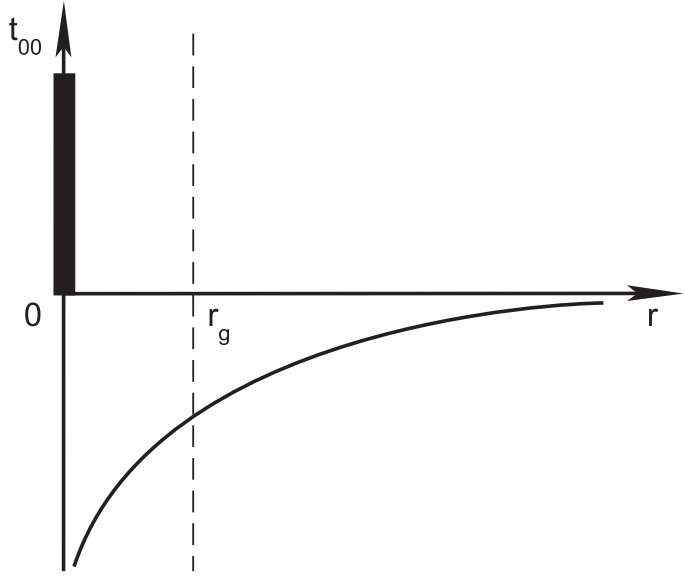}\\
            \caption{The energy distribution for the field configuration associated  with the Schwarzschild solution. The case: $\alf \in (0,\,2]$.}
      \label{F8}%
       \end{figure*}

\begin{figure*}[t]
     \centering
     \includegraphics*%
     [width=9cm]%
      {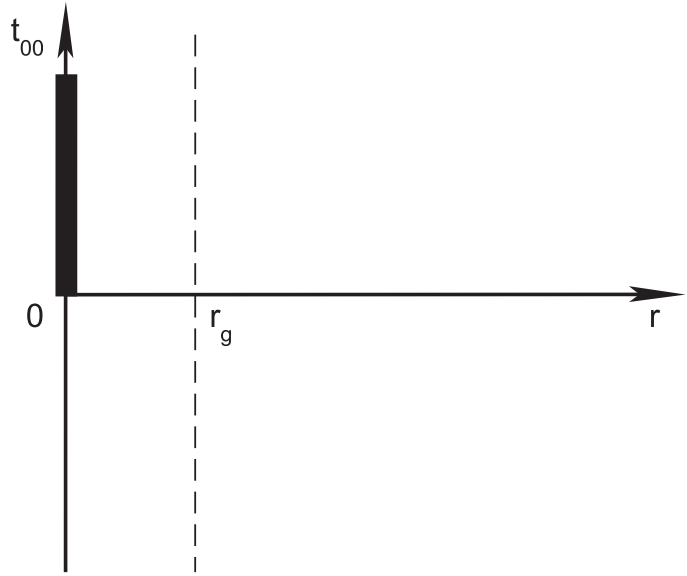}\\
            \caption{The energy distribution for the field configuration associated with the Schwarzschild solution. The case: $\alf =0$.}
      \label{F9}%
       \end{figure*}

Both the gravitational potentials and the components of the total
energy-momentum tensor for the field configurations with $\alf \in
[0,2]$ have no a break at $r = r_g$.  The energy distribution is
described by the $00$-component of the energy-momentum tensor, and
for $\alf \in (0,2]$ qualitatively is presented on the figure
\ref{F8}. Then the total energy of the system can be calculated both
by the volume integration and  by the surface integration over the
2-sphere with $r \goto \infty$ giving $mc^2$. For the case $\alf =
0$ the components of the total energy-momentum tensor for the field
configuration have the simplest form:
 \bea
 t^{tot}_{00} & = & mc^2 \delta({\bf r})\, ,\nonumber\\
t^{tot}_{11} & = & - mc^2\delta({\bf r})\, ,\nonumber\\
t^{tot}_{AB} & = & - \half \Bar g_{AB}\, mc^2\delta({\bf r})\,.
 \m{t-tot-EF}
 \eea
This energy-momentum is concentrated {\it only} at $r=0$, see the
energy distribution on the figure \ref{F9}. The other component
$t^{tot}_{11}$ and the angular ones $t^{tot}_{AB}$ formally could be
interpreted as related to the ``inner'' properties of the point.
Indeed, they are proportional only to $\delta({\bf r})$ and, thus,
describe the point ``inner radial'' and ``inner tangent'' pressure.
At last, again the total energy for (\ref{t-tot-EF}) is $mc^2$.

The field configurations for $\alf \in [0,2]$ satisfy the Einstein
equations (\ref{(a2.23)}) at all the points of the Minkowski space
including $r=0$. Then, keeping in mind the presentations on the
figures \ref{F8} and \ref{F9} one can
 conclude that an appropriate
description of a point particle in GR is approached. It is directly
and simply continues the Newtonian derivation and, at the same time,
does not contradict GR corresponding its principles.

\subsubsection{Integral constraints in FRW models} \m{intFRW}

In this subsection we consider the linear perturbations on FRW
backgrounds from the point view of the conservation laws
(\ref{(2.24)}), or the same (\ref{B32}), for the detail calculations
see \cite{PK}. Write the background metric $d\Bar s^2$ in
dimensionless coordinates $x^0=\eta, x^k$ as follows
\cite{Weinberg-book}:
 \be d\Bar s^2= \Bar g_{\mu\nu}dx^\mu dx^\nu=
a^2(\eta)(d\eta^2-f_{kl} dx^kdx^l)= a^2(\eta) e_{\mu\nu}dx^\mu
dx^\nu, \m{FRWback}
 \ee
 $a(\eta)$ is the scale factor, $f_{kl} =
\delta_{kl} + k {{\delta_{km}\delta_{ln}x^m x^n}} (1-kr^2)^{-1}$,
$f= det\,f_{kl} =(1-kr^2)^{-1}$, $k=0$ or $\pm 1$, and
$r^2=\delta_{kl}x^k x^l$. We define the perturbations by
$g_{\mu\nu}= a^2(e_{\mu\nu}+ \~ h_{\mu\nu})$ as often used in
cosmology (see, e.g., \cite{KRMS,Bertschinger96}), which are
connected with the the usual definition (\ref{(a2.4)}) in linear
approximation:
 \be \hat l^{\mu\nu} =
a^2\sqrt{f}(-e^{\mu\rho}e^{\nu\sig}+
{\half}e^{\mu\nu}e^{\rho\sig})\~h_{\rho\sig}\, . \m{FRWperturb}
 \ee
Because the FRW spacetimes are conformal to Minkowski space the
conformal Killing vectors of FRW spacetimes are thus those of
Minkowski space written in $\eta,\, x^k$ coordinates (see
\cite{FultonRW} - \cite{KeaneBarrett}). There are 15 conformal
Killing vectors, ${\xi}_A$, $A = 1,\,2,\,\ldots,\, 15$, some of
which are pure Killing ones.

These vectors we use as displacement vectors in the expression
(\ref{(2.24)}). Next, we make linear combinations of $\xi_A$'s with
functions of $\eta$, say $c^A(\eta)\xi_A$. In general the
$c^A\xi_A$'s are not conformal Killing vectors but some linear
combinations turn out to be rather simple with clear physical
interpretations in appropriate gauge conditions. Integrating the
conservation law (\ref{(2.24)}) we construct the integral relations
connecting perturbations inside of a volume with perturbations at
its surface. Thus we consider a sphere $r= \const$ at constant time
$\eta$ and integrate zero components of (\ref{(2.24)}) $\hat
{J}^0_{(B)}=\di_l\hat{J}^{0l}_{(B)}$ over such a ball. After
cumbersome calculations for perturbations (\ref{FRWperturb}) one has
15 volume integrals connected with surface integrals.

Besides the matter perturbations $\delta T^0_\mu$ the volume
integrands contain two field quantities ${\cal Q} = \delta(-D_\mu
n^\mu)$ --- the perturbation of the external curvature scalar of
$\eta=\const$, and $\~h^m_m$ --- the space trace of the metric
petrurbations. There are four linear combinations which do not
involve $\~h^m_m$; they are associated with some linear combinations
of the conformal Killing vectors, which are just Traschen's
\cite{Traschen} ``integral constraint vectors".

Of course, we  must  still impose four gauge conditions to fix the
mapping of  the perturbed spacetime onto the background. One gauge
condition that simplifies almost all volume integrands is the
``uniform Hubble expansion" gauge  ${\cal Q}=0$ discussed, e.g., by
Bardeen \cite{Bardeen80}. Then, 14 of the 15 volume integrands
reduce to linear combinations of $\delta T^0_\mu$ {\it only}, and
thus the 14 relations present ``integral constraints'', among which
the 4 are new. The integrals represent momenta of the matter
energy-momentum tensor of order 0, 1 and 2 in powers of $x^a$ when
$k=0$ and with similar interpretations when $k= \pm1$.  There
remains one integral that contains both $\delta T^0_\mu $ and
$\~h^m_m$: it is  related to so-called conformal time translations
for $k=\pm 1$ or to so-called time accelerations for $k=0$.

Another gauge condition often used (see, e.g.,
\cite{Bertschinger96}) is $\nabla_l \vtop{{\hbox{$\~ h^l_k$}}
\vskip-7pt {\hbox{$\scriptscriptstyle T$}}}=0$ in which
$\vtop{{\hbox{$\~ h^l_k$}} \vskip-7pt {\hbox{$\scriptscriptstyle
T$}}}$ is the traceless part of $\~h^l_k$; these gauges remove the
longitudinal modes of the gravitational waves.  Combining $\nabla_l
\vtop{{\hbox{$\~ h^l_k$}} \vskip-7pt {\hbox{$\scriptscriptstyle
T$}}}=0$ with ${\cal Q}=0$ one finds four relations that are
independent of the gravitational radiation.

\section{$D$-dimensional
metric theories of gravity} \m{D-dimensions}
\setcounter{equation}{0}

As was discussed in Introduction,  multidimensional gravitational
models become more and more popular, their solutions (for example,
generalized $p$-branes and brane-world black hole solutions
\cite{branesW2} - \cite{branesW8}) induce an arising interest. Other
generalizations, say,  scalar-tensor theories of gravity
\cite{IgorVlasov}, are developed intensively also. In this section,
keeping in mind such generalizations we construct a perturbed scheme
and conservation laws for perturbations in a generic metric theory.
The presentation is based on the works
\cite{PP87,[15],PK2003a,Petrov2005b}. Besides, here the approach is
presented in the united scheme, and more details are given.

\subsection{The main identities}
\m{NIdentities}

In this subsection, we present necessary identities and stress their
important properties. Let the system of fields, set of tensor
densities $Q^A$, be described by the Lagrangian
 \be
\lag = \lag (Q^A; Q^A{}_{,\alf}; Q^A{}_{,\alf\beta}) \m{lagQ}
 \ee
including derivatives up to the second order. We assume that it is
an {\it arbitrary metric} theory of gravity in $D$ dimensions with
the generalized gravitational variables $(g^a, \Psi^B)$ and the
matter sources $\Phi^C$. Thus $Q^A = \{(g^a, \Psi^B),\, \Phi^C\}$;
the variables $\Psi^B$ are included keeping in mind, say,
scalar-tensor or vector-tensor theories; the metric variables $g^a$
are thought as defined in (\ref{(1)}), only in $D$ dimensions. Then
we include $\Bar g_{\mu\nu}$ as usual. The ordinary derivatives
$\di_{\alf}$ are rewritten over the covariant $\BD_\alf$ ones by
changing $Q_{B,\tau} \equiv \Bar D_\tau Q_B - \Bar
\Gamma^\sig_{\tau\rho} \l. Q_B \r|_\sig^\rho$. Then the Lagrangian
(\ref{lagQ}) takes an explicitly covariant form:
 \be \lag = \lag_c = {\lag}_c (Q_B; \Bar D_\alf
Q_{B}; \Bar D_\beta \Bar D_\alf  Q_{B})\, .
 \m{(+1+)}
 \ee
After that indexes  are shifted by $\Bar g_{\mu\nu}$ and $\Bar
g^{\mu\nu}$.

Now we use the standard technique (see, for example, \cite{Mitzk}).
For the Lagrangian (\ref{(+1+)}), as a scalar density,  we write out
again the identity: $
 {\pounds}_\xi \lag_c + (\xi^\alf \lag_c){}_{,\alf} \equiv 0$,
 which can be rewritten in the form:
 \be - \l[{{\delta \lag_c} \over {\delta Q_B}} \Bar
D_\alf Q_{B} + \Bar D_\beta \l( {{\delta \lag_c} \over {\delta Q_B}}
\l.Q_B\r|^\beta_\alf\r)\r]\xi^\alf
+\Bar D_\alf \l[\hat u^\alf_\sig\xi^\sig + \hat
m^{\alf\tau}_{\sig}\Bar D_\tau \xi^\sig + \hat
n^{\alf\tau\beta}_\sig \Bar D_{\beta} \Bar D_{\tau}\xi^\sig\r]
\equiv 0\, .
 \m{(+2+)}
\ee In (\ref{(+2+)}), the coefficients are defined by the Lagrangian
in unique way:
 \bea \hat u^\alf_\sig & \equiv & \lag_c
\delta^\alf_\sig +
 {{\delta \lag_c} \over {\delta Q_B}} \l.Q_B\r|^\alf_\sig -
 \l[{{\di \lag_c} \over {\di (\Bar D_\alf Q_{B})}} -
 \Bar D_\beta
\l({{\di \lag_c} \over {\di (\Bar D_\beta \Bar D_\alf Q_{B})}}\r)
\r] \Bar D_\sig Q_{B} \nonumber \\ & - & {{\di \lag_c} \over {\di
(\Bar D_\beta \Bar D_\alf Q_{B})}} \Bar D_\sig \Bar D_\beta Q_{B}-
\hat n^{\alf\tau\beta}_\lam \Bar R^\lam_{~\tau\beta\sig}\, ,
\m{(+3+)}
 \eea
\bea \hat m^{\alf\tau}_\sig & \equiv &
 \l[{{\di \lag_c} \over {\di (\Bar D_\alf Q_{B})}} -
 \Bar D_\beta \l({{\di \lag_c} \over {\di (\Bar D_\beta \Bar D_\alf Q_{B})}}\r)\r]
 \l.Q_{B}\r|^\tau_\sig \nonumber \\ & - &
 {{\di \lag_c} \over {\di (\Bar D_\tau \Bar D_\alf Q_{B})}}
\Bar D_\sig Q_B +
 {{\di \lag_c} \over {\di (\Bar D_\beta \Bar D_\alf Q_{B})}}
 \Bar D_\beta (\l.Q_{B}\r|^\tau_\sig)\, ,
\m{(+4+)}
 \eea
\be \hat n^{\alf\tau\beta}_\sig \equiv \half \l[{{\di \lag_c} \over
{\di (\Bar D_\beta \Bar D_\alf Q_{B})}}
 \l.Q_{B}\r|^\tau_\sig +
 {{\di \lag_c} \over {\di (\Bar D_\tau \Bar D_\alf Q_{B})}}
 \l.Q_{B}\r|^\beta_\sig\r].
\m{(+5+)}
 \ee

The coefficient at $\xi^\sig$ in the first term in  (\ref{(+2+)}) is
identically equal to zero (generalized Bianchi identity). Thus the
identity (\ref{(+2+)}) has the form of the differential conservation
law:
 \be
 \BD_\alf \hat \imath^\alf \equiv
 \di_\alf \hat \imath^\alf \equiv 0
 \m{(+6+)}
 \ee
with a generalized current
 \be \hat \imath^\alf \equiv
 -\l[\hat u^\alf_\sig\xi^\sig + \hat
m^{\alf\tau}_{\sig}\Bar D_\tau \xi^\sig + \hat
n^{\alf\tau\beta}_\sig \Bar D_{\beta} \Bar D_{\tau}\xi^\sig\r]\, .
 \m{++7++}
 \ee
We also use the helpful form:
 \be
\hat \imath^\alf \equiv - \l[(\hat u^{\alf}_\sig + \hat
n^{\alf\beta\gamma}_\lam \Bar R^\lam_{~\beta\gamma\sig})\xi^\sig +
\hat
 m^{\alf\beta}_{\sig}\bar g^{\sig\rho} \di_{[\beta}\xi_{\rho]} +
 \hat z^{\alf}\r]
 \m{(+7+)}
 \ee
with $z$-term  defined as
 \be
 \hat z^{\alf}(\xi) \equiv   \hat
m^{\alf\beta}_{\sig}\zeta^{\sig}_{\beta}+ \hat
n^{\alf\beta\gamma}_\sig \Bar g^{\sig\rho} \l(2
\BD_{\gamma}\zeta_{\beta\rho} - \BD_\rho \zeta_{\beta\gamma}\r)\, .
 \m{(+8+)}
 \ee
Again, if $\xi^\alf = \lam^\alf $, then $\hat z^{\alf} = 0$ and the
current (\ref{(+7+)}) is determined by the energy-momentum $(u +
n\Bar R)$-term and the spin $m$-term. Opening the identity
(\ref{(+6+)}) and, since $\xi^\sig$,  $ \di_{\alf}\xi^\sig$, $
\di_{\beta\alf}\xi^\sig$ and $\di_{\gamma\beta\alf}
 \xi^\sig$
are arbitrary at every world point, equating independently to zero
the coefficients at $\xi^\sig$,  $\BD_{\alf} \xi^\sig$,
$\BD_{(\beta\alf)} \xi^\sig$ and $\BD_{(\gamma\beta\alf)} \xi^\sig$
we get a ``cascade'' (in terminology by Julia and Silva
\cite{JuliaSilva98}) of identities:
 \bea
 &{}& \BD_\alf  \hat u^{\alf}_{\sig} + \half
 \hat m^{\alf\rho}_\lam \Bar R^{~\lam}_{\sig~\rho\alf}
 +{\textstyle{1\over 3}} \hat n^{\alf\rho\gamma}_\lam
\BD_\gamma \Bar R^{~\lam}_{\sig~\rho\alf}
  \equiv 0, \nonumber \\
&{}&    \hat u^{\alf}_\sig + \BD_\lam \hat m^{\lam \alf}_{\sig} +
\hat n^{\tau\alf\rho}_\lam
 \Bar R^{~\lam}_{\sig~\rho\tau} +{\textstyle{2\over 3}} \hat
 n^{\lam\tau\rho}_\sig \Bar R^{\alf}_{~\tau\rho\lam} \equiv 0,
 \nonumber \\ &{}&
 \hat m^{(\alf\beta)}_\sig +
\BD_\lam  \hat n^{\lam(\alf\beta)}_{\sig} \equiv 0, \nonumber\\ &{}&
 \hat
 n^{(\alf\beta\gamma)}_\sig \equiv 0.
 \m{(+9+)}
\eea The above constructions are the generalization to arbitrary
curved backgrounds of the expressions given by Mitzkevich
\cite{Mitzk}.

Since Eq. (\ref{(+6+)}) is identically satisfied, the current
(\ref{(+7+)}) must be a divergence of a superpotential $ \hat
\imath^{\alf\beta}$, for which $\di_{\beta\alf} \hat
\imath^{\alf\beta} \equiv 0$, that is
 \be
 \hat \imath^\alf \equiv \BD_{\beta} \hat \imath^{\alf\beta} \equiv
\di_{\beta}   \hat \imath^{\alf\beta}.
 \m{(+10+)}
 \ee
Indeed,  substituting $\hat u^\alf_\sig$ and $\hat
m^{\alf\beta}_\sig$ from Eqs.~(\ref{(+9+)}) {\it directly} into the
current and using algebraic properties of $n^{\alf\beta\gamma}_\sig$
and $\Bar R^{\alf}_{~\beta\rho\sig}$, and the third identity in
Eqs.~(\ref{(+9+)}) we reconstruct (\ref{(+7+)}) into the form
(\ref{(+10+)}) where  the superpotential is
 \be
 \hat \imath^{\alf\beta}  \equiv \l({\textstyle{2\over 3}}
 \BD_\lam  \hat n^{[\alf\beta]\lam}_{\sig}  - \hat
 m^{[\alf\beta]}_\sig\r)\xi^\sig   -
 {\textstyle{4\over 3}} \hat n^{[\alf\beta]\lam}_\sig
 \BD_\lam  \xi^\sig.
 \m {(+11+)}
 \ee
It is explicitly antisymmetric in $\alf$ and $\beta$. Of course, Eq.
(\ref{(+10+)}) has also a sense of the differential conservation law
Eq. (\ref{(+6+)}).

Let us find contributions into currents and superpotentials from a
divergence in the Lagrangian $\delta_d \lag_c = div = \hat
d^\nu{}_{,\nu}$. For the scalar density $ \hat d^\nu{}_{,\nu}$ one
has the identity $ ({\Lix} \hat d^\alf + \xi^\alf
 \hat d^\nu{}_{,\nu})_{,\alf} \equiv 0 $, which gives contributions
$\delta_d \hat \imath^\alf$ into the current (\ref{++7++}) and $
\delta_d\hat \imath^{\alf\beta}$ into the superpotential
(\ref{(+11+)}). The additional quantities are
 \bea
 \delta_d\hat u^\alf_\sig & =& 2\BD_\beta
(\delta^{[\alf}_\sig \hat d^{\beta]} )\,, \m{du}
\\
\delta_d\hat m^{\alf\beta}_\sig &=&
 2 \delta^{[\alf}_\sig \hat d^{\beta]}\, ,\m{dm}\\
 \delta_d\hat n^{\alf\beta\gamma}_\sig &=& 0\, , \m{dn} \\
\delta_d\hat \imath^{\alf\beta}  &=& -2 \xi^{[\alf} \hat
 d^{\beta]}\, .\m{dPhi}
 \eea
Then, for $\lag_c \goto \lag_c +
 \delta_d\lag_c$  Eq. (\ref{(+10+)}) changes as
\be
 \hat \imath^\alf+ \delta_d \hat \imath^\alf \equiv
\BD_{\beta}  \l(\hat \imath^{\alf\beta} + \delta_d \hat
 \imath^{\alf\beta}\r),
 \m{(+12+)}
\ee where the changes do not depend on a structure of $\hat
d^{\nu}$.

Next, using the general  Belinfante rule (\ref{(2.17)}) we define a
tensor density
 \be
\hat s^{\alf\beta\sig} \equiv - \hat s^{\beta\alf\sig} \equiv -
 \hat m^{\sig[\alf}_\lam \bar g^{\beta]\lam} -
 \hat m^{\alf[\sig}_\lam \bar g^{\beta]\lam} + \hat m^{\beta[\sig}_\lam
 \bar g^{\alf]\lam}\, ,
 \m{(+13+)}
 \ee
add  $ \BD_{\beta}(\hat s^{\alf\beta\sig}\xi_{\sig})$ to both sides
of (\ref{(+10+)}), and obtain a new identity:
 \be
 \hat \imath^\alf_B \equiv\BD_{\beta}  \hat \imath^{\alf\beta}_B
 \equiv \di_{\beta} \hat \imath^{\alf\beta}_B.
 \m{(+14+)}
 \ee
This modification cancels the spin term from the current
(\ref{(+7+)}):
 \be
\hat \imath^\alf_B \equiv \l(- \hat u^{\alf}_\sig -\hat
n^{\alf\beta\gamma}_\lam \Bar R^\lam_{~\beta\gamma\sig} +
\BD_{\beta} \hat s^{\alf\beta}{}_{\sig}\r)  \xi^\sig+
 \hat z^{\alf}_B(\xi)  \equiv
\hat u_{B\sig}^\alf  \xi^\sig+
 \hat z^{\alf}_B(\xi),
 \m{(+15+)}
 \ee
and a new $z$-term disappears also on Killing vectors of the
background:
 \be
 \hat z^{\alf}_B(\xi) =
\l(   \bar g^{\lam\tau}\hat
 m^{\beta\alf}_\lam +  \hat m^{\alf\tau}_\lam \bar
 g^{\beta\lam} - \hat  m^{\tau\beta}_\lam \bar g^{\alf\lam}\r)
 \zeta_{\tau\beta} +
 \hat n^{\alf\tau\beta}_\lam \l(2\Bar D_{(\beta} \zeta_{\tau)}^\lam -
 \Bar D_{\sig} \zeta_{\beta\tau}\bar g^{\lam\sig}\r)\, .
 \m{Zmu-2}
 \ee
Thus, the  current $\hat i^\alf_B$ is defined, in fact, by the
modified energy-momentum tensor density
 $\hat u_{B\sig}^\alf$.
Because the new superpotential depends on  the $n$-coefficients
  only:
 \be
 \hat
 i^{\alf\beta}_{B}  \equiv 2
\l({\textstyle{1\over 3}}\BD_\rho \hat n^{[\alf\beta]\rho}_{\sig}+
\BD_\tau\hat
 n^{\tau\rho[\alf}_{\lam} \bar g^{\beta]\lam} \bar g_{\rho\sig}\r)
\xi^\sig - {\textstyle{4\over 3}} \hat
 n^{[\alf\beta]\lam}_\sig\BD_{\lam}\xi^\sig,
 \m{(+16+)}
 \ee
then due to the definition (\ref{(+5+)}) it vanishes for Lagrangians
with only the first order derivatives. On the other hand, the
superpotential (\ref{(+16+)}) is well adapted to theories with
second derivatives in Lagrangians, like GR or the
Einstein-Gauss-Bonnet gravity. It is also important to note that for
the superpotentials (\ref{(+11+)}) and (\ref{(+16+)}) their forms do
not depend {\em explicitly} on  a dimension $D$; the same property
is related to the Deser and Tekin superpotentials \cite{DT2}.

It is important to  note that the Belinfante procedure applied to
(\ref{(+12+)}) cancels the quantities $\delta_d \hat \imath^{\alf}$
and $\delta_d \hat \imath^{\alf\beta}$ and gives again
(\ref{(+14+)}). Let us show this. The quantity (\ref{(+13+)})
constructed for $\delta_d \hat m^{\alf\beta}_{\sig}$ in (\ref{dm})
gives
 \be
 \delta_d \hat s^{\alf\beta\sig}\xi_\sig =
 2\xi^{[\alf}\hat d^{\beta]}\, .
 \m{s-dm}
 \ee
Then, adding  $\BD_{\beta}(\hat s^{\alf\beta\sig}\xi_{\sig} +
\delta_d \hat s^{\alf\beta\sig}\xi_{\sig})$ to Eq. (\ref{(+12+)})
one cancels completely the spin term and suppresses $\delta_d \hat
u^{\alf}_\sig$ and $\delta_d \hat \imath^{\alf\beta}$. Indeed,
combining (\ref{du}) and (\ref{dPhi}) with (\ref{s-dm}) one has
$\delta_d \hat u^{\alf}_\sig - \BD_{\beta}(\delta_d \hat
s^{\alf\beta\rho}\bar g_{\rho\sig}) = 0$~and~ $ \delta_d \hat
\imath^{\alf\beta} + \delta_d \hat s^{\alf\beta\sig}\xi_\sig = 0$
for an arbitrary $\hat d^\nu$. This just supports the claim that
$\hat{J}^{\mu}_{(B)}$, $\hat{\Theta}^{\mu}{}_{(B)\nu}$ and
$\hat{J}^{\mu\nu}_{(B)}$ in (\ref{(2.24)}) are independent on
divergences.

\subsection{The field-theoretical formulation for perturbations}
\m{PerturbedDTheory}

In this subsection, following the method of section
\ref{decompositions} we develop a perturbed derivation. The field
equations corresponding to the system (\ref{lagQ}) are derived as
usual:
 \be
 {\delta \lag}/{\delta Q^A} = 0\, .
 \m{Q-Eq}
 \ee
Let us decompose $Q^A$ onto the background part $\Bar Q^A$ and the
dynamic part $q^A$:
 \be
 Q^A = \Bar Q^A + q^A\, .
 \m{Q-Dec}
 \ee
The background fields satisfy the background equations
 \be
 {\delta \Bar \lag}/{\delta \Bar Q^A} = 0\,
 \m{Q-Back}
 \ee
where $\Bar \lag = \lag (\Bar Q)$. Taking into account the form of
the Lagrangian (\ref{(2.10)}) (or (\ref{(2.10-a)})) we will describe
the perturbed system by the Lagrangian
 \be
\lag^{dyn}(\Bar Q,q) = \lag (\Bar Q+q) - q^A \frac{\delta \Bar
\lag}{\delta \Bar Q^A} - \Bar \lag + div
 \m{lag}
 \ee
instead of the original Lagrangian $\lag(Q) = \lag(\Bar Q+q)$. As
before, the background equations should not  be taken into account
before variation of $\lag^{dyn} (\Bar Q,q)$ with respect to $\Bar
Q^A$. Using the evident property ${\delta \lag (\Bar Q+q)}/{\delta
\Bar Q^A} = {\delta \lag (\Bar Q+q)}/{\delta q^A}$, the equations of
motion related to the Lagrangian (\ref{lag}) are presented as
 \be
\frac{\delta \lag^{dyn}}{\delta q^A} = \frac{\delta }{\delta \Bar
Q^A}\l[\lag(\Bar Q + q) - \Bar \lag\r] = 0\, .
 \m{PERTeqs1}
 \ee
It is clear that they are equivalent to the equations (\ref{Q-Eq})
if the background equations (\ref{Q-Back}) hold.

Defining the ``background current''
 \be
t^q_A \equiv \frac{\delta \lag^{dyn}}{\delta \Bar Q^A} =
\frac{\delta \lag^{dyn}}{\delta q^A} -\frac{\delta }{\delta \Bar
Q^A}q^B \frac{\delta \Bar \lag}{\delta \Bar Q^B}\,
 \m{PERTcurrent}
 \ee
and combining this expression with (\ref{PERTeqs1}) one obtains
another form for the equation (\ref{PERTeqs1}):
 \be
 G^{Lq}_A + \Phi^{Lq}_A\equiv
-\frac{\delta }{\delta \Bar Q^A}q^B \frac{\delta \Bar \lag}{\delta
\Bar Q^B} = t^q_A\,.
 \m{PERTeqs2}
 \ee
The left hand side in (\ref{PERTeqs2}) is a linear perturbation of
the expression ${\delta \lag}/{\delta Q^A}$ in (\ref{Q-Eq}), $G^L_A
$ is a pure gravitational part. Really gravitational variables $g^a$
and $\Psi^B$ have a different nature. Conservation laws are
connected with symmetries of a geometry related directly to $g^a$,
not with $\Psi^B$. Thus, we will consider $\Psi^B$ as included in
$\Phi^B$, setting thus in calculations $\Psi^B = 0$ and $Q^A =
\{g^a,\, \Phi^B\}$. Later we will consider the lagrangian
 \be
 \lag = \lag_{Dg}(g^a) + \lag_{Dm}(g^a,\Phi^B)
 \m{lag-g}
 \ee
with the pure metric gravitational part $\lag_{Dg}$. Then the
equations (\ref{PERTcurrent}) are separated into gravitational and
matter parts as follows
 \be
G^{Lq}_a + \Phi^{Lq}_a\equiv - \frac{\delta }{\delta \Bar g^a}q^B
\frac{\delta \Bar {\lag}}{\delta \Bar Q^B} = t^q_a\,,
 \m{PERTeqs2+}
 \ee
 \be
\Phi^{Lq}_C \equiv - \frac{\delta }{\delta \Bar \Phi^C}q^B
\frac{\delta \Bar {\lag}_{Dm}}{\delta \Bar Q^B} = t^q_C\,.
 \m{PERTeqs2++}
 \ee
As is seen, the form of the perturbed equations (\ref{PERTeqs2}),
(\ref{PERTeqs2+}) and (\ref{PERTeqs2++}) is a quite universal form
with the ``background current'' as a source on the right hand side.
Contracting (\ref{PERTeqs2+}) with $2\di \Bar g^a/\Bar g^{\mu\nu}$
one obtains
  \be
\hat G^{Lq}_{\mu\nu} + \hat \Phi^{Lq}_{\mu\nu} = \hat t^q_{\mu\nu}\,
 \m{PERT-munu}
 \ee
where the  linear operator is defined as
 \be
 \hat G^{Lq}_{\mu\nu}+ \hat \Phi^{Lq}_{\mu\nu} \equiv
 -2\frac{\delta }{\delta \Bar g^{\mu\nu}}\l(\hat
 l^{\alf\beta}_{(a)}
\frac{\delta \Bar \lag_{Dg}}{\delta \Bar {\hat g}^{\alf\beta}} + q^B
\frac{\delta \Bar \lag_{Dm}}{\delta \Bar Q^B} \r)
 \m{GLPhiL-q}
 \ee
with independent gravitational variables $\hat
 l^{\alf\beta}_{(a)}$ defined as in (\ref{B40}).
The right hand side in (\ref{PERT-munu}) is defined by variation of
(\ref{lag})
 \be
 \hat t^q_{\mu\nu} \equiv 2\frac{\delta\lag^{dyn}(\Bar Q,q)}{\delta \Bar g^{\mu\nu}}
 \m{em-q}
 \ee
and is the generalized symmetric energy-momentum.

The equations (\ref{PERT-munu}) generalize the equations of GR
(\ref{(a2.17)}), they generalize also the Deser-Tekin equations
\cite{DT2,DT1,DT3} constructed for the quadratic theories by direct
calculations. Below we will show that for the vacuum backgrounds
(when $\hat \Phi^{Lq}_{\mu\nu}= 0$) $\Bar D^\mu\hat
G^{Lq}_{\mu\nu}\equiv 0$, and thus the energy momentum tensor $\hat
t^q_{\mu\nu}$ is differentially conserved, like in the Petrov spaces
(\ref{(2.23')}) in GR. The presented here model has the same
properties for expansions as was described in subsection
\ref{Expansions} and the same gauge properties as was described in
detail in subsection \ref{fieldgauge}.

\subsection{Currents and superpotentials in the field formulation}
\m{CSinFF}

In the recent series of the current works
\cite{DT2,Desercom,DT1,DT3} Deser with coauthors develop a
construction of conserved charges for perturbations about vacua in
metric quadratic (in curvature) gravity theories in $D$ dimensions.
They apply the Abbott and Deser procedure \cite{AbbottDeser82} and
develop it. The aim of the present subsection is to suggest an
approach, which generalizes the Deser and Tekin constructions. We
construct conserved currents and superpotentials corresponding to
the equations (\ref{PERT-munu}) derived for an arbitrary metrical
gravitational theory, and not only on a vacuum background. Below
taking into account the definition of the gravitational part of
linear operator (\ref{GLPhiL-q}) we demonstrate that conserved
quantities of the system and their properties can be obtained and
described analyzing {\it only} the scalar density $\lag_1 \equiv
\hat l_{(a)}^{\alf\beta} ({\delta \Bar \lag_{Dg}}/{\delta \Bar {\hat
g}^{\alf\beta}})$, which is the gravitational part of  second term
in the Lagrangian (\ref{lag}). As an important case  we consider
explicitly only such theories where $\lag_1$ has derivatives not
higher than of second order, like the Einstein-Gauss-Bonnet gravity.
In principle, our results can be repeated when $\lag_1$ has
derivatives of higher orders, like in \cite{DT2}.

Keeping in mind that $\lag_1$ is the scalar density we again follow
the standard technique and use the results of the subsection
\ref{NIdentities}, which are universal. One transforms the identity
${\pounds}_\xi \lag_1 + \di_\alf (\xi^\alf\lag_1) \equiv 0 $ into
the identity:
 \be
 \Bar D_\mu \hat \imath_1^\mu \equiv \di_\mu \hat \imath_1^\mu
\equiv 0\, \m{nabla-m2}
 \ee
where
 \bea
\hat \imath_1^\mu &\equiv &-\l(\lag_1\xi^\mu + \xi^\nu \l.\hat
l^{\rho\sig}_{(a)}\r|^\mu_\nu \frac{\delta \Bar \lag_{Dg}}{\delta
\Bar {\hat g}^{\rho\sig}}  + 2\xi^\sig \frac{\delta \lag_1}{\delta
\Bar g^{\rho\sig}}\Bar g^{\rho\mu}\r) + \hat {\cal Z}^\mu_{(s)}\,,
 \label{Jmu}\\
 \hat{\cal Z}^\mu_{(s)}&\equiv  &2\frac{\di\lag_1}{\di \Bar
g_{\rho\sig,\mu\nu}}\Bar D_\nu \zeta_{\rho\sig} -
2\zeta_{\rho\sig}\Bar D_\nu
 \frac{\di\lag_1}{\di \Bar
g_{\rho\sig,\mu\nu}}\,.\label{ZDmu}
 \eea
It is just the current (\ref{++7++}) for the Lagrangian  $\lag_1$.

In the case of a vacuum background one has ${\delta \Bar \lag_{Dg}}/
{\delta \Bar {\hat g}^{\rho\sig}} = 0$ and $\lag_1 =0$. Then
assuming arbitrary Killing vectors $\xi^\alf=\lam^\alf$ one
transforms the identity (\ref{nabla-m2}) into
 \be \Bar D_\mu \l(\frac{\delta
\lag_1}{\delta \Bar g_{\mu\nu}}\r) \equiv \Bar D_\mu \l(\frac{\delta
}{\delta \Bar g_{\mu\nu}}\hat l^{\alf\beta}_{(a)} \frac{\delta \Bar
\lag_{Dg}}{\delta \Bar {\hat g}^{\alf\beta}}\r) \equiv 0\, .
 \m{nabla-m1}
 \ee
Recalling that for the vacuum case $\hat \Phi^{Lq}_{\mu\nu}=0$ and
taking into account the identity (\ref{nabla-m1}) with the
definition (\ref{GLPhiL-q}) in the equations (\ref{PERT-munu}) one
gets $\Bar D_\mu \hat t_q^{\mu\nu}= 0$. This generalizes the results
in \cite{DT1, DT2} for the quadratic theories. Recall also that in
the Lagrangian (\ref{(a2.25)}) $\Lambda$-term was interpreted as
``degenerated'' matter, then with this assumption the identity
(\ref{DGLl=0}) has been approached. However, the Lagrangian
(\ref{(a2.25)}) fully is, of course, a kind of metric Lagrangians,
therefore the identity (\ref{DGLl=0}) also is interpreted in the
terms of the general formula (\ref{nabla-m1}).

Again, because  (\ref{nabla-m2}) is the identity the current $\hat
\imath^\mu_1$ has to be presented through a divergence of a
superpotential. To construct the last we use the results of the
subsection \ref{NIdentities} adopted for the Lagrangian $\lag_1$. We
set $\Bar g_{\mu\nu} \goto g_{\mu\nu}$ and construct the
coefficients (\ref{(+4+)}) - (\ref{(+5+)}) with $\lag_1 =
\lag_1(Q;\, \di_\mu Q;\, \di_{\mu\nu} Q)\equiv \lag_1^{c}(Q;\, \Bar
D_\mu Q;\, \Bar D_{\mu\nu} Q)$, where $Q^A = \{\hat l_a^{\mu\nu},\,
g_{\mu\nu} \}$. Then we go back, $ g_{\mu\nu} \goto \Bar
g_{\mu\nu}$, and obtain simple expressions
 \be
\hat m_{1\sig}{}^{\mu\nu}  =  2 \Bar D_\lam \l(\frac{\di \lag_1}{\di
\Bar g_{\rho\nu,\mu\lam}}\r)\Bar g_{\rho\sig}\, ,\qquad \hat
n_{1\sig}^{\lam\mu\nu}  =  - 2 \frac{\di \lag_1}{\di \Bar
g_{\rho(\mu,\nu)\lam}}\Bar g_{\rho\sig}\, .
 \m{mnL1}
 \ee
Then, substituting these into the expression (\ref{(+11+)}) we can
define the superpotential $\hat {\cal I}_{(s)}^{\mu\nu}$, which is
evidently linear in $\hat l_{(a)}^{\mu\nu}$. In the case of the
Einstein gravity it is the superpotential (\ref{B54}). Thus, we have
the identity
 \be
 \hat \imath^\mu_1 \equiv \di_\nu \hat{\cal I}_{(s)}^{\mu\nu}\, ,
 \m{i1=I1}
 \ee
which generalizes the identity (\ref{B50}) and the superpotential
has the form:
  \be
\hat {\cal I}_{(s)}^{\alf\beta}  \equiv \l({\textstyle{2\over 3}}
 \BD_\lam  \hat n^{[\alf\beta]\lam}_{1\sig}  - \hat
 m^{[\alf\beta]}_{1\sig}\r)\xi^\sig   -
 {\textstyle{4\over 3}} \hat n^{[\alf\beta]\lam}_{1\sig}
 \BD_\lam  \xi^\sig.
 \m{(+16+A)}
\ee

Using in (\ref{Jmu}) the field equations in the form
(\ref{PERT-munu}) and the definition (\ref{GLPhiL-q}) we write out
the physically real current: $\hat \imath_1^\mu \goto \hat {\cal
I}_{(s)}^\mu$:
 \be
\hat {\cal I}_{(s)}^\mu  \equiv  \hat {\cal T}^\mu_{(s)\nu}\xi^\nu +
\hat {\cal Z}^\mu_{(s)}\,
 \m{current}
 \ee
where the generalized energy-momentum is
 \be
 \hat {\cal T}^\mu_{(s)\nu} \equiv
\l(\hat t_{q\nu}^{\mu} - \hat \Phi_{Lq\nu}^{\mu} \r)  - \l(
\delta^\mu_\nu\hat l^{\rho\sig}_{(a)} \frac{\delta \Bar
\lag_{Dg}}{\delta \Bar {\hat g}^{\rho\sig}} + \l.\hat
l^{\rho\sig}_{(a)}\r|^\mu_\nu \frac{\delta \Bar \lag_{Dg}}{\delta
\Bar {\hat g}^{\rho\sig}} \r)\, .
 \m{TauD}
 \ee
Thus finally the identity (\ref{i1=I1}) transforms to the
conservation law:
 \be
 \hat {\cal I}^\mu_{(s)} = \di_\nu \hat {\cal I}_{(s)}^{\mu\nu}\,.
 \m{nablacurrent}
 \ee

The expressions (\ref{current}) and (\ref{TauD}) are based on the
symmetrical energy-momentum (\ref{em-q}) and generalize the
correspondent expressions in GR: (\ref{B32}) and (\ref{B34}),
therefore we choose the subscript ``$(s)$''. Thus, this subsection,
generalizing the results of the subsection \ref{NewCLinFieldGR}
develop the idea to use the identity (\ref{B21'}) in GR. The
expression (\ref{nablacurrent}), on the one hand, generalizes the
conservation law (\ref{B51}) in GR, on the other hand, it
generalizes the Deser-Tekin expressions \cite{DT2,DT1,DT3}. In the
case of a vacuum background and using the Killing vectors $\xi^\alf
= \lam^\alf$ one has for the current
  \be
\hat {\cal I}_{(s)}^\mu = \hat t_{q\nu}^{\mu}\lam^\nu\, ,
 \m{current+1}
 \ee
that also is in correspondence with the general definition
(\ref{Spsi-current}) in a  field theory.

\subsection{Canonical N{\oe}ther and Belinfante symmetrized currents and\\ superpotentials}
\m{NBCS}

The expressions presented in subsection \ref{NIdentities} are
maximally adopted to construct both Noether canonical conserved
quantities in the framework of the bimetric formulation and
Belinfante corrected quantities. To construct such quantities one
has to consider a pure metric part $\lag_{Dg}$ of the Lagrangian
(\ref{lag-g}). At the beginning we construct the N{\oe}ther
canonical quantities. We construct the corresponding to $\lag_{Dg}$
coefficients (\ref{(+3+)}) - (\ref{(+5+)}), $\hat u^{\alf}_{g\sig}$,
$\hat m^{\alf\beta}_{g\sig}$, $\hat n^{\alf\beta\gamma}_{g\sig}$,
with the use of which we present the related identity
(\ref{(+10+)}):
 \be
 \hat \imath^{\alf}_g \equiv  \di_\beta \hat \imath^{\alf\beta}_g\, .
 \m{jg=dJg}
 \ee
Next, following to the KBL ideology (\ref{(1.27)}) we construct a
metric Lagrangian for the perturbed system:
 \be
 \lag_{DG} = \lag_{Dg} - \Bar\lag_{Dg} + \di_\alf \hat d^\alf\, .
 \m{ArbitraryLagPert}
 \ee
Then, we apply the barred procedure to the identity (\ref{jg=dJg})
and take into account the divergence keeping in mind (\ref{du}) -
(\ref{(+12+)}). As a result we obtain the identity corresponding to
(\ref{ArbitraryLagPert}):
 \be
 \hat \imath^\alf_g - \Bar{\hat \imath^\alf_g} +\delta_d \hat \imath^\alf_g
 \equiv \Bar D_\beta \l[\hat
\imath_g^{\alf\beta}(\xi) - \Bar{\hat \imath^{\alf\beta}_g}(\xi)+
\delta_d{\hat \imath}^{\alf\beta}_g(\xi)\r]\, . \m{(1.48+)}
 \ee

After substituting the dynamical equations (\ref{Q-Eq}) and the
background equations (\ref{Q-Back}) in the form
 \bea
\frac{\delta \lag_{Dg}}{\delta g^a} &=& - \frac{\delta
\lag_{Dm}}{\delta
g^a}\, , \m{ddd}\\
\frac{\delta \Bar\lag_{Dg}}{\delta \Bar g^a} &=& - \frac{\delta \Bar
\lag_{Dm}}{\delta \Bar g^a}
 \m{bbb}
 \eea
into $\hat u^{\alf}_{g\sig}$ and $\Bar{\hat u}^{\alf}_{g\sig}$ one
obtains $\hat U^{\alf}_{g\sig}$ and $\Bar{\hat U}^{\alf}_{g\sig}$,
respectively. Then the identity  (\ref{(1.48+)}) transforms into a
real conservation law:
  \be
\hat {\cal I}^\alf_{(c)}(\xi) = \di_\beta \hat {\cal
I}^{\alf\beta}_{(c)} (\xi)\, .
 \m{(1.29Q)}
  \ee
The left hand side (current), in correspondence with (\ref{(+7+)}),
is
 \be
\hat {\cal I}^\alf_{(c)}(\xi) \equiv \hat {\cal
T}^\alf_{\sig(c)}\xi^\sig + \hat {\cal
S}^{\alf\beta\rho}_{(c)}\di_{[\beta}\xi_{\rho]} + \hat {\cal
Z}^\alf_{(c)}(\xi)
  \m{(+7+A)}
 \ee
where the generalized canonical energy-momentum, spin and $Z$-term
are
 \bea
\hat {\cal T}^\alf_{\sig(c)} &\equiv &-  \l[\l(\delta\hat
U^{\alf}_{g\sig} + 2\BD_\beta(\delta^{[\alf}_\sig\hat d^{\beta]})\r)
+ \delta\hat n^{\alf\beta\gamma}_{g\lam} \Bar
R^\lam_{~\beta\gamma\sig}\r]\m{7A'}\\
\hat {\cal S}^{\alf\rho\beta}_{(c)} &\equiv & \delta\hat
 m^{\alf\beta}_{g\sig}+2\delta^{[\alf}_\sig\hat d^{\beta]}
 \bar g^{\sig\rho}\m{7A''}\\
\hat {\cal Z}^\alf_{(c)}(\xi) &\equiv &
  -\l(\delta \hat z^{\alf}_g + 2\delta^{[\alf}_\sig\hat d^{\beta]}
\zeta^\sig_\beta\r) \, .
 \m{(7A''')}
 \eea
A more detailed expression for the superpotential is
 \bea
 \hat {\cal I}^{\alf\beta}_{(c)}(\xi)& \equiv & \hat
\imath_g^{\alf\beta}(\xi) - \Bar{\hat \imath^{\alf\beta}_g}(\xi) +
\delta_d \hat
\imath_g^{\alf\beta}(\xi)\nonumber\\
&=&
 \l({\textstyle{2\over 3}}
 \BD_\lam  \delta\hat n^{[\alf\beta]\lam}_{\sig}  - \delta\hat
 m^{[\alf\beta]}_{\sig}\r)\xi^\sig   -
 {\textstyle{4\over 3}} \delta\hat n^{[\alf\beta]\lam}_{g\sig}
 \BD_\lam  \xi^\sig  -
2\xi^{[\alf}\hat d^{\beta]}\, .
 \m{IsupD}
 \eea
Here the perturbed expressions are used:
 \bea
 \delta\hat U^{\alf}_{g\sig} &=& \hat U^{\alf}_{g\sig} -
 \Bar{\hat U}^{\alf}_{g\sig}\, ,\m{difU}\\
 \delta\hat m^{\alf\beta}_{g\sig} & =& \hat m^{\alf\beta}_{g\sig}-
 \Bar{\hat m}^{\alf\beta}_{g\sig}\, ,\m{difm}\\
 \delta\hat
 n^{\lam\alf\beta}_{g\sig} &=& \hat
 n^{\lam\alf\beta}_{g\sig}- \Bar{\hat
 n}^{\lam\alf\beta}_{g\sig}\, .
 \m{difn}
 \eea
The perturbation $\delta \hat z^{\alf}_g$ is defined by the the
definition (\ref{(+8+)}) and by the perturbations (\ref{difm}) and
(\ref{difn}). Of course, if a displacement vector is a Killing
vector in the background spacetime then $\hat {\cal Z}^\alf_{(c)}$
disappears. In the case of GR the conservation law (\ref{(1.29Q)})
goes to the KBL conservation law (\ref{(1.29+)}), if $\hat d^\alf$
is defined as
 $\hat k^\alf$ in (\ref{k-KBL}).

The presented here procedure gives well defined currents and
superpotentials in the following sense. Without changing the
identity (\ref{(+6+)}) one can add to the current an arbitrary
quantity $\Delta \hat \imath^\alf(\xi)$ satisfying $ [\Delta \hat
\imath^\alf(\xi)]_{,\alf} \equiv 0 $. Analogously, without changing
  $\hat \imath^\alf$ in (\ref{(+10+)}) the superpotential can be
added by $\Delta \hat \imath^{\alf\beta}(\xi)$ with the property
$[\Delta \hat \imath^{\alf\beta}(\xi)]_{,\beta} \equiv 0 $. However,
the ``broken'' current and superpotential can be ``restored'' by the
same way because the quantities  $\Delta \hat \imath^\alf(\xi)$ and
$\Delta \hat \imath^{\alf\beta}(\xi)$ are not connected {\it at all}
with the procedure applied to the given Lagrangian in a non-explicit
form. Whereas, in the sense of the  procedure, the current and the
superpotential in Eq. (\ref{(+10+)}) are given by the coefficients
(\ref{(+3+)}) - (\ref{(+5+)}) {\it uniquely} defined by the
Lagrangian.  This claim develops also the criteria by Szabados
\cite{Szabados91} who suggested to consider a connection of
pseudotensors with Lagrangians ``as a selection rule to choose from
the mathematically possible pseudotensors''. Thus one can assert
that the current and superpotential in (\ref{(1.29Q)}) are defined
by the unique way in the above sense. Of course, the same claim is
related to the KBL quantities in (\ref{(1.29+)}). They {\em
uniquely} are defined by the Lagrangian (\ref{(1.27)}) in the sense
of the N{\oe}ther procedure.

To construct the Belinfante corrected conserved quantities for the
perturbed system (\ref{ArbitraryLagPert}) we again turn to
subsection \ref{NIdentities}. For the coefficients $\hat
u^{\alf}_{g\sig}$, $\hat m^{\alf\beta}_{g\sig}$ and  $\hat
n^{\alf\beta\gamma}_{g\sig}$ we construct the identity
(\ref{(+14+)}) and subtract the same barred identity
  \be
 \hat \imath^\alf_{gB} - \Bar{\hat \imath^\alf_{gB}}
 \equiv \Bar D_\beta \l[\hat
\imath_{gB}^{\alf\beta}(\xi) - \Bar{\hat
\imath^{\alf\beta}_{gB}}(\xi)\r]\, . \m{identity-gB}
 \ee
After substitution the equations (\ref{ddd}) and (\ref{bbb}) into
the left hand side of (\ref{identity-gB}) one gets the conservation
law:
  \be
 \hat {\cal I}^\alf_{(B)}(\xi) = \di_\beta \hat {\cal I}_{(B)}^{\alf\beta} (\xi)\, .
 \m{IB=supIB}
  \ee
The current is
 \be
\hat {\cal I}^\alf_{(B)} \equiv \hat {\cal
T}^\alf_{\sig(B)}\xi^\sig+ \hat {\cal Z}^\alf_{(B)}(\xi) \, .
 \m{(+15+A)}
 \ee
where the Belinfante corrected energy-momentum and $Z$-term are
 \bea
\hat {\cal T}^\alf_{\sig(B)} &\equiv & - \delta\hat U^{\alf}_{g\sig}
-\delta\hat n^{\alf\beta\gamma}_{g\lam} \Bar
R^\lam_{~\beta\gamma\sig} + \BD_{\beta}\delta \hat
s^{\alf\beta}_g{}_{\sig}\, ,
\m{7A+}\\
\hat {\cal Z}^\alf_{(B)}(\xi) &\equiv & \delta\hat
z^{\alf}_{gB}(\xi)\, .
 \m{(7A++)}
 \eea
The perturbations $\delta \hat z^{\alf}_{gB}$ and $\delta \hat
s^{\alf\beta}_g{}_{\sig}$ in (\ref{(+15+A)}) are defined by the
definitions (\ref{Zmu-2}) and  (\ref{(+13+)}) and  by the
perturbations (\ref{difm}) and (\ref{difn}). A detailed expression
for the superpotential is
  \bea
 \hat {\cal I}_{(B)}^{\alf\beta}(\xi) &\equiv & \hat
\imath_{gB}^{\alf\beta}(\xi) - \Bar{\hat
\imath^{\alf\beta}_{gB}}(\xi) \nonumber \\&=&
 2
\l({\textstyle{1\over 3}}\BD_\rho \delta\hat
n^{[\alf\beta]\rho}_{g\sig}+ \BD_\tau \delta\hat
 n^{\tau\rho[\alf}_{g\lam} \bar g^{\beta]\lam} \bar g_{\rho\sig}\r)
\xi^\sig - {\textstyle{4\over 3}} \delta\hat
 n^{[\alf\beta]\lam}_{g\sig}\BD_{\lam}\xi^\sig\, .
 \m{(supIB+}
 \eea
In the case of GR the conservation law (\ref{IB=supIB}) goes to the
Belinfante corrected conservation law (\ref{(2.24)}). We conclude
also that  in the sense of the united N{\oe}ther-Belinfante
procedure the current  and the superpotential in (\ref{IB=supIB})
are defined in unique way by the Lagrangian
(\ref{ArbitraryLagPert}).

Now, rewrite the conservation laws (\ref{nablacurrent}),
(\ref{(1.29Q)}) and (\ref{IB=supIB}) in the united form:
  \be
 \hat {\cal I}^\alf_D(\xi) = \di_\beta \hat {\cal I}^{\alf\beta}_D (\xi)\, .
 \m{generalCLs}
  \ee
This allows us to construct the conserved charges in generalized
form  in $D$-dimensions:
 \be
{\cal P}(\xi) = \int_\Sigma d^{D-1} x\,\hat {\cal I}^0_D(\xi) =
\oint_{\di\Sigma} dS_i \,\hat {\cal I}^{0i}_D(\xi)\,
 \m{charges}
 \ee
where $\Sigma$ is a spatial $(D-1)$ hypersurface $x^0 = \const$ and
$\di\Sigma$ is its  $(D-2)$ dimensional boundary.

\subsection{The Einstein-Gauss-Bonnet gravity. \\
The mass of the Schwarzschild-anti-de Sitter black hole}
 \m{BHinEGB}

To illustrate the above theoretical results we apply them to the
Einstein-Gauss-Bonnet (EGB) gravity. The action of the Einstein
$D$-dimensional theory with a bare cosmological term $\Lambda_0$
corrected by the Gauss-Bonnet term (see, for example, \cite{DT2}) is
 \bea
&{}& S  =  \int d^D x\l\{\lag_{EGB} + {\cal
L}_{Dm}\r\}\nonumber\\&=&
 \int d^D x\l\{\!\!-\frac{\sqrt{-g}}{2} \l[{\k}^{-1}\!\!\l(R - 2\Lambda_0\r) +
 \gamma\l(R^2_{\mu\nu\rho\sig}\! - 4 R^2_{\mu\nu} + R^2\r)\r] + {\cal L}_{Dm}\r\}
 \,
 \m{EGBaction}
 \eea
where $\k = 2\Omega_{D-2}G_D> 0$  and $\gamma >0$; $G_D$ is the
$D$-dimension Newton's constant. The equations of motion that follow
from (\ref{EGBaction}) are
 \be
 \hat {\cal E}^{\mu\nu} = \hat {T}^{\mu\nu}\,  .
 \m{EGBequations}
 \ee
The metric part is defined as
 \bea
\hat {\cal E}^{\mu\nu} \equiv
  \frac{\delta}{\delta
g_{\mu\nu}}\lag_{EGB}
  &{\equiv}& \frac{\sqrt{-g}}{2}\l\{\frac{1}{\k}\l(R^{\mu\nu} - \half g^{\mu\nu}
 R + g^{\mu\nu}\Lambda_0\r)\r.\nonumber\\ &+& \l.2\gamma\l[RR^{\mu\nu} -
 2 R^{\mu}{}_{\sig}{}^\nu{}_{\rho} R^{\sig\rho} +
 R^{\mu}{}_{\sig\rho\tau}R^{\sig\nu\rho\tau} - 2 R^{\mu}{}_{\sig}
 R^{\sig\nu}  \r. \r.\nonumber\\ &  - & \l.\l.{\textstyle \frac{1}{4}} g^{\mu\nu}
 \l(R^2_{\tau\lam\rho\sig} - 4 R^2_{\rho\sig} + R^2\r)\r] \r\}\,,
 \m{EGBequationsE}
 \eea
In the vacuum case,  $T^{\mu\nu}=0$, and the equations are
 \be
 \hat {\cal E}^{\mu\nu} = 0\,  .
 \m{EGBequationsV}
 \ee

In the present subsection, as a background we consider the AdS
solution, which is a solution to the equations (\ref{EGBequationsV})
and is described by the metric:
 \be
 d\Bar s^2 = -(1+\Bar f)dt^2 + (1+\Bar f)^{-1}dr^2 +
 r^2\sum_{a,b}^{D-2}q_{ab}dx^adx^b\, .
 \m{AdS}
 \ee
The last term describes $(D-2)$-dimensional sphere of  the radius
$r$, and $q_{ab}$ depends on coordinates on the sphere only; for the
other components $\Bar g_{00} = -(1+\Bar f)$ and $\Bar g_{11} =
(1+\Bar f)^{-1}$ where
 \be
\Bar f(r) = - r^2\frac{2\Lambda_{eff}}{(D-1)(D-2)}\, .
 \m{fBar}
 \ee
The background Christoffel symbols corresponding (\ref{AdS}) are
 \bea
 \Bar \Gamma^1_{00} &=& \half (1+\Bar f) \,{\Bar f}' ,\qquad
 \Bar \Gamma^0_{10} = \frac{{\Bar f}'}{2(1+{\Bar f})} ,\qquad
 \Bar \Gamma^1_{11} = -\frac{{\Bar f}'}{2(1+{\Bar f})} ,\qquad
 \nonumber\\
 \Bar \Gamma^a_{1b}& =& \frac{1}{r}\, \delta^a_b ,\qquad
 \Bar \Gamma^1_{ab} = -r\,(1+\Bar f) q_{ab}\,.
 \m{ChristAdS}
 \eea

The effective cosmological constant (see \cite{DT2}):
 \be
 \Lambda_{eff} = \frac{\Lambda_{EGB}}{2} \l(1\pm \sqrt{1 -
 \frac{4\Lambda_{0}}{\Lambda_{EGB}}}\r)\,
 \label{Leff+}
 \ee
is the solution of the equation $
 \Lambda^2_{eff} - \Lambda_{eff}\Lambda_{EGB} +
 \Lambda_{EGB}\Lambda_{0} = 0$, where
 \be
 \Lambda_{EGB} = -\frac
{(D-2)(D-1)}{{2\k\gamma}(D-4)(D-3)}\,
 \m{LambdaEGB}
 \ee
is defined {\em only} by the Gauss-Bonnet term. Thus $\Lambda_{eff}$
is negative, and the background Riemannian, Ricci tensors and
curvature scalar are
  \be \Bar R_{\mu\alf\nu\beta} = 2\Lambda_{eff}\frac{(\Bar
g_{\mu\nu}\Bar g_{\alf\beta} - \Bar g_{\mu\beta}\Bar
g_{\nu\alf})}{(D-2)(D-1)},\qquad \Bar R_{\mu\nu} =
2\Lambda_{eff}\frac{\Bar g_{\mu\nu}}{D-2},\qquad \Bar R =
2\Lambda_{eff}\frac{D}{D-2}\, . \label{R}
 \ee
Then equations (\ref{EGBequationsV}) are transformed into
 \be
 \Bar R_{\mu\nu} - \half\Bar g_{\mu\nu}\Bar R + \Lambda_{eff} \Bar
 g_{\mu\nu} = 0\,.
 \m{AdSequations}
\ee

The Schwarzschild-AdS (S-AdS) solution \cite{BD+} can be also
considered as a solution to (\ref{EGBequationsV}):
  \be
 d s^2 = -(1+ f)dt^2 + (1+ f)^{-1}dr^2 +
 r^2\sum_{a,b}^{D-2}q_{ab}dx^adx^b\,
 \m{S-AdS}
 \ee
where
 \be
 f(r) =  \frac{r^2}{2\k\gamma (D-3)(D-4)}
\l\{1 \pm \sqrt{1 - \frac{4\Lambda_{0}}{\Lambda_{EGB}} + 4\k\gamma
(D-3)(D-4)\frac{r_0^{D-3}}{r^{D-1}}} \r\}\, .
 \label{39bis}
 \ee
For the metrics (\ref{AdS}) and (\ref{S-AdS}) the relation  $-g =
-\Bar g = r^{D-2}\det q_{ab}$ has a place and is important for
future calculations. The Riemannian, Ricci tensors and curvature
scalar corresponding to (\ref{S-AdS}) are
 \bea
R_{0101}&=& \half f''\,, \qquad R_{0a0b}= \half r(1+f)f'q_{ab}\,
,\qquad  R_{1a1b}= - \frac{rf'}{2(1+f)}q_{ab}\, ,\nonumber\\
R_{abcd}&=& - r^2f (q_{ac}q_{bd}-q_{ad}q_{bc})\, ;
 \m{SAdS-RRR}\\
R_{00} &=& \frac{1+f}{2}\l(f'' + f'\frac{D-2}{r}\r)\, ,\qquad R_{11}
= - \frac{1}{2(1+f)}\l(f'' + f'\frac{D-2}{r}\r)\, ,\nonumber\\
R_{ab} &=& -\l[f(D-3)+ rf'\r]q_{ab}\, ; \m{SAdS-RR}\\
R &=& -\l(f'' + f'\frac{D-2}{r}\r)-\frac{D-2}{r^2}\l[f(D-3)+
rf'\r]\, .
  \m{SAdS-R}
 \eea

Of course, the barred solution (\ref{S-AdS}) goes to (\ref{AdS}),
and the barred expressions (\ref{SAdS-RRR}) - (\ref{SAdS-R}) go to
(\ref{R}). It is evidently, in the case of the solutions (\ref{AdS})
and (\ref{S-AdS}) perturbations can be described only by $\Delta f =
f-\Bar f$. In linear approximation it is
 \be
 \Delta f = \pm \l(\sqrt{1 - \frac{4\Lambda_{0}}{\Lambda_{EGB}}}\r)^{-1}
\l(\frac{r_0}{r}\r)^{D-3}\, .
 \m{Deltaf}
 \ee

\subsubsection{The field-theoretical prescription}
\m{FTP}

Here, we turn to the results of subsection \ref{CSinFF}. To
concretize and to have a possibility to compare with \cite{DT2}, we
define the gravitational perturbations from the set $h^a$, like in
(\ref{B38}), as $h_{\alf\beta} = g_{\alf\beta} - \Bar
g_{\alf\beta}$. Thus $\lag_1 = \lag_1^{EGB} = h_{\alf\beta} ({\delta
\Bar \lag_{EGB}}/{\delta \Bar g_{\alf\beta}}) = h_{\alf\beta}\Bar
{\hat{\cal E}}^{\alf\beta}$, where $\Bar {\hat{\cal E}}^{\alf\beta}$
is the barred expression (\ref{EGBequationsE}). Then for the AdS
background (\ref{AdS}) the equations (\ref{EGBequations}) can be
rewritten in the form of the equations (\ref{PERT-munu}):
 \be
 { G}_{Lq}^{\mu\nu} \equiv \l[{1}_{{(E)}} -{\l(1
\pm \sqrt{1 - \frac{4\Lambda_{0}}{\Lambda_{EGB}}}\r)}_{{(GB)}}\r]
{G}_L^{\mu\nu}
 \equiv
\mp \sqrt{1 -
 \frac{4\Lambda_{0}}{\Lambda_{EGB}}}~~ {G}_L^{\mu\nu} = \k
 t^{\mu\nu}_q\,
 \m{GL=T}
 \ee
where subscripts ${}_{(E)}$ and ${}_{(GB)}$ are related to the
Einstein and the Gauss-Bonnet part in (\ref{EGBaction}), which are
with the coefficients ``$\k$'' and ``$\gamma$'', respectively. The
left hand side in (\ref{GL=T}) is calculated using (\ref{GLPhiL-q})
and
 \bea
2{ G}_L^{\mu\nu} \equiv & - &  \Bar D_\sig\Bar D^\sig h^{\mu\nu} -
\Bar D^{\mu}\Bar D^{\nu} h^\sig_\sig + \Bar D_\sig\Bar D^\nu
h^{\sig\mu} + \Bar D_\sig
 \Bar D^\mu h^{\sig\nu} - \frac{4\Lambda_{eff}}{D-2}h^{\mu\nu}
 \nonumber\\
  &-&  \Bar g^{\mu\nu}\l(- \Bar D_\sig\Bar D^\sig h^\rho_\rho + \Bar D_{\sig}\Bar D_{\rho}
 h^{\sig\rho} - \frac{2\Lambda_{eff}}{D-2}h^\sig_\sig\r)\,.
 \m{GL}
 \eea
The right hand side of (\ref{GL=T}) generalizes the energy-momentum
in (\ref{(a2.27)}). It could be presented as
 \be
  t^q_{\mu\nu} \equiv - \frac{2}{\sqrt{-\Bar g}}\l[\hat {\cal E}_{\mu\nu}(\Bar
 g_{\alf\beta}+ h_{\alf\beta}) - \hat T_{\mu\nu}(\Bar
 g_{\alf\beta}+ h_{\alf\beta})\r] + \k^{-1}{ G}^{Lq}_{\mu\nu}\,
 \m{tq-munu}
 \ee
where indexes for $\hat {\cal E}_{\mu\nu}$ and $ \hat T_{\mu\nu}$
were lowered by $g_{\mu\nu}$, and they are thought as depending on
the sum $\Bar g_{\alf\beta}+ h_{\alf\beta}$.

Constructing the charges for the EGB system one has to use the
generalized expression (\ref{charges}). The current could be used.
However, looking at (\ref{current+1}), (\ref{EGBequationsE}) and
(\ref{tq-munu}) one can see that it is very complicated. Evidently
that the superpotential expression (\ref{(+16+A)}) is significantly
simpler. We set $\lag_1=\lag^{EGB}_1$ in (\ref{mnL1}) and obtain
 \bea
 &{}&\hat m_{1\sig}{}^{\mu\nu}  = \mp \frac{\sqrt{-\Bar g}}{\k}
 \sqrt{1 -
 \frac{4\Lambda_{0}}{\Lambda_{EGB}}}
\Bar g_{\sig\rho} \Bar D_\lam H^{\mu(\nu\rho)\lam},\nonumber\\
 &{}&\hat n_{1\sig}^{\rho\mu\nu}  =  \mp \frac{\sqrt{-\Bar g}}{\k} \sqrt{1 -
 \frac{4\Lambda_{0}}{\Lambda_{EGB}}}\Bar
g_{\sig(\lam}\delta^{(\mu}_{\pi)}H^{\nu)\pi\rho\lam};\nonumber\\
 H^{\mu\nu\rho\lam}\!\! & \equiv &
\!\! h^{\mu\rho} \Bar g^{\nu\lam} + h^{\lam\nu} \Bar g^{\rho\mu} -
h^{\mu\lam} \Bar g^{\rho\nu} - h^{\rho\nu} \Bar g^{\mu\lam}
+h^\sig_\sig\l( \Bar g^{\mu\lam} \Bar g^{\rho\nu} - \Bar g^{\mu\rho}
\Bar g^{\nu\lam}\r). \m{mnL2}
 \eea
The substitution of the quantities (\ref{mnL2}) into the expression
(\ref{(+16+A)})  gives the superpotential  related to the general
EGB case:
 \bea
&{}&\hat {\cal I}_{(s)}^{\mu\rho} \equiv \pm \frac{\sqrt{-\Bar
g}}{\k} \sqrt{1 -
 \frac{4\Lambda_{0}}{\Lambda_{EGB}}}\times\nonumber\\
 &{}&\times\l(\xi^{[\mu} \Bar D_{\nu}h^{\rho]\nu}  - \xi_\nu
\Bar D^{[\mu}h^{\rho]\nu} - \xi^{[\mu} \Bar D^{\rho]}h -
h^{\nu[\mu}\Bar D^{\rho]}\xi_\nu - \half h \Bar D^{[\mu}
\xi^{\rho]}\r)\, .
 \m{DTsuperpotential}
 \eea
It is expressed through the Abbott-Deser superpotential in the
Einstein theory \cite{AbbottDeser82,DT2}, $\hat I_{AD}^{\mu\rho}$,
as $\hat {\cal I}_{(s)}^{\mu\rho} = \mp \sqrt{1 -
{4\Lambda_{0}}/{\Lambda_{EGB}}}~ \hat I_{AD}^{\mu\rho}$. Recently,
developing the results \cite{DT2} Deser with co-authors
\cite{Desercom} have reached the same expression.  Paddila
\cite{Paddila} reffering to \cite{DT2} also analyzed it. Keeping in
mind that different definitions for the metric perturbations $h^a$
can be used we find that in fact the superpotential
(\ref{DTsuperpotential}) belongs to the set
 \be
 \hat {\cal I}_{(s)}^{\mu\rho} \equiv {1
\over \k}
 \l[{1}_{{(E)}} -{\l(1
\pm \sqrt{1 - \frac{4\Lambda_{0}}{\Lambda_{EGB}}}\r)}_{{(GB)}}\r]
\l(\hat l^{\sig[\mu}_{(a)}\Bar D_\sig\xi^{\rho]}+ \xi^{[\mu}\Bar
D_\sig \hat l^{\rho]\sig}_{(a)}-\bar D^{[\mu}\hat l^{\rho
]}_{(a)\sig} \xi^\sig \r)
 \m{DTsuperpotential+}
 \ee
where the Einstein part defined in $D$ dimensions formally coincides
with (\ref{B54}) in GR.

For the S-AdS solution (\ref{S-AdS}) considered as perturbations
with respect to the AdS spacetime (\ref{AdS}) one has in linear
approximation:
 \be
h_{00} = h^{11} = - \Delta f \approx \mp \l(\sqrt{1 -
\frac{4\Lambda_{0}}{\Lambda_{EGB}}}\r)^{-1}
\l(\frac{r_0}{r}\r)^{D-3}\,.
 \label{h}
 \ee
To calculate the total conserved energy of the S-AdS solution we use
the formula (\ref{charges}) where we substitute $(01)$-component of
the superpotential (\ref{DTsuperpotential}). The last is calculated
with (\ref{h}) and the time-like Killing vector $\lam^\mu = (-1,\,
{\bf 0})$; covariant derivatives are defined by (\ref{ChristAdS}).
Finally (\ref{charges}) gives:
 \be
 E = \frac{(D-2)r_0^{D-3}}{4G_D}\,
 \label{E}
 \ee
that is the result of \cite{DT2}, and is the standard  accepted
result obtained with using the various approaches (see
\cite{Paddila} - \cite{Okuyama} and references therein).

\subsubsection{The canonical N{\oe}ther  mass}
\m{CNChs}

The results of this subsubsection, in fact, repeat the results by
Deruelle, Katz and Ogushi \cite{DerKatzOgushi}. However, there is a
difference between the approach in \cite{DerKatzOgushi} and our
approach in \cite{PK2003a}. In \cite{DerKatzOgushi} conserved
quantities are constructed in the way, when the charges for the
physics system and for the background system are constructed
separately. Only after that  they are compared  and differences are
interpreted as charges for perturbations. On the other hand, in
\cite{PK2003a} the conserved charges from the starting are
constructed in the perturbed form. This approach is in the spirit of
the present paper and is presented explicitly here in significantly
more detail than in \cite{PK2003a}.

To construct global conserved quantities in the above prescription
we again turn to the generalized integral (\ref{charges}), only this
time with the superpotential (\ref{IsupD}). Then it is necessary to
calculate the perturbations (\ref{difm}) and (\ref{difn}). For this
one has to use the general formulae (\ref{(+4+)}) and (\ref{(+5+)})
with the metric Lagrangian of the EGB gravity in (\ref{EGBaction}).
Also the expressions in $D$-dimensions:
 \bea
\Del^\alpha_{\mu\nu} & = &\Gamma^\alpha_{\mu\nu} - \Bar
{\Gamma}^\alpha_{\mu\nu} = \half g^{\alf\rho}\l( \BD_\mu g_{\rho\nu}
+ \BD_\nu g_{\rho\mu} - \BD_\rho g_{\mu\nu}\r)\, , \m{DeltaDefD}\\
 \BD_\rho g_{\mu\nu} & = & g_{\tau\mu}\Delta^\tau_{\rho\nu} +
 g_{\tau\nu}\Delta^\tau_{\rho\mu}\, ,
 \m{BDgDelta}\\
  R^\lam{}_{\tau\rho\sig} & =&
\BD_\rho \Delta^\lam_{\tau\sig} -  \BD_\sig\Delta^\lam_{\tau\rho} +
 \Delta^\lam_{\rho\eta} \Delta^\eta_{\tau\sig} -
 \Delta^\lam_{\eta\sig} \Delta^\eta_{\tau\rho}
 + \Bar R^\lam{}_{\tau\rho\sig}\,
 \m{(17-DmD)}
 \eea
are useful. For calculation of the superpotential (\ref{IsupD}) we
need in antisymmetric part of (\ref{difm}) only. Thus after
prolonged calculations we have
 \bea
 &{}&{\hat m}^{[\alf\beta]}_{g\sig} = {}_{(E)}{\hat m}^{[\alf\beta]}_{\sig}
 +{}_{(GB)}{\hat m}^{[\alf\beta]}_{\sig}
 \nonumber\\ &=&
\frac{\sqrt{- g}}{2\k}\l\{2 \Delta^{[\alf}_{\sig\rho}g^{\beta]\rho}
+ \delta^{[\alf}_{\sig}\Delta^{\beta]}_{\rho\tau} g^{\rho\tau}  \r\}
 \nonumber\\ &+ & \gamma\sqrt{-g}
 \l\{2R_{\sig}{}^{\rho\tau[\alf}\Delta^{\beta]}_{\rho\tau}  -
  6 R_{\tau}{}^{[\alf\beta]\rho}\Delta^{\tau}_{\sig\rho}+
 6 R^{\rho[\alf\beta]}{}_\sig\Delta^{\tau}_{\tau\rho} -
 4 R^{\rho[\alf}\Delta^{\beta]}_{\sig\rho} +
 2 R^{\rho}_{\sig}\Delta^{[\alf}_{\rho\tau}g^{\beta]\tau}\r.
 \nonumber\\&+& \l.
 4 R^{\rho\tau}\delta^{[\alf}_{\sig}\Delta^{\beta]}_{\rho\tau} -
 6 R^{\rho[\alf}g^{\beta]\tau}\Delta^{\pi}_{\rho\tau}g_{\sig\pi} -
 12 R^{[\alf}_{\rho}g^{\beta]\tau}\Delta^{\rho}_{\sig\tau} -
 8 R^{[\alf}_{\sig}g^{\beta]\rho}\Delta^{\tau}_{\tau\rho} +
  2 R \Delta^{[\alf}_{\sig\rho}g^{\beta]\rho}
  \r.
 \nonumber\\&+& \l.
 R \Delta^{[\alf}_{\rho\tau}\delta^{\beta]}_{\sig}g^{\rho\tau} +
 3 R \Delta^{\tau}_{\tau\rho}\delta^{[\alf}_{\sig}g^{\beta]\rho} -
 2 g_{\sig\rho}g^{\tau[\alf}\BD_\tau R^{\beta]\rho} -
 \delta^{[\alf}_\sig g^{\beta]\rho}\BD_\rho R\r\}\, .
 \m{mEGB}
 \eea
The expression for (\ref{(+5+)}) is more simple:
 \bea
 &{}&{\hat n}^{\lam\alf\beta}_{g\sig} ={}_{(E)}{\hat
 n}^{\lam\alf\beta}_{\sig} + {}_{(GB)}{\hat
 n}^{\lam\alf\beta}_{\sig}=
 \nonumber\\ &=&\!\!\!
\frac{\sqrt{- g}}{2\k}\l\{g^{\alf\beta}\delta^\lam_\sig
-g^{\lam(\alf}\delta^{\beta)}_\sig \r\}
 \nonumber\\ &+&\!\!\! \gamma\sqrt{-g}
 \l\{-2R_\sig{}^{(\alf\beta)\lam} - 4 R_\sig^{\lam}g^{\alf\beta} +
 4 R_\sig^{(\alf}g^{\beta)\lam} + R\l(g^{\alf\beta}\delta^\lam_\sig
-g^{\lam(\alf}\delta^{\beta)}_\sig\r)\r\}
 \m{nEGB}
 \eea
For calculation of the superpotential (\ref{IsupD}) we need in the
next part of (\ref{nEGB})\footnote{In \cite{PK2003a} an analogous to
(\ref{nEGB}) expression has a missprint.}:
  \bea
 {\hat n}^{[\alf\beta]\lam}_{g\sig} &= &{}_{(E)}{\hat
 n}^{[\alf\beta]\lam}_{\sig} + {}_{(GB)}{\hat
 n}^{[\alf\beta]\lam}_{\sig}
 \nonumber\\ &=& \frac{3\sqrt{- g}}{4\k}\delta^{[\alf}_\sig g^{\beta]\lam}
 + \frac{3\gamma\sqrt{-g}}{2}\l\{R_\sig{}^{\lam\alf\beta} +
 4 g^{\lam[\alf}R^{\beta]}_{\sig} +
 \delta_\sig^{[\alf}g^{\beta]\lam}R
 \r\}.
 \m{nEGBantis1}
 \eea
Perturbations of (\ref{mEGB}) and (\ref{nEGBantis1}) are obtained
following the recommendation given in (\ref{difm}) and (\ref{difn}):
 \bea
\delta\, {\hat m}^{[\alf\beta]}_{g\sig} &= & \delta[{}_{(E)}{\hat
m}^{[\alf\beta]}_{\sig}]
 +\delta[{}_{(GB)}{\hat m}^{[\alf\beta]}_{\sig}]\, , \m{d-mEGB}\\
\delta\, {\hat n}^{[\alf\beta]\lam}_{g\sig} &= &\delta[{}_{(E)}{\hat
 n}^{[\alf\beta]\lam}_{\sig}] + \delta[{}_{(GB)}{\hat
 n}^{[\alf\beta]\lam}_{\sig}]\, .
 \m{d-nEGBantis1}
 \eea
We do not present explicit expressions for these perturbations
because they are evident due to (\ref{mEGB}) and (\ref{nEGBantis1}).
We note that only the two last terms in (\ref{mEGB}) contribute into
$\Bar {\hat m}^{[\alf\beta]}_{g\sig}$ because $\Bar {\hat
\Delta^\gamma}_{\alf\beta}\equiv 0$.

Recall that in the canonical prescription one has to define a
divergence in the Lagrangian (\ref{ArbitraryLagPert}).  We exactly
follow the recommendation in \cite{DerKatzOgushi}. In our notations
it is the divergence of
 $\hat d^\lam = {\hat n}^{\lam\alf\beta}_{g\sig}\Delta^\sig_{\alf\beta}$.
In the GR case this leads to the choice of KBL, that is to $\hat
k^\alf$ in (\ref{KBLLagrangian}) that leads to variation of the
action under Dirichlet boundary conditions. A more detail discussion
on a choice of a divirgence one can found in \cite{DerKatzOgushi}.
Thus, we define
 \bea
&{}&\hat d^\lam = \l({}_{(E)}{\hat n}^{\lam\alf\beta}_{\sig}+
{}_{(GB)}{\hat n}^{\lam\alf\beta}_{\sig}\r)\Delta^\sig_{\alf\beta} =
\frac{\sqrt{-g}}{2\k}\l(\Delta^\lam_{\alf\beta}g^{\alf\beta}-
\Delta^\alf_{\alf\beta}g^{\lam\beta} \r) \nonumber\\ &+&
 \gamma\sqrt{-g} \l(
-2R_\sig{}^{\alf\beta\lam} + 4R^\alf_\sig g^{\beta\lam} -
4R^\lam_\sig g^{\alf\beta} + \delta^\lam_\sig g^{\alf\beta} -
\delta^\alf_\sig g^{\beta\lam} \r)\Delta^\sig_{\alf\beta}\, .
 \m{divd}
 \eea
Substitution of (\ref{d-mEGB}), (\ref{d-nEGBantis1}) and
(\ref{divd}) into (\ref{IsupD}) gives the canonical N{\oe}ther
superpotential $\hat {\cal I}^{\alf\beta}_{(c)}= \hat {\cal
I}^{\alf\beta}_{(c)E} + \hat {\cal I}^{\alf\beta}_{(c)GB}$ for
perturbations in EGB gravity. Its Einstein part formally coincides
with the KBL superpotential (\ref{(1.31)}):
 \be
 \hat {\cal
I}^{\alf\beta}_{(c)E} = {1\over \k}\big({\hat g^{\rho[\alf}\Bar
D_{\rho} \xi^{\beta]}} + \hat
g^{\rho[\alf}\Delta^{\beta]}_{\rho\sig}\xi^\sig- \Bar {D^{[\alf}\hat
\xi^{\beta]}\big)}- 2\xi^{[\alf} \hat d^{\beta]}_E\, . \m{(1.31+A)}
 \ee
 only one has to keep in mind that it
is presented in $D$-dimensions \cite{DerKatzOgushi}.

To calculate the mass of the S-AdS black hole we use the formulae
(\ref{AdS}) - (\ref{Deltaf}) for background and perturbed systems.
We use the component $\hat {\cal I}^{01}_{(c)}$ in the general
formula (\ref{charges}) under the requirement $r\goto \infty$. For
this it is enough to calculate $\hat {\cal I}^{01}_{(c)}$ in linear
approximation with respect to the perturbation $\Delta f$ in
(\ref{Deltaf}). Next, for calculating the mass we again need in the
Killing vector $\lam^\alf = \{-1,\,\bf 0\}$. In linear approximation
the symbols (\ref{DeltaDefD}) are
  \bea
 \Delta^1_{00} = -\frac{1}{2r}{\Delta f \Bar f} (D-5)\, ,~~
 \Delta^0_{10} = -\frac{1}{2r}\frac{\Delta f }{\Bar f} (D-1) \, ,\nonumber\\~~
 \Delta^1_{11} = \frac{1}{2r}\frac{\Delta f }{\Bar f} (D-1)\, ,~~
 \Delta^1_{ab} = -r\,\Delta f q_{ab}\,.
 \m{DChristAdS}
 \eea
In calculations the simple relations $(\Delta f)' = -(D-3)\Delta
f/r$,  $~~(\Bar f)' = 2\Bar f/r$ are used. Then in linear
approximation the Einstein part (\ref{(1.31+A)})  gives
 \be
 \hat {\cal I}^{01}_{(c)E} = -\frac{\sqrt{-\Bar g}}{2\k r}\Delta f (D-2)\, .
 \m{KBL-D-sup}
 \ee
The contribution from the Gauss-Bonnet part of (\ref{IsupD}) we
write out in part. Thus in linear approximation
 \be
 -\delta[{}_{(GB)}{\hat
m}^{[01]}_{\sig}]\lam^\sig = \frac{\gamma \sqrt{-\Bar
g}}{2r^3}\Delta f \Bar f(D-2)(D-3)\l[3 +2(D-5)\r]\, .
  \m{d-mEGB01}
  \ee
The contribution from (\ref{d-nEGBantis1}) is
 \be
 {\textstyle{2\over 3}}\lam^\sig
 \BD_\lam  \delta[{}_{(GB)}\hat n^{[01]\lam}_{\sig}]    -
 {\textstyle{4\over 3}} \delta[{}_{(GB)}\hat n^{[01]\lam}_{\sig}]
 \BD_\lam \lam^\sig = -\frac{3\gamma \sqrt{-\Bar g}}{2r^3}\Delta
f \Bar f(D-2)(D-3)\, .
 \m{d-nEGBantis1-01}
 \ee
At last, the contribution from the Gauss-Bonnet part of (\ref{divd})
is
 \be
 -2\lam^{[0}\hat d^{1]}_{GB} = \frac{\gamma \sqrt{-\Bar g}}{r^3}\Delta
f \Bar f(D-2)(D-3)\, .
 \m{divd01}
 \ee
Summing (\ref{d-mEGB01}) - (\ref{divd01}) we obtain the Gauss-Bonnet
part of the superpotential (\ref{IsupD}) in linear approximation
 \be
 \hat {\cal I}^{01}_{(c)GB} = \frac{\sqrt{-\Bar g}}{2\k r}\Delta f
 (D-2)\l[1\pm \sqrt{1 -
 \frac{4\Lambda_{0}}{\Lambda_{EGB}}} \r]\, .
 \m{GBsup01}
 \ee
The definitions (\ref{fBar}) - (\ref{LambdaEGB}) were used. Summing
(\ref{KBL-D-sup}) and (\ref{GBsup01}) we obtain the full
$01$-component of the EGB canonical superpotential necessary for
calculating the mass of the S-AdS black hole:
 \be
 \hat {\cal I}^{01}_{(c)} = \frac{\sqrt{-\Bar g}}{2\k r} \l(\frac{r_0}{r}
 \r)^{D-3}(D-2).
 \m{EGBsup01}
 \ee
The definition (\ref{Deltaf}) was used. Substitution of
(\ref{EGBsup01}) into (\ref{charges}) gives again the standard
result (\ref{E}).

\subsubsection{The Belinfante corrected mass}
\m{BelMass}

Comparing the Belinfante corrected superpotential (\ref{(supIB+})
and the canonical one (\ref{IsupD}) we find that for constructing
the first we need (together with (\ref{nEGBantis1})) in the
expression $\hat n^{\tau\rho[\alf}_{g\lam}\Bar g^{\beta]\lam}\Bar
g_{\rho\sig}$ instead of (\ref{mEGB}) and (\ref{divd}). Thus in EGB
gravity
 \bea
&{}&\hat n^{\tau\rho[\alf}_{g\lam}\Bar g^{\beta]\lam}\Bar
g_{\rho\sig}=\nonumber\\&{}&
\frac{\sqrt{-g}}{4\k}\l(2g^{\rho[\alf}\Bar g^{\beta]\tau}\Bar
g_{\rho\sig} -g^{\tau[\alf}\delta^{\beta]}_{\sig}\r) +
\frac{\gamma\sqrt{-g}}{2}\l\{2R_\lam{}^{\rho\tau[\alf}
-2R^{\rho\tau}{}_\lam{}^{[\alf} - 8R^\tau_\lam g^{\rho[\alf}\r.
 \nonumber\\&+& \l.4
R^\rho_\lam g^{\tau[\alf} +4 g^{\rho\tau} R^{[\alf}_\lam + R\l(
2\delta^\tau_\lam g^{\rho[\alf}- \delta^\rho_\lam
g^{\tau[\alf}\r)\r\}\Bar g^{\beta]\lam}\Bar g_{\rho\sig}\, .
 \m{nEGBantis2}
 \eea
As is seen, it significantly simplify the expression (\ref{(supIB+})
with respect to (\ref{IsupD}). Perturbation of (\ref{nEGBantis2}) is
again obtained following the recommendation given in (\ref{difn}):
 \be
 \delta[\hat n^{\tau\rho[\alf}_{g\lam}\Bar g^{\beta]\lam}\Bar
g_{\rho\sig}] = \delta[{}_{(E)}\hat n^{\tau\rho[\alf}_{\lam}\Bar
g^{\beta]\lam}\Bar g_{\rho\sig}] + \delta[{}_{(GB)}\hat
n^{\tau\rho[\alf}_{\lam}\Bar g^{\beta]\lam}\Bar g_{\rho\sig}]\, .
 \m{d-nEGBantis2}
 \ee
Combining (\ref{d-nEGBantis1}) and (\ref{d-nEGBantis2}) in the
definition (\ref{(supIB+}) we obtain the Belinfante corrected
superpotential $\hat {\cal I}^{\alf\beta}_{(B)}= \hat {\cal
I}^{\alf\beta}_{(B)E} + \hat {\cal I}^{\alf\beta}_{(B)GB}$ in EBG
gravity. Its Einstein part
 \be
 \hat {\cal I}^{\alf\beta}_{(B)E} = {1
\over \k} \l(\hat l^{\rho[\alf}\Bar D_\rho\xi^{\beta]}+
\xi^{[\alf}\Bar D_\rho \hat l^{\beta]\rho}-\bar D^{[\alf}\hat
l^{\beta ]}_{\rho} \xi^\rho\r)
 \m{BEinstein+}
 \ee
formally coincides with the superpotential (\ref{alaAbbottDeser}),
only one has to keep in mind that it is presented in $D$-dimensions.

To calculate the mass of the S-AdS black hole we use again the
formulae (\ref{AdS}) - (\ref{Deltaf}) for background and perturbed
systems. We use the component $\hat {\cal I}^{01}_{(B)}$ in the
general formula (\ref{charges}) under the requirement $r\goto
\infty$ and for $\lam^\alf = \{-1,\,\bf 0\}$. We calculate a linear
approximation of $\hat {\cal I}^{01}_{(B)}$ in $\Delta f$. The
$(01)$-component of the Einstein part (\ref{BEinstein+}) gives the
same result as in (\ref{KBL-D-sup}). The contribution from the
Gauss-Bonnet part of (\ref{(supIB+}) we again write out in part.
Thus in linear approximation
 \be
2 \lam^\sig \BD_\tau \delta[{}_{(GB)}\hat n^{\tau\rho[0}_{\lam}\Bar
g^{1]\lam}\Bar g_{\rho\sig}] = \frac{\gamma \sqrt{-\Bar
g}}{2r^3}\Delta f \Bar f(D-2)(D-3)\l[3 +2(D-4)\r]\, .
  \m{d-nEGBantis2-01}
  \ee
Summing (\ref{d-nEGBantis1-01}) and (\ref{d-nEGBantis2-01}) we
obtain the Gauss-Bonnet part of the $(01)$-component of the
superpotential (\ref{(supIB+}) in linear approximation that is
exactly (\ref{GBsup01}). Finally,  thus, adding the Einstein part we
again approach the standard result (\ref{E}).

\subsection{Discussion}
\m{Discussion}

The results of subsections \ref{PerturbedDTheory} and \ref{CSinFF},
first, generalize the corresponding results in GR presented in
sections \ref{decompositions} and \ref{CL}, second, they generalize
the Deser and Tekin approach \cite{DT2} in quadratic theories that
is particulary presented in the framework of EGB gravity in
subsection \ref{FTP}. Thus really we answer the criticism in
\cite{Paddila} related to the Deser and Tekin approach. Indeed, in
subsections \ref{PerturbedDTheory} and \ref{CSinFF} we have extended
the approach in \cite{DT2} to perfectly arbitrary backgrounds, which
can be without symmetries at all, and arbitrary displacement vectors
(no necessity in Killing vectors) can be used.

The results of subsection \ref{NBCS} generalize the corresponding
results in GR presented in section \ref{CL}. Also, our canonical
N{\oe}ther expressions generalize the Deruelle, Katz and Ogushi
results \cite{DerKatzOgushi} in EGB gravity, which in fact are
presented in subsection \ref{CNChs}, but only in a preferred here
perturbed form. The Belinfante corrected expressions both in a
general form in subsection \ref{NBCS} and in application to EGB
gravity in subsection \ref{BelMass} are quite new.

A choice for more suitable and successful expressions for conserved
quantities is not so simple  problem even in the framework of GR.
Therefore, only we discuss general proposals for the generalized
superpotentials (\ref{(+16+A)}), (\ref{IsupD}) and (\ref{(supIB+})
from a general viewpoint. However, a relation to GR, and to EGB
gravity (subsection \ref{BHinEGB}), of course, could be useful.

Let us turn to the superpotentials of the family (\ref{(+16+A)})
with the substitution of (\ref{mnL1}). They differ from each other
starting from the second order for different $\hat
l^{\mu\nu}_{(a)}$, like the superpotentials (\ref{B54}) in GR. The
choice $h^a \goto\hat l^{\mu\nu} = \hat g^{\mu\nu} - \Bar{ \hat
g}^{\mu\nu}$ in GR was based on the two points (see the end of
subsection \ref{BelOnArbitrary}). First, only for this choice the
superpotential from (\ref{B54}) gives the standard energy-momentum
and its flux at null infinity  for the Bondi-Sachs solution. Second,
only the superpotential related to $\hat l^{\mu\nu}$ from the family
(\ref{B54}) coincides with the Belinfante correction of the KBL
superpotential. We do not know a test model for a choice of
perturbations $h^a$ in the family (\ref{(+16+A)}) in EBG gravity.
For example, an arbitrary choice $h^a$ for the solution
(\ref{S-AdS}) leads to the same result (\ref{E}) if we use a
correspondent superpotential from the family
(\ref{DTsuperpotential+}) instead of (\ref{DTsuperpotential}).
Indeed, for this calculations only the linear order is crucial, but
it is the same for all of them. Considering the Belinfante corrected
expression (\ref{(supIB+}) and the family (\ref{(+16+A)}) we see
that, in general, they cannot be identified. Indeed, the
superpotentials (\ref{(+16+A)}) are linear in $\hat
l^{\mu\nu}_{(a)}$ for an arbitrary metric theory, whereas
(\ref{(supIB+}) does not. However many modified $D$-dimensional
gravities, as a rule, are the Einstein theories with corrections
(quadratic and others, like string and M-theory corrections). Among
such theories  the Lanczos-Lovelock theory (in particulary, EGB
gravity) plays an important role. Then, comparing the Einstein parts
{\em only} (in the EGB gravity they are (\ref{BEinstein+}) and the
one in (\ref{DTsuperpotential+})) we again can prefer $\hat
l^{\mu\nu}$ from the set $\hat l^{\mu\nu}_{(a)}$. Of course, it is
not an absolute choice, but could be a recommendation.

Comparing (\ref{(+16+A)}), (\ref{IsupD}) and (\ref{(supIB+}) we
remark that the first is in the framework of the field-theoretical
approach,  where perturbations are examined explicitly. At the same
time, the expressions (\ref{IsupD}) and (\ref{(supIB+})  are
presented rather as bimetric ones. Moreover, the expression
(\ref{(+16+A)}), being a linear one, is significantly simpler. Next,
the superpotential (\ref{IsupD}), as a canonical quantity,
essentially depends on a choice of a divergence in the Lagrangian.
But for an arbitrary metric theory,  as we know, there is no a
crucial principle for the definition of such a divergence; there are
only reasonable recommendations (see \cite{DerKatzOgushi} and
subsection \ref{BHinEGB}). On the other hand, expressions
(\ref{(+16+A)}) and (\ref{(supIB+}) do not depend on divergences at
all, that could be an advantage. Comparing (\ref{IsupD}) and
(\ref{(supIB+}) we can also note that the Belinfante corrected
expressions are simpler significantly. This is demonstrated clearly
by the corresponding expression in EGB gravity and by the
calculation of mass of the S-AdS black hole in subsection
\ref{BelMass}.

The applications in subsection \ref{BHinEGB} related to calculation
of mass of the S-AdS black hole in EGB gravity are, in fact, the
test of the presented here approach. Recalling the papers
\cite{Paddila} - \cite{Okuyama} where the S-AdS mass was calculated
we note that all of them are in agreement giving the acceptable
result, although they use different approaches. In the most of the
papers the definitions are based on the first law of the black hole
thermodynamics and the connection with the surface terms of
Hamiltonian dynamics. The surface terms are defined following the
recommendation by Regge and Teitelboim \cite{ReggeTeitelboim}. On
the other hand, considering the AdS background as an arena for
perturbations one can connect conserved charges with AdS symmetries
expressed by Killing vectors and define them in the quite classical
form. Such approach  was used in \cite{DerKatzOgushi} and in just
the present paper. In \cite{AFrancavigliaR} the charges are
associated to the diffeomorphism symmetries of Lovelock gravities in
any odd dimensions. In \cite{Okuyama} asymptotic symmetries of AdS
spacetime were used with employing the conformal completion
technique.

Sometimes it is important to interpret some special situations with
the use of developing approaches. In the works
\cite{Desercom,Paddila,Cai,DerMor05,Der05} such a case, when
 \be
 \Lambda_{EGB}= 4\Lambda_{0}\, ,
 \m{LEGB=4L0}
 \ee
is remarked. Indeed, this case looks as degenerated one.  The reason
is that the condition (\ref{LEGB=4L0}) leads to a zero coefficient
at the linear approximation of the EGB gravity equations around the
AdS background. Applying the generalized KBL approach in EGB gravity
\cite{DerKatzOgushi} Deruelle and Morisava \cite{DerMor05,Der05}
have  found that mass and angular momentum expressions for the
Kerr-AdS solution in EGB gravity have the same coefficient and have
to be defined only as zero. In the framework of the
field-theoretical formalism we approach the analogous conclusion.
Turn to the equations (\ref{GL=T}). Under the condition
(\ref{LEGB=4L0}) their linear left hand side disappears: the
Einstein part is canceled by the Gauss-Bonnet part. Consequently the
total energy-momentum for perturbations at the right hand side
disappears also. This means that the matter energy-momentum is
compensated by the energy-momentum of the metric perturbations. In
the vacuum case the energy-momentum of the metric perturbations is
equal to zero. Thus the conserved integrals have to be equal to
zero, like the charges in \cite{DerMor05,Der05}. This situation on
the AdS background is similar to the definition of zero energy of
the S-dS black hole in a so-called Nariari space  in
\cite{CNojiriO}. The last is interpreted as a real ground state
whose energy is lower than energy of the pure dS space. To escape
the difficulties with the situation (\ref{LEGB=4L0}) Paddila
\cite{Paddila} using his approach suggests to use a modified
background without all the symmetries, unlike AdS one. All three the
approaches presented in the present paper have this possibility
also.

In future we plan to check the formulae of the subsection
\ref{BHinEGB}  applying them to the Kerr-AdS solution in EGB gravity
keeping in mind, firstly,  like in \cite{DerMor05,Der05}, that
asymptotically in linear approximation it has to go to the solution
in the Einstein $D$-dimensional theory
\cite{GLPPope,GPPope,DerKatz05}, secondly, checking the just now
appeared exact solutions \cite{KEGB1,KEGB2}. Moreover, the
expressions related both to the field-theoretical and to the
canonical N{\oe}ther prescriptions already were used and discussed
for examination of higher $D$ Kerr-AdS spacetimes
\cite{Desercom,DerMor05,Der05,DerKatz05} including EGB gravity.
Planning to check the formulae in EGB gravity of subsections
\ref{FTP} and \ref{CNChs} for the Kerr-AdS solution we anticipate
that the results will be identical with the ones in
\cite{Desercom,DerMor05,Der05,DerKatz05}. Applications of the
Belinfante corrected expressions of subsection \ref{BelMass} for
calculating mass and angular momentum for the Kerr-AdS black hole in
EGB gravity have to be testable.

Gravitational theories in $D$ dimensions are very intensively
developed. New solutions with interesting properties appear
following one by one (see, for example, \cite{Aliev,Aliev2}). To
understand and describe their properties, which could be quite
unusual, many of new solutions need in calculating conserved
charges. Therefore all of known and new possibilities to define and
calculate conserved quantities, if they are non-contradictive and
satisfy all the acceptable tests, could be interesting for future
studies.

\subsection*{Acknowledgments} Author is very grateful to Chiang-Mei
Chen, Nathalie Deruelle, Stanley Deser, Jacek Jezierski, Sergei
Kopeikin, James Nester, L\'aszl\'o Szabados, Brian Pitts, Bayram
Tekin and Shwetketu Vibhadra for their useful conversations,
comments, remarks related to the problems of this review and
explanations of their works. Author expresses a special gratitude to
Leonid Grishchuk and Alla Popova, the earlier collaboration with
whom became a basis of the review; and to coauthors Joseph Katz,
Stephen Lau and Deepak Baskaran for very fruitful collaboration in
last years. Author also is grateful to Deepak Baskaran for a big
help in checking English, and to Maria Grigorian and Michael
Prokhorov for help in preparing figures.

\label{lastpage-01}


\begin{thebibliography}{999}

\bibitem{LL} Landau L D and  Lifshitz E M 1975
{\em The Classical   Theory of Fields}, (Oxford, Pergamon)

\bibitem{Fock} Fock V A 1964 {\em The Theory of Space, Time and Gravitation}
(Oxford, Pergamon)

\bibitem{[12]}
Misner Ch W, Thorne K S and Wheller J A 1973 {\em Gravitation} (San
Francisco, Freeman)

\bibitem{DeWitt-book}  DeWitt B S 1965 {\em Dynamical
Theory of Groups and Fields} (New York, Gordon and Breach)

\bibitem {Lifshitz46} Lifshitz E M 1946 On a gravitational stability of an
expanding world {\em J. of Phys., Moscow} {\bf 10} 116 [{\it Zh.
Eksp. Teor. Fiz.}  {\bf 16} 587 (1946)]

\bibitem {Mukhanov} Mukhanov V F,  Feldman H A and
Brandenberger R H 1992 Theory of cosmological perturbations {\em
Phys. Rep.} {\bf 215} 203

\bibitem{Lukash-rev} Lukash V N 2006 On the relation between tensor
and scalar perturbation modes in Friedmann cosmology {\em Phys.
 Usp.} {\bf 49} 103 [2006  {\em Usp. Fiz. Nauk} {\bf 176} 113]
{\em (Preprint astro-ph/0610312)}

\bibitem{Grishchuk74} Grishchuk L P 1974 An amplification of gravitational waves
in isotropic world {\it Zh. Eksp. Teor. Fiz.} {\bf 67} 825

\bibitem{Lukash1} Lukash V N 1980 The birth of sound waves in an
isotropic universe {\em Zh. Eksp. Teor. Fiz.} {\bf 79} 1601

\bibitem{Bardeen80}  Bardeen J M 1980 Gauge invariant cosmological
perturbations {\em Phys. Rev.} {\bf D22} 1882

\bibitem{KKS02} Khalatnikov I M, Kamenshchik A Yu
and Starobinsky A A 2002 Comment about quasi-isotropic solution of
Einstein equations near cosmological singularity  {\it Class.
Quantum Grav.} {\bf 19}, 3845 {\em (Preprint gr-qc/0204045)}

\bibitem{KKMS03}  Khalatnikov I M, Kamenshchik A Yu, Martellini M
and Starobinsky A A 2003 Quasi-isotropic solution of the Einstein
equations near a cosmological singularity for a two-fluid
cosmological model {\it JCAP},  {\bf 0303} 001 {\em (Preprint
gr-qc/0301119)}

\bibitem{Chernin1} Chernin A D 2002
Physical vacuum and cosmic coincidence problem {\em New Astron.},
{\bf 7}, 113 {\em (Preprint astro-ph/0107071)}

\bibitem{Bartolo} Bartolo N, Komatsu E,  Matarrese S and
Riotto A 2004 Non-Gaussianity from inflation: Theory and
observations {\em Phys. Rep.}, {\bf 402} 103 {\em (Preprint
astro-ph/0406398)}

\bibitem{Grishchuk01} Grishchuk L P 2001 Relic gravitational waves
and their detection {\em Lect. Notes Phys.} {\bf 562}, 167  {\em
(Preprint gr-qc/0002035)}

\bibitem{Grishchuk05} Grishchuk L P 2005 Relic gravitational waves and cosmology
{\em Phys.Usp.} {\bf 48}, 1235 [2005  {\em Usp. Fiz. Nauk} {\bf 175}
1289] {\em (Preprint gr-qc/0504018)}

\bibitem{DolgovZS-book} Dolgov A D, Sazhin M V and
Zel'divich Ya B  1990 {\em Basics of Modern Cosmology} (Editions
Fronti\'eres, France).

\bibitem{Starobinsky1} Kiefer C, Polarski D  and Starobinsky A A
2000 Entropy of gravitons produced in the early Universe {\em Phys.
Rev. D} {\bf 62} 043518 {\em (Preprint gr-qc/9910065)}

\bibitem{Starobinsky2} Starobinsky A A, Tsujikawa S and Yokoyama J 2001
Cosmological perturbations from multi-field inflation in generalized
Einstein theories {\em Nucl. Phys. B} {\bf 610} 383 {\em (Preprint
astro-ph/0107555)}

\bibitem{Starobinsky3} Kiefer C, Polarski D  and Starobinsky A A
2006 Pointer states for primordial fluctuations in inflationary
cosmology {\em (Preprint astro-ph/0610700)}

\bibitem{Lukash-lect} Lukash V N 1996 The very early Universe, in Australian:
{\em  Cosmology: The physics of the universe} eds. Robson B A et al.
(World Scientific) 213 [in Barsilian: {\em Cosmology and
gravitation, II.} ed. Novello M (Editions Frontieres) 288] {\em
(Preprint astro-ph/9910009)}

\bibitem{Linde1} Linde A D 2005 Inflation and String Cosmology
{\em J. Phys. Conf. Ser.} {\bf 24} 151 {\em (Preprint
hep-th/0503195)}

\bibitem{Linde2} Linde A D 2005 Inflation and String Cosmology
{\em Contemp. Concepts Phys.} {\bf 5} 1 {\em (Preprint
hep-th/0503203)}

\bibitem{Chernin}  Chernin A D 2001 Cosmic vacuum
 {\em Uspekhi Fiz. Nauk} {\bf 171} 1153

\bibitem{Chernin2} Chernin A D, Santiago D I and
Silbergleit A S  2002 Interplay between gravity and quintessence: A
set of new GR solutions {\em Phys. Lett. A} {\bf 294} 79 {\em
(Preprint astro-ph/0106144)}

\bibitem{Chandrasekhar} Chandrasekhar S 1983 {\em The Mathematical Theory of Black Holes}
(Oxfrod Univ. Press, New York)

\bibitem{BlackHole0} Kokkotas K D and Schmidt B G 1999
Quasi-Normal Modes of Stars and Black Holes {\it Living Rev.
Relativity} {\bf 2} 2 {\em (Preprint gr-qc/9909058)}

\bibitem{BlackHole1} Sarbach O and Tiglio M 2001
Gauge invariant perturbations of Schwarzschild black holes in
horizon-penetrating coordinates {\em Phys. Rev. D} {\bf 64} 084016
{\em (Preprint gr-qc/0104061)}

\bibitem{BlackHole2}  Glampedakis K and  Andersson N 2001
Scattering of scalar waves by rotating black holes {\em Class.
Quant. Grav.} {\bf 18} 1939 {\em (Preprint gr-qc/0102100)}

\bibitem{BlackHole3}  Karkowski J, Malec E and \'Swierczy\'nski Z
2002 Backscattering of electromagnetic and gravitational waves off
Schwarzschild geometry  {\em Class. Quantum Grav.} {\bf 19} 953
(2002) {\em (Preprint gr-qc/0105042)}

\bibitem{BlackHole4} Karlovini M 2002 Axial perturbations of
general spherically symmetric spacetimes {\em Class. Quantum Grav.}
{\bf 19} 2125  {\em (Preprint gr-qc/0111066)}

\bibitem{BlackHole5}  Karkowski J, Roszkowski K,
\'Swierczy\'nski Z and   Malec E 2003 Waves in Schwarzschild
spacetimes: How strong can imprints of the spacetime curvature be
{\em Phys. Rev. D} {\bf 67} 064024 {\em (Preprint gr-qc/0210041)}

\bibitem{BlackHole6} Frolov V P and  Lee H K 2005 Observable form
of pulses emitted from relativistic collapsing objects {\em Phys.
Rev. D} {\bf 71} 044002 {\em (Preprint gr-qc/0412124)}

\bibitem{obzor} Grishchuk L P, Lipunov V M,
Postnov K A, Prokhorov M E and Sathyaprakash B S 2001 Gravitational
wave astronomy: In anticipation of first sources to be detected {\em
Phys. Usp.} {\bf  44}, 1 [2001 {\em Usp. Fiz. Nauk} {\bf 171}, 3]
{\em (Preprint astro-ph/0008481)}

\bibitem{Alekseev1} Alekseev G A 1987 The exact solutions in general relativity
{\it Transactions of the Steklov institute} {\bf 176}, 211, in
Russian

\bibitem{Alekseev2} Alekseev G A 2005  Monodromy-data parameterization
of spaces of local solutions of integrable reductions of Einstein's
field equations  {\em Theor. Math. Phys.} {\bf 143} 720 [2005 {\em
Teor. Mat. Fiz.} {\bf 143} 278] {\em (Preprint gr-qc/0503043)}

\bibitem{Alekseev3} Alekseev G A and  Griffiths J B 2000  Infinite
hierarchies of exact solutions of the Einstein and Einstein-Maxwell
equations for interacting waves and inhomogeneous cosmologies {\em
Phys. Rev. Lett.}  {\bf 84} 5247 {\em (Preprint gr-qc/0004034)}

\bibitem{Alekseev4}  Alekseev G A and  Griffiths J B 2001 Solving the
characteristic initial value problem for colliding plane
gravitational and electromagnetic waves {\em Phys. Rev. Lett.} {\bf
87}, 221101 {\em (Preprint gr-qc/0105029)}

\bibitem{Alekseev5}  Alekseev G A and  Griffiths J B 2004 Collision of plane
gravitational and electromagnetic waves in a Minkowski background:
solution of the characteristic initial value problem {\em Class.
Quantum Grav.} {\bf 21} 5623 {\em (Preprint gr-qc/0410047)}

\bibitem{RubakovSh} Rubakov V A and Shaposhnikov M E 1983 Do we
live inside a domain wall? {\it Phys. Lett. B} {\bf 125} 136

\bibitem{Akama} Akama K 2000 An early proposal of ``Brane World''
{\em (Preprint hep-ph/0001113)}

\bibitem{RS1} Randall L and Sundrum R 1999 A large mass hierarchy
from a small extra dimension {\it Phys. Rev. Lett.} {\bf 83} 3370
{\em (Preprint hep-ph/9905221)}

\bibitem{RS2} Randall L and Sundrum R 1999 An alternative to
compactification  {\it Phys. Rev. Lett.} {\bf 83} 4690 {\em
(Preprint hep-th/9906064)}

\bibitem{Rubakov} Rubakov V A 2001 Large and infinite extra dimensions
{\it Phys. Usp.} {\bf 44} 871 [2001 {\it Usp. Fiz. Nauk} {\bf 171}
913] {\em (Preprint hep-ph/0104152)}

\bibitem{branaPert1}  Kodama H,  Ishibashi A and Seto O 2000
Brane world cosmology --- gauge-invariant formalism for perturbation
{\em  Phys. Rev. D} {\bf 62} 064022 {\em (Preprint hep-th/0004160)}

\bibitem{branaPert2}  Van de Bruck C,  Dorca M,
Brandenberger R H and  Lukas A 2000 Cosmological perturbations in
brane-world theories: Formalism {\em  Phys. Rev. D} {\bf 62} 123515
{\em (Preprint hep-th/0005032)}

\bibitem{branaPert3} Deruelle N, Dolezel T and  Katz J 2001
Perturbations of brane worlds {\em  Phys. Rev. D} {\bf 63} 083513
{\em (Preprint hep-th/0010215)}

\bibitem{branaPert4} Gorbunov D S, Rubakov V A and Sibiryakov S M
2001 Gravity waves from inflating brane or mirrors moving in
AdS${}_5$ {\em JHEP} {\bf 0110} 015 {\em (Preprint hep-th/0108017)}

\bibitem{Nojiri} Nojiri S, Odintsov S D and  Ogushi S 2002
Friedmann-Robertson-Walker brane cosmological equations from the
five-dimensional bulk (A)dS black hole {\it Int. J. Mod. Phys. A}
{\bf 17} 4809 ({\em Preprint} hep-th/0205187)

\bibitem{branaPert5} Binetruy P, Bucher M and  Carvalho C 2004
Models for the brane-bulk interaction: Toward understanding
braneworld cosmological perturbation {\em Phys. Rev. D} {\bf 70}
043509 {\em (Preprint hep-th/0403154)}


\bibitem{branaPert6} Libanov M V and  Rubakov V A 2005
Lorentz-violating brane worlds and cosmological perturbations {\em
Phys. Rev. D} {\bf 72} 123503 {\em (Preprint hep-th/0509148)}

\bibitem{Einstein15a} Einstein A 1915 On general theory of
relativity {\em Sitzungsber. preuss. Akad. Wiss.} {\bf 44} 778

\bibitem{Einstein15b} Einstein A 1915 On general theory of
relativity (Comment) {\em Sitzungsber. preuss. Akad. Wiss.} {\bf 46}
799

\bibitem{Einstein15c} Einstein A 1915 Equations of gravitational field
{\em Sitzungsber. preuss. Akad. Wiss.} {\bf 48} 844

\bibitem{Einstein16} Einstein A 1916 The Hamiltonian principle and general theory of
relativity {\em Sitzungsber. preuss. Akad. Wiss.} {\bf 2} 1111

\bibitem{Einstein16a} Einstein A 1916 Aproximate integration of
equations of gravitational field {\em Sitzungsber. preuss. Akad.
Wiss.} {\bf 1} 688

\bibitem{[1]} Einstein A 1918 On gravitational waves
{\em Sitzungsber. preuss. Acad. Wiss.} {\bf 1} 154

\bibitem{Einstein18a} Einstein A 1918 An energy conservation law in
general theory of relativity  {\it Sitzungsber. preuss. Akad. Wiss.}
{\bf 1} 448

\bibitem{Einstein18b} Einstein A 1918 A remark to the work by E
Schr\"odinger ``Components of energy of gravitational field'' {\em
Phys. Z.} {\bf 19} 115

\bibitem{Szabados04}
Szabados L B 2004 Quasi-local energy-momentum and angular momen\-tum
in GR: A review article {\it Living Rev. Relativity} {\bf 7} 4
               [Online article: cited on 7 June 2004,
               http://www.livingreviews.org/lrr-2004-4]

\bibitem{York80}  York J W 1980 Energy and momentum of the gravitational field
{\em Essays in General Relativity} ed. Tipler F J (Academic, New
York, 1980) 39

\bibitem{Ashtekar80} Ashtekar A 1980 Asymptotic structure of
the gravitational field at spatial infinity {\em General Relativity
and Gravitation} ed. Held A (Plenum, New York) {\bf 2}  37

\bibitem{AshtekarBicakSchmidt} Ashtekar A, Bicak J and  Schmidt B G
1997 Asymptotic structure of symmetry reduced general relativity
{\em Phys. Rev. D} {\bf 55} 669 {\em (Preprint gr-qc/9608042)}

\bibitem{Jacek1998a} Chru\'sciel P T,
Jezierski J and MacCallum M A H  1998 Uniqueness of the
Trautman-Bondi mass {\em Phys. Rev. D} {\bf 58} 084001 {\em
(Preprint gr-qc/9803010)}

\bibitem{DainFriedrich} Dain S and Friedrich H 2001
Asymptotically flat initial data with prescribed regularity at
infinity {\em  Commun. Math. Phys.} {\bf 222} 569 {\em (Preprint
gr-qc/0102047)}

\bibitem{BeigSchmidt} Beig R and Schmidt B G 2000 Time-independent
gravitational fields {\em Lect. Notes Phys.} {\bf 540} 325 {\em
(Preprint gr-qc/0005047)}

\bibitem{CJK-book} Chru\'sciel P T,  Jezierski J and  Kijowski J
2001  Hamiltonian field theory in the radiating regime {\em Lect.
Notes Phys.} {\bf  570} (Springer, Berlin, Heidelberg, New York)

\bibitem{AbbottDeser82}
Abbott L F  and Deser S 1982 Stability of gravity with a
cosmological constant  {\it Nucl. Phys. B} {\bf 195} 76

\bibitem{AshtekarDas} Ashtekar A and  Das S 2000 Asymptotically
anti-de Sitter space-times: Conserved quantities {\em Class. Quantum
Grav.} {\bf 17} L17 {\em (Preprint hep-th/9911230)}

\bibitem{SchoenYau}  Schoen R and  Yau S-T 1979 On the proof of
the positive mass conjecture in general relativity {\em Commun.
Math. Phys.} {\bf 65} 45

\bibitem{SchoenYau1} Schoen R and  Yau S-T 1981 Proof
of the positive mass theorem. II {\em Commun. Math. Phys.} {\bf 79}
231

\bibitem{Witten81} Witten E 1981 A new proof of the positive
energy theorem {\em Commun. Math. Phys.} {\bf 80} 381

\bibitem{Nester81}  Nester J M 1981 A new gravitational
energy expression with a simple positive proof {\em Phys. Lett. A}
{\bf 83} 241

\bibitem{Faddeev} Faddeev L D 1982 The problem of energy in the
Einstein gravitational theory {\it Uspekhi Fiz. Nauk} {\bf 136} 433

\bibitem{DT2} Deser S and Tekin B 2003 Energy in generic
higher curvature gravity theories {\em Phys. Rev. D} {\bf 67} 084009
{\em (Preprint hep-th/0212292)}

\bibitem{Desercom} Deser S, Kanik I and Tekin B 2005 Conserved
charges in higher D Kerr-AdS spacetimes {\em Class. Quantum Grav.}
{\bf 22} 3383 {\em (Preprint gr-qc/0506057)}

\bibitem{Olea1}  Kofinas G and Olea R 2006  Vacuum energy in
Einstein-Gauss-Bonnet AdS gravity  {\em Phys. Rev. D} {\bf 74}
084035 {\em (Preprint hep-th/0606253)}


\bibitem{Olea2} Miskovic O and Olea R 2006 On boundary
conditions in three-dimensional AdS gravity {\em Phys. Lett. B} {\bf
640}  101 {\em (Preprint hep-th/0603092)}

\bibitem{BY93} Brown J D and  York J W 1993 Quasilocal energy and
conserved charges derived from the gravitational action {\em Phys.
Rev. D} {\bf 47} 1407 {\em (Preprint gr-qc/9209012)}

\bibitem{Lau93}  Lau S R 1993 Canonical variables
and quasilocal energy in general relativity {\em Class. Quantum
Grav.} {\bf 10} 2379 {\em (Preprint gr-qc/9307026)}

\bibitem{Lau95} Lau S R 1995  Spinors and the
reference point of quasilocal energy {\em Class. Quantum Grav.} {\bf
12} 1063 {\em (Preprint gr-qc/9409022)}

\bibitem{Lau96} Lau S R 1996  On the canonical
reduction of spherically symmetric gravity {\em Class. Quantum
Grav.} {\bf 13} 1509  {\em (Preprint gr-qc/9508028)}

\bibitem{BLY97} Brown J D, Lau S R and York J W 1997 Energy of isolated
systems at retarded times as the null limit of quasilocal energy
{\em Phys. Rev. D} {\bf 55} 1977  {\em (Preprint gr-qc/9609057)}


\bibitem{BLY99} Brown J D, Lau S R and York J W 1999 Canonical
quasilocal energy and small spheres {\em Phys. Rev. D} {\bf 59}
064028  {\em (Preprint gr-qc/9810003)}

\bibitem{Lau99} Lau S R 1999 Lightcone reference
for total gravitational energy {\em Phys. Rev. D} {\bf 60} 104034
{\em (Preprint gr-qc/9903038)}

\bibitem{BLY02} Brown J D, Lau S R and York J W 2002 Action and energy
of the gravitational field {\em Ann. Phys.}  {\bf 297} 175 {\em
(Preprint gr-qc/0010024)}

\bibitem{ADM}
Arnowitt R, Deser S  and  Misner C W 1962 The dynamics of general
relativity  {\em Gravitation: an Introduction to Current Research}
ed. Witten L (Wiley, New York) 227

\bibitem{Tulczyiew2} Kijowski J and Tulczyiew W M 1979 A symplectic
framework for field theories {\em Lect. Notes Phys.} {\bf 107}
(Springer, Berlin)

 \bibitem{Jacek1} Jezierski J and Kijowski J 1990 The localization
 of energy in gauge field theories and in linear gravitation
 {\em Gen. Relat. Grav.} {\bf 22} 1283

 \bibitem{Kijowski2} Kijowski J 1997 A simple derivation of
 canonical structure in quasilocal Hamiltonian in general
 relativity   {\em  Gen. Relat. Grav.} {\bf 29} 307

\bibitem{Jacek2} Jezierski J 1999 Energy and angular momentum of the
weak gravitational waves on the Schwarzschild background ---
quasi-local gauge-invariant formulation {\em  Gen. Relat. Grav.}
{\bf 31} 1855 {\em (Preprint gr-qc/9801068)}

  \bibitem {Nester8}  Nester J M 1991 Covariant Hamiltonian for
  gravity theories {\em Mod. Phys. Lett. A} {\bf 6} 2655

 \bibitem {Nester9} Chen C-M,  Nester J M  and
 Tung R-S 1995 Quasilocal energy-momentum for geometric gravity theories
{\em Phys. Lett. A} {\bf 203}, 5 {\em (Preprint gr-qc/9411048 )}

 \bibitem {ChenNester} Chen C-M and Nester J M  1999 Quasilocal
 quabtities fo GR and other gravity theories
{\em Class. Quantum Grav.} {\bf 16} 1279 {\em (Preprint
gr-qc/9809020)}

\bibitem {ChangNesterChen} Chang C-C,  Nester J M and Chen C-M 1999
Pseudotensors and quasilocal energy-momentum {\em Phys. Rev. Lett.}
{\bf 83} 1897 {\em (Preprint gr-qc/9809040)}

 \bibitem {Nester10}  Chen C-M and  Nester J M 2000 A simplectic
 Hamiltonian derivation of quasilocal energy-momentum for GR
{\em  Grav. Cosmol.} {\bf 6}  257 {\em (Preprint gr-qc/0001088)}

\bibitem {Nester2004} Nester J M 2004 General pseudotensors amd
quasilocal quantities {\em Class. Quantum Grav.} {\bf 21} S261

\bibitem {ChenNesterTung2005} Chen C-M, Nester J M and Tung R-S 2005
The Hamiltonian boundary term and quasilocal energy flux {\em Phys.
Rev. D} {\bf 72} 104020 {\em (Preprint gr-qc/0508026)}

\bibitem{[11]}  Deser S 1970 Self-interaction and gauge invariance
{\em Gen. Relat. Grav.} {\bf 1} 9 {\em (Preprint gr-qc/0411023)}

\bibitem{GPP} Grishchuk L P, Petrov A N  and Popova A D 1984
Exact theory of the (Ein\-stein) gravitational field in an arbitrary
background space-time {\em Commun. Math. Phys.} {\bf 94} 379

\bibitem{GP86} Grishchuk L P and Petrov A N 1986 Closed worlds as
gravitational fields {\em Sov. Astron. Lett.} {\bf 12} 179 [1986
{\em Pis'ma Astron. Zh.} {\bf 12}, 429]

\bibitem{GP87} Grishchuk L P and Petrov A N 1987 The Hamiltonian description
of the gravitational field and gauge symmertries {\em Sov. Phys.:
JETP} {\bf 65} 5 [1986 {\em Zh. Eksp. Teor. Fiz.} {\bf 92}, 9]

\bibitem{PP87}  Petrov A N and Popova AD 1987 On exact dynamic theories acting
on a given background {\em Vestnik Mosk. Univ. Fiz. Astron.} {\bf 28
No 6} 13

\bibitem{[15]} Popova A D and   Petrov A N 1988
The  dynamic  theories  on  a fixed  background in gravitation {\em
Int. J. Mod. Phys. A} {\bf 3} 2651

\bibitem{Petrov-H2} Petrov A N 1990 New harmonic coordinates for the Schwarzschild
geometry {\em Vestnik Mosk. Univ. Fiz. Astron.} {\bf 31 No 5} 88


\bibitem{Petrov-H3} Petrov A N 1992 New harmonic coordinates for the
Schwarzschild geometry and the field approach {\em  Astronom.
Astrophys. Trans.} {\bf  1} 195


\bibitem{[16]} Petrov A N 1993 General  relativity  from  `localization'
of Killing vector fields {\em Class. Quantum Grav.} {\bf 10} 2663

\bibitem{PP1} Popova A D and  Petrov A N 1993
Nonlinear  quantum  mechanics with nonclassical  gravitational
self-interaction.  II.  Nonstationary situation {\em Intern. J. Mod.
Phys. A} {\bf  8} 2683

\bibitem{PP1a} Popova A D and Petrov A N 1993 Nonlinear  quantum
mechanics with nonclassical  gravitational  self-interaction.  III.
Related topics  {\em Intern. J. Mod. Phys. A} {\bf 8} 2709

\bibitem{PP2} Petrov A N and Popova A D 1994
Associated length  and  inflation in quantum mechanics with
gravitational self-interaction {\em Intern. J. Mod. Phys. D} {\bf 3}
461

\bibitem{PP3} Petrov A N and Popova A D 1994
The associated length and inflation in quantum mechanics with
gravitational coupling {\em Gen. Relat. Grav.} {\bf 26} 1153


\bibitem{Petrov95} Petrov A N 1995
Asymptotically flat spacetimes at spatial infinity: The field
approach and the Lagrangian description {\em Int. J. Mod. Phys. D}
{\bf 4} 451

\bibitem{Petrov97} Petrov A N 1997
Asymptotically flat spacetimes at spatial infinity: II. Gauge
invariance of  the integrals of motion in the field approach {\em
Int. J. Mod. Phys. D} {\bf 6} 239

\bibitem{PetrovNarlikar1}  Petrov A N and Narlikar J V 1996
The energy distribution for a spherically symmetric isolated system
in general relativity {\em Found. Phys.} {\bf{26}} 1201; Erratum
1998 {\em Found. Phys.} {\bf{28}} 1023

\bibitem{PK-lett} Petrov A N and  Katz J 1999
Conservation laws for large perturbations on curved backgrounds {\em
Fundamental Interactions: From Symmetries to Black Holes} eds. Frere
J M et al. (Universite de Bruxelles, Belgium) 147  {\em (Preprint
gr-qc/9905088)}

\bibitem{PK} Petrov A N and  Katz J 2002
Conserved currents, superpotentials and cosmological perturbations
{\it Proc. R. Soc. A, London} {\bf 458} 319  {\em (Preprint
gr-qc/9911025)}

\bibitem{BLP03} Baskaran D, Lau S R and Petrov A N 2003
Center of mass integral in canonical general relativity {\em Ann.
Phys.} {\bf 307} 90  {\em (Preprint gr-qc/0301069)}

\bibitem{PK2003b} Petrov A N 2004 Perturbations in the Einstein
theory of gravity: Conserved currents {\em Mosc. Univ. Phys. Bull.}
{\bf 59 No 1} 24 [2004 {\em Vestnik Mosk. Univ. Fiz. Astron.} {\bf
No 1} 18]  {\em (Preprint gr-qc/0402090)}

\bibitem{PK2003a} Petrov A N 2004 Conserved currents in
$D$-dimensional gravity and brane cosmology {\em Mosc. Univ. Phys.
Bull.} {\bf 59 No 2} 11 [2004 {\em Vestnik Mosk. Univ. Fiz. Astron.}
{\bf No 2} 10]  {\em (Preprint gr-qc/0401085)}

\bibitem{Petrov2005a} Petrov A N 2005 The Schwarzschild black
hole as a point particle {\em Found. Phys. Lett.} {\bf 18} 477 {\em
(Preprint gr-qc/0503082)}

\bibitem{Petrov2005b} Petrov A N 2005 A note on the Deser-Tekin charges
{\em Class. Quantum Grav.} {\bf 22} L83  {\em (Preprint
gr-qc/0504058)}

\bibitem{BabakGrishchuk} Babak S V and Grichshuk L P 2000
The energy-momentum tensor for the gravitational field {\em  Phys.
Rev. D} {\bf 61} 24038 {\em (Preprint gr-qc/9907027)}

\bibitem{BabakGrishchuk1} Babak S V and Grichshuk L P 2003
Finite-range gravity and its role in gravitational waves, black
holes and cosmology {\em Int. J. Mod. Phys. D} {\bf 12} 1905 {\em
(Preprint gr-qc/0209006)}

\bibitem{Petrov2004} Petrov A N 2004 The field theoretical
formulation of general relativity and gravity with non-zero masses
of gravitons {\em Searches for a mechanism of gravity} eds. Ivanov M
A and Savrov L A (Nizhny Novgorod, Nickolaev Publisher) 230 [a
compressed version: 2004 {\em Proceedings to PIRT-IX, London}  ed.
Duffy M {\bf 2} 433] {\em (Preprint gr-qc/0505058)}


\bibitem{PintoNetoSilva} Pinto-Neto N and Silva R R 2000
Generalized field theoretical approach to general relativity and
conserved quantities in anti-de Sitter spacetimes   {\em Phys. Rev.
D} {\bf 61} 104002 {\em (Preprint gr-qc/0005101)}

\bibitem{KRMS} Kopeikin S, Ramirez J, Mashhoon B and  Sazhin M 2001
Cosmological perturbations: A new gauge-invariant approach {\em
Phys. Lett. A} {\bf 292} 173 {\em (Preprint gr-qc/0106064)}

\bibitem{RK} Ramirez J and Kopeikin S 2002
A decoupled system of hyperbolic equations for linearized
cosmological perturbations {\em Phys. Lett. B} {\bf 532}, 1  {\em
(Preprint gr-qc/0110071)}

\bibitem{PittsSchive2001a}  Pitts J B and Schieve W C 2001
Slightly Bimetric Gravitation  {\em Gen. Relat. Grav.} {\bf 33} 1319
{\em (Preprint gr-qc/0101058)}

\bibitem{PittsSchive2001b} Pitts J B and Schieve W C 2001
Null cones in Lorentz-covariant general relativity {\em (Preprint
gr-qc/0111004)}

\bibitem{Pitts2004}  Pitts J B and Schieve W C 2004
Null cones and Einstein's equations in Minkowski spacetime {\em
Found. Phys.} {\bf 34} 211
 {\em (Preprint gr-qc/0406102)}

\bibitem{Pitts2003}  Pitts J B and Schieve W C 2003
Nonsingularity of flat Robertson-Walker models in the special
relativistic approach to Einstein's equations {\em Found. Phys.}
{\bf 33} 1315 {\em (Preprint gr-qc/0406103)}

\bibitem{ZG} Zeldovich Ya B and Grishchuk L P  1986 Gravity, general
relativity and altrernative theories {\em Sov. Phys. Usp.} {\bf 29}
780  [1986 {\em Usp. Fiz. Nauk} {\bf 149} 695]

\bibitem{G} Grishchuk L P 1990 The general theory of relativity:
Familiar and unfamiliar {\em Sov. Phys. Usp.} {\bf 33} 669 [1990
{\em Usp. Fiz. Nauk} {\bf 160}, 147]

\bibitem{Grishchuk92} Grishchuck L P   1992 Gravity-wave astronomy:
Some mathematical aspects {\em Carrent Topics in Astrofundamental
Physics} eds. Sanches N et al. (World Scientific) 435

\bibitem{Barnebey}
Barnebey T A 1974 Gravitational waves: the nonlinearized theory {\em
Phys. Rev D} {\bf 10} 1741

\bibitem{[2]} Rosen N 1940 General relativity and flat space
{\em Phys. Rev.} {\bf 57} 147

\bibitem{Katz85}  Katz J 1985 A note on Komar's anomalous factor
{\it Class. Quantum Grav.} {\bf 2} 423

\bibitem{Eisenkh}
Eisenkhart L P 1933 {\em Continuous Groups of Transformations}
(Princeton Univ. Press, Princeton)

\bibitem{Schouten} Schouten J A 1951 {\em Tensor Analysis for Physicists}
(Claredon Press, Oxford)

\bibitem{Mitzk} Mitzkevich N V 1969  {\em Physical Fields in
General Theory of Relativity} (Nauka, Moscow), in Russian.

\bibitem{Bruni} Bruni M, Matarrese S, Mollerqash S and
Sonego S 1997 Perturbations of spacetime: Gauge transformations and
gauge invariance at second order and beyond {\em Class. Quantum
Grav.} {\bf 14} 2585 {\em (Preprint gr-qc/9609040)}

\bibitem{Abramo} Abramo L R W, Branderberg R H and
Mukhanov V F 1997 Energy-momentum tensor for cosmological
perturbations {\em Phys. Rev. D} {\bf 56} 3248 {\em (Preprint
gr-qc/9704037)}

\bibitem{Petrov}  Petrov A Z 1969  {\em Einstein Spaces}
(Pergamon Press, London)

\bibitem{Deser87} Deser S 1987 Gravity from self-interaction
in a curved background {\em Class. Quantum Grav.} {\bf 4} 99

\bibitem{B-Deser} Boulware D C and   Deser S 1975
Classical general relativity derived from quantum gravity {\em Ann.
Phys.} {\bf 89} 193


\bibitem {Trautman62} Trautman A 1962 Conservation laws in general relativity
{\em Gravitation: an Introduction to Current Research}, ed. Witten L
(Wiley, New York)


\bibitem {Tolman-book}
Tolman R C 1934 {\em Relativity, Thermodynamics and Cosmology}
(Oxford, Oxford Univ. Press)


\bibitem {Freud39} Von Freud Ph 1939
\"Uber die ausdr\"ucke der gesamtenergie und des gesamtimpulses eins
materiellen systems in der allgemeinen relativit\"atstheorie {\em
Ann. of Math.} {\bf 40} 417

\bibitem {Bergman49} Bergmann P G 1949
Non-linear field theories {\em Phys. Rev.} {\bf 75} 680

\bibitem {Goldberg58} Goldberg J N 1958
Conservation laws in general relativity {\em Phys. Rev.} {\bf 111}
315

\bibitem {Moller58} M{\o}ller C  1958 On the localization
of the energy of a physical system in the general theory of
relativity {\it Ann. Phys.} {\bf 4} 347

\bibitem {Papapetrou48} Papapetrou A 1948 Einstein's theory of
gravitation and flat space {\em Proc. R. Irish Ac.} {\bf 52} 11

\bibitem{Weinberg-book} Weinberg S 1972
{\em Gravitation and Cosmology} (Wiley, New York)

\bibitem{Komar59} Komar A 1959
Covariant conservation laws in general relativity {\em Phys. Rev.}
{\bf 113} 934

\bibitem{Garecki01}  Garecki J 2001
Remarks on the Bergmann-Thomsom expression on angular momentum in
general relativity {\em Grav. Cosmol.} {\bf 7} 131 {\em (Preprint
gr-qc/0102091)}

\bibitem{Vibhadra1} Rosen N and Virbhadra K S 1993 Energy and momentum of
cylindrical gravitational waves {\em Gen. Relat. Grav.} {\bf 25} 429

\bibitem{Vibhadra1+}  Virbhadra K S 1995 Energy and momentum of
cylindrical gravitational waves. II {\em Pramana J. Physics} {\bf
45} 215 {\em (Preprint gr-qc/9509034)}

\bibitem{Favata01} Favata M 2001
Energy localization invariance of tidal work in general relativity
{\em Phys. Rev. D} {\bf 63} 064013 {\em (Preprint gr-qc/0008061)}

\bibitem{Vibhadra2} Aguirregabiria J M, Chammorro A and  Virbhadra K
S 1996 Energy and angular momentum of charged rotating black holes
{\em Gen. Relat. Grav.} {\bf 28} 1393 {\em (Preprint gr-qc/9501002)}

\bibitem{Vibhadra3} Virbhadra K S 1999 Naked singularities and
Seifert's conjecture {\em Phys. Rev. D} {\bf 60} 104041 {\em
(Preprint gr-qc/9809077 )}

\bibitem{Radinschi1}  Radinschi I 2000
Energy of a conformal scalar dyon black hole {\em Mod. Phys. Lett.
A} {\bf 15} 2171 {\em (Preprint gr-qc/0010094)}

\bibitem{Radinschi2}  Radinschi I 2000
The energy distribution of the Bianchi type I universe {\em Acta.
Phys. Slov.} {\bf 50}  609 {\em (Preprint gr-qc/0008034)}

\bibitem{Radinschi3}  Radinschi I 2001
Energy distribution of a charged regular black hole {\em Mod. Phys.
Lett. A} {\bf 16}, 673 {\em (Preprint gr-qc/0011066)}

\bibitem{Radinschi4}  Radinschi I 2005
On the Moller energy-momentum complex of the Melvin magnetic
universe {\em Fizika B} {\bf 14} 311 {\em (Preprint gr-qc/0202075)}

\bibitem{Xulu1}  Xulu S S 2000
Moller energy for the Kerr-Newman metric {\em Mod. Phys. Lett. A}
{\bf 15} 1511 {\em (Preprint gr-qc/0010062)}

\bibitem{Xulu2}  Xulu S S 2003
Moller energy of the nonstatic spherically symmetric metrics {\em
Astrophys. Space  Sci.} {\bf 283} 23 {\em (Preprint gr-qc/0010068)}


\bibitem{Belinfante} Belinfante F J 1939
On the spin angular momentum and mesons {\em Physica} {\bf 6} 887

\bibitem{Szabados91}  Szabados L B 1991
Canonical pseudotensor Sparling's form and Noether currents {\em
Preprint:} {KFKI}-1991-29/B.

\bibitem{Berezin92} Berezin V T 1992
Phenomenological foundations for a theory of gravity {\em Teor. Mat.
Fiz.} {\bf 93} 154

\bibitem{Szabados92} Szabados L B 1992
On canonical pseudotensor Sparling's form and Noether currents {\em
Class. Quantum Grav.}
 {\bf 9} 2521

\bibitem{Borokhov} Borokhov V 2002
Belinfante tensors induced by matter-gravity couplings {\em Phys.
Rev. D} {\bf 65} 125022 {\em (Preprint hep-th/0201043)}


\bibitem{Traschen} Traschen J 1985
Constraints on stress-energy perturbations in general relativity
{\em Phys. Rev. D} {\bf 31} 283

\bibitem{TraschenEardley} Traschen J and  Eardley D M 1986
Large-scale anisotropy of the cosmic background radiation in
Friedmann universes {\em Phys. Rev. D} {\bf 34} 1665

\bibitem{Stebbin} Veeraraghavan S and Stebbin A 1990
Causal compensated perturbations in cosmology {\em Ap. J.} {\bf 365}
37
\bibitem{UzanDerTurok}  Uzan J P,  Deruelle N and Turok N 1998
Conservation laws and cosmological perturbations in curved universes
{\em Phys. Rev. D} {\bf 57} 7192 {\em (Preprint gr-qc/9805020)}

\bibitem {KBL} Katz J, Bi\v c\'ak J and  Lynden-Bell D 1997
Relativistic conservation laws and integral constraints for large
cosmological perturbations {\em Phys. Rev. D} {\bf 55} 5957 {\em
(Preprint gr-qc/0504041)}

\bibitem {Chrusciel85}  Chru\'sciel P T 1985
On the relation between the Einstein and the Komar expressions for
the energy of the gravitational field {\em Ann. Inst. Henri
Poincar\'e} {\bf 42} 267

\bibitem{Katz-LB-Israel} Katz J,  Lynden-Bell D and
Israel W  1988 Quasilocal energy in static gravitational fields {\em
Class. Quantum Grav.} {\bf 5} 971

\bibitem{KatzOri} Katz J and  Ori A 1990 Localization of field
energy {\em Class. Quantum Grav.} {\bf 7} 787

 \bibitem{KatzLerer} Katz J and  Lerer D 1997
 On global conservation laws at null infinity
{\em Class. Quantum Grav.} {\bf 14} 2297 {\em (Preprint
gr-qc/9612025)}

 \bibitem{DerKatzUzan} Deruelle N,  Katz J and  Uzan J P 1997
 Integral constraints on cosmological perturbations and their energy
 {\em Class. Quantum Grav.} {\bf 14} 421
 {\em (Preprint gr-qc/9608046)}

\bibitem{DerKatzOgushi}  Deruelle N, Katz J and Ogushi S 2004
Conserved charges in Einstein Gauss-Bonnet theory {\it Class.
Quantum Grav.} {\bf 21} 1971 {\em (Preprint gr-qc/0310098)}

\bibitem {JuliaSilva98} Julia B and  Silva S 1998
Currents and superpotentials in classical invariant theories. I.
Local results with applications to perfect fluids and general
relativity  {\em Class. Quantum Grav.} {\bf 15} 2173 {\em (Preprint
gr-qc/9804029)}

 \bibitem {Silva99}  Silva S 1999
 On superpotentials and charge algebras of gauge theories
{\em Nucl. Phys. B} {\bf 558}, 391 {\em (Preprint hep-th/9809109)}

\bibitem{BMS} Bondi H,  Metzner A W K and Van der Berg M J C 1962
Gravitational waves in general relativity. VII. Waves from
axi-symmetrical isolated systems  {\em Proc. R. Soc. A London} {\bf
269} 21

\bibitem{Soloviev} Soloviev V O 1985 The generator algebra
of asymptotic Poincr\'e group in general relativity {\em Teor. Mat.
Fiz.} {\bf 65} 400

\bibitem{Bartnic} Bartnic R 1986
The mass of an asymptotically flat manifold {\em Commun. Pure Appl.
Math.} {\bf 39} 661

\bibitem{OMurchadha86} \'OMurchadha N 1986
Total energy-momentum in general realtivity {\em J. Math. Phys.}
{\bf 27} 2111

\bibitem{Chrusciel87} Chru\'sciel P T 1987
On angular momentum at spatial infinity {\em Class. Quantum Grav.}
{\bf 4} L205


\bibitem{ReggeTeitelboim} Regge T and Teitelboim T 1974 Role of
surface integrals in the Hamiltonian formulation {\em Ann. Phys.}
{\bf 88} 286


\bibitem{BOM} Beig R and  \'OMurchadha N 1987
The Poincar\'e group as the symmetry group of canonical general
relativity {\em Ann. Phys.} {\bf 174} 463


\bibitem{Szabados03} Szabados L B 2003
On the roots of the Poincar\'e structure of asymptotically flat
spacetimes {\em Class. Quantum Grav.} {\bf 20} 2627

\bibitem{NesterHoChen}  Nester J M, Ho  F-H and
Chen C-M  2003 Quasilocal center-of-mass for teleparallel gravity
{\em In the proceedings of the 10th Marcel Grossman meeting} (Rio de
Janeiro, 2003) {\em (Preprint gr-qc/0403101)}


\bibitem{NesterMengChen}  Nester J M, Meng F-F and
Chen C-M  2004 Quasilocal center-of-mass {\em J. Korean Phys. Soc.}
{\bf 45}  S22 {\em (Preprint gr-qc/0403103)}

\bibitem{AshtekarHansen} Ashtekar A and Hansen R O 1978
A unified treatment of null and spatial infinity in general
relativity. I. Universal structure of asymptotic symmetries and
conserved quantities at spatial infinity {\em  J. Math. Phys.} {\bf
19} 1542

\bibitem{JVWitten} Joshi P S,  Vaz C and Witten L 2004 A
time-like naked singularity {\em Phys. Rev. D} {\bf 70} 084038 {\em
(Preprint gr-qc/0410041)}


\bibitem{Fiziev} Fiziev P The gravitational field
of massive non-charged point source in general relativity {\em
(Preprint gr-qc/0412131)}

\bibitem{GJoshi} Goswami R and Joshi P S A resolution of spacetime
singularity and black hole paradoxes through avoidance of trapped
surface formation in Einstein gravity {\em (Preprint gr-qc/0504019)}

\bibitem{GolubevKelner} Golubev M B and Kelner S R 2005 Point charge
self-energy in the general relativity {\em Int. J. Mod. Phys. A}
{\bf 20} 2288 {\em (Preprint gr-qc/0504097)}

\bibitem{Narlikar2} Narlikar J V 1985 Some conceptual problems in general
relativity and cosmology {\em In A Random Walk in Relativity and
Cosmology} eds N Dadhich et al. (Viley Eastern Limited, New Delhi,
1985) 171

\bibitem{Damour-1} Damour T Jaranowski P and
Sch\"afer G 2001 Dimensional regularization of the gravitational
interaction of point masses  {\em Phys. Lett. B} {\bf 513} 147 {\em
(Preprint gr-qc/0105038)}

\bibitem{Damour-4}Schaefer G 2003 Binary black holes and gravitational wave production:
Post-newtonian analytic treatment {\it Current Trends in
Relativistic Astrophysics},  eds: Fernendez-Jambrina L and
Gonzolez-Romero L M (Springer's {\em Lecture Notes in Physics}, Vol.
617 2003) 195

\bibitem{GSh} Gelfand I M and Shilov G E 1964 {\em Generalized
functions. Vol. 1. Properties and Operations} (Academic Press, New
York)

\bibitem{Bertschinger96} Bertschinger E 1996 Cosmological dynamics
{\em In Cosmology and Large Scale Structures} eds Schaefer R, Silk
J, Spiro M and zin-Justin J  (North Holland, Amsterdam) 273

\bibitem{FultonRW} Fulton Y, Rohrlich F and and Witten L 1976
Conformal invariance in physics {\em Rev. Mod. Phys.} {\bf 34} 442

\bibitem{PenroseRindler} Penrose R and  Rindler W 1988
{\em Spinors and Space-Time: Spinor and Twistor Methods in
Space-Time Geometry} (Cambridge Univ. Press, Cambridge)

\bibitem{KeaneBarrett} Keane A J and  Barrett R K 2000
{\em Class. Quantum Grav.}  {\bf 17} 201 {\em (Preprint
gr-qc/9907002)}

\bibitem{branesW2} Bronnikov K A, Grebeniuk M A,
Ivashchuk V D and  Melnikov V N 1997 Integrable multidimensional
cosmology for intersecting $p$-branes {\em Grav. Cosmol.} {\bf 3}
105 {\em (Preprint gr-qc/9709006)}

\bibitem{branesW3} Grebeniuk M A, Ivashchuk V D and Melnikov V N
1998 Multidimensional cosmology for intersecting p-branes with
static internal spaces {\em  Grav. Cosmol.} {\bf 4} 145  {\em
(Preprint gr-qc/9804042)}

\bibitem{branesW8a} Ivashchuk V D and Melnikov V N 2000
Billiard representation for multidimensional cosmology with
intersecting p-branes near the singularity {\em  J. Math. Phys.}
{\bf 41} 6341 {\em (Preprint hep-th/9904077)}

\bibitem{branesW8} Bronnikov K A, Dehnen H and Melnikov V N 2003
On a general class of brane-world black holes {\em Phys. Rev. D},
{\bf 68} 024025 {\em (Preprint gr-qc/0304068)}

\bibitem{IgorVlasov} Kopeikin S and Vlasov I 2004
Parametrized post-Newtonian theory of reference frames, multipolar
expansions and equations of motion in the N-body problem {\em Phys.
Rep.} {\bf 400} 209 {\em (Preprint gr-qc/0403068)}

\bibitem{DT1} Deser S and Tekin B 2002 Gravitational energy in
quadratic curvature gravities {\it Phys. Rev. Lett.} {\bf 89} 101101
{\em (Preprint hep-th/0205318)}

\bibitem{DT3} Deser S and Tekin B 2003 Energy in topologically
massive gravity {\it Class. Quantum Grav.} {\bf 20} L259 {\em
(Preprint gr-qc/0307073)}

\bibitem{BD+}
Boulware D C and Deser S 1985 String-generated gravity models {\it
Phys. Rev. Lett.} {\bf 55} 2656

\bibitem{Paddila} Paddila A 2003 Surface terms and
Gauss-Bonnet Hamiltonian {\it Class. Quantum Grav.} {\bf 20} 3129
({\em Preprint} gr-qc/0303082)

\bibitem{JacobsonMyers} Jacobson T and Myers R C 1993 Entropy of Lovelock
black holes {\em Phys. Rev. Lett.} {\bf 70} 3684 {\em (Preprint
hep-th/9305016)}

\bibitem{BanadosTZ} Banados M, Teitelboim C and Zanelli J 1993 Black hole
entropy and the dimensional continuation of the Gauss-Bonnet theorem
{\em (Preprint gr-qc/9309026)}

\bibitem{LoukoSW-H} Louko J, Simon J Z and Winter-Hilt S N 1996
Hamiltonian thermodynamics of a Lovelock black hole {\em (Preprint
gr-qc/9610071)}

\bibitem{NojiriOO} Nojiri S, Odintsov S D and Ogushi S 2002
Cosmological and black hole brane-world Universes in higher
derivative gravity {\em Phys. Rev. D} {\bf 65} 023521 {\em (Preprint
hep-th/0108172)}

\bibitem{Cai} Cai R-G 2002  Gauss-Bonnet black holes in AdS spaces
 {\it Phys. Rev. D} {\bf 65} 084014
{\em (Preprint hep-th/01092133)}

\bibitem{CNojiriO} Cvetic M, Nojiri S and Odintzov S D 2002 Black hole
thermodynamics and negative entropy in de Sitter and anti-de Sitter
Einstein-Gauss-Bonnet gravity {\it Nucl. Phys. B} {\bf 628} 295 {\em
(Preprint hep-th/0112045)}

\bibitem{ChoN} Cho Y M and Neupane I P 2002 Anti-de Sitter black
holes, thermal phase trnsition and holography in higher curvature
gravity {\it Phys. Rev. D} {\bf 66} 024044 {\em (Preprint
hep-th/0202140)}

\bibitem{AFrancavigliaR} Allemandi G, Francaviglia M and Raiteri M
Charges and energy in Chern-Simons theories and Lovelock gravity
{\em (Preprint gr-qc/0308019)}

\bibitem{Okuyama} Okuyama N and Koga J-I 2005 Asymptotically anti-de
Sitter spacetimes and conserved quantities in higher curvature
gravitational theories {\em Phys. Rev. D} {\bf 71}  084009 {\em
(Preprint hep-th/0501044)}

\bibitem{DerMor05}  Deruelle N and Morisawa 2005  Mass and angular
momenta of Kerr anti-de Sitter spacetimes in Einstein-Gauss-Bonnet
theory {\it Class. Quantum Grav.} {\bf 22} 933 ({\em Preprint}
gr-qc/0411135)

\bibitem{Der05} Deruelle N 2005 Mass and angular momentum of a
Kerr-anti-de Sitter spacetimes ({\em Preprint} gr-qc/0502072)

\bibitem{GLPPope}  Gibbons G W, Lu H, Page D N and Pope C N 2005
The General Kerr-de Sitter Metrics in All Dimensions {\em J. Geom.
Phys.} {bf 53} 49 {\em (Preprint hep-th/0404008)}

\bibitem{GPPope} Gibbons G W, Perry M J and Pope C N 2005
The first law of thermodynamics for Kerr-anti-de Sitter black holes
{\em Class. Quantum Grav.} {\bf 22} 1503 {\em (Preprint
hep-th/0408217)}

\bibitem{DerKatz05} Deruelle N and   Katz J 2005 On the mass of a
Kerr-anti-de Sitter spacetime in $D$ dimensions {\it Class. Quantum
Grav.} {\bf 22} 421 ({\em Preprint} gr-qc/0411035)

\bibitem{KEGB1} Alexeyev S, Popov N, Startseva M, Barrau and Grain J
2007 Kerr-Gauss-Bonnet black holes: Exact analitical solution ({\em
Preprint} arXiv:0712.3546[gr-qc])

\bibitem{KEGB2} Brihaye Y and Radu E 2008 Five-dimensional rotating black
holes in Einstein-Gauss-Bonnet theory ({\em Preprint}
arXiv:0801.1021[hep-th])

\bibitem{Aliev}  Aliev A N 2006 A slowly rotating
charged black hole in five dimensions {\em Mod. Phys. Lett. A} {\bf
21} 751 ({\em Preprint} gr-qc/0505003)

\bibitem{Aliev2}  Aliev A N 2007 Electromagnetic Properties of
Kerr-Anti-de Sitter Black Holes
({\em Preprint} hep-th/0702129)


\end{thebibliography}
\end{document}